\documentclass[10pt,aps,prx,twocolumn,english,superscriptaddress,floatfix,longbibliography]{revtex4-2}
\usepackage[colorlinks=true, allcolors=blue]{hyperref}
\usepackage{braket}
\usepackage[T1]{fontenc}
\usepackage{graphicx,subfigure}
\usepackage{amsmath,amssymb,mathrsfs,bbm,amsthm}
\usepackage[normalem]{ulem}
\usepackage{tikz}
\usetikzlibrary{quantikz2}
\usepackage{enumerate}
\usepackage{enumitem}

\providecommand{\EE}{\mathcal{E}}

\providecommand{\LL}{\mathcal{L}}

\providecommand{\II}{\mathcal{I}}
\providecommand{\HH}{\mathcal{H}}

\providecommand{\UU}{\mathcal{U}}

\providecommand{\tr}{\operatorname{tr}}

\providecommand{\Tb}{\mathbf{T}}
\providecommand{\Cb}{\mathbf{C}}
\providecommand{\Fbb}{\mathbb{F}}

\providecommand{\xtilde}{\widetilde{x}}
\providecommand{\ptilde}{\widetilde{p}}
\providecommand{\Ntilde}{\widetilde{N}}

\providecommand{\T}{\mathsf{T}}

\providecommand{\Hgate}{\operatorname{H}}
\providecommand{\Pgate}{\operatorname{P}}
\providecommand{\CZgate}{\mathrm{CZ}}
\providecommand{\CXgate}{\mathrm{CX}}
\providecommand{\Rb}{\mathrm{Rb}}
\providecommand{\Cs}{\mathrm{Cs}}
\providecommand{\measset}{\mathcal{M}}

\providecommand{\Tpre}{T_{\mathrm{pre}}}
\providecommand{\TM}{T_{\mathrm{M}}}
\providecommand{\tLS}{t_{\mathrm{LS}}}
\providecommand{\tCZ}{t_{\mathrm{CZ}}}
\providecommand{\toneq}{t_{1q}}

\begin{document}

\title{Efficient Entanglement Purification Circuit Design for Dual-Species Atom Arrays}

\author{Bikun Li}
\email{bikunli@uchicago.edu}
\affiliation{Pritzker School of Molecular Engineering, University of Chicago, Chicago, Illinois 60637, USA}

\author{Daniel Dilley}
\affiliation{Mathematics and Computer Science Division, Argonne National Laboratory, Lemont, Illinois 60439, USA}

\author{Alvin Gonzales}
\affiliation{Mathematics and Computer Science Division, Argonne National Laboratory, Lemont, Illinois 60439, USA}

\author{Thomas A. Hahn}
\affiliation{The Center for Quantum Science and Technology, Department of Complex Systems, Weizmann Institute of Science, Rehovot, Israel}

\author{Ryan White}
\affiliation{Department of Physics, University of Chicago, Chicago, Illinois 60637, USA}

\author{Rotem Arnon}
\affiliation{The Center for Quantum Science and Technology, Department of Complex Systems, Weizmann Institute of Science, Rehovot, Israel}

\author{Hannes Bernien}
\affiliation{Pritzker School of Molecular Engineering, University of Chicago, Chicago, Illinois 60637, USA}

\author{Zain Saleem}
\affiliation{Mathematics and Computer Science Division, Argonne National Laboratory, Lemont, Illinois 60439, USA}

\author{Liang Jiang}
\email{liangjiang@uchicago.edu}
\affiliation{Pritzker School of Molecular Engineering, University of Chicago, Chicago, Illinois 60637, USA}

\begin{abstract}
Entanglement purification protocols (EPPs) are essential for generating high-fidelity entangled states in noisy quantum systems, enabling robust quantum networking and computation. Building on the circuit of the foundational recurrence protocol, we generalize two-way EPPs to arbitrary stabilizer codes. Through analytical derivations and noisy circuit simulations incorporating circuit-level noise, we demonstrate enhanced purification performance, with fidelity improvements and finite distillation rates for distillable input states. We propose efficient circuit designs for EPPs tailored to dual-species Rydberg atom arrays, leveraging species-specific laser control and interspecies Rydberg interactions. Introducing a low-overhead operation set, the dual-species atom convenient operation set, we facilitate straightforward compilation of EPP circuits without the need for ancillary atoms or complex atom rearrangements. Our framework provides practical guidance for near-term implementations on dual-species platforms, advancing towards scalable entanglement distribution in neutral atom systems and paving the way for fault-tolerant quantum technologies.
\end{abstract}

\maketitle

\section{Introduction}
Recent advances in physics and engineering have significantly accelerated the development of quantum computing technologies, fostering innovation across diverse quantum architectures. Various quantum computing platforms have emerged, each characterized by distinct features and operational strategies~\cite{Bruzewicz_2019, Iqbal_2024, Siddiqi_2021, Giustina_2023, Henriet_2020, Graham_2022}. These platforms push the boundaries of quantum computation and communication, yet significant challenges remain, including noise, scalability limitations, and complexities in integrating quantum systems. To fully harness quantum advantages, particularly in quantum networks spanning long distances, maintaining high-fidelity entanglement is critical. Although numerous experiments have successfully demonstrated entanglement distribution across various quantum hardware platforms~\cite{Tamara2021Science, Covey2023, Saha2025}, inevitable decoherence and operational imperfections degrade entangled states, limiting practical performance.

Entanglement purification protocols (EPPs), first introduced by Bennett \textit{et al.}~\cite{bennett_1996_PRL}, provide insight into extracting fewer high-fidelity entangled states from multiple noisy entangled pairs. Due to the complexity of local operations and classical communication (LOCC), distilling entanglement from noisy ensembles has been a compelling theoretical topic~\cite{Horodecki1998PRL, bennett_1996_PRA, Watrous2004PRL_distillation, PhysRevA.97.062333, DistillRank4_2016, Chitambar_2020, Lami2024, Ataides2025}. Experimentally, simplified EPPs have been demonstrated on various quantum platforms, including photonic qubits~\cite{Pan2003, Hu2021PRL_photonicEPP, yu2025_integratedphotonics_EPP, Zhou2025PRL}, superconducting qubits~\cite{Yan2022PRL}, trapped atoms and ions~\cite{Reichle2006}, and solid-state spins~\cite{Kalb_2017_science}. These experiments confirm the feasibility of EPPs while highlighting the need to tailor protocols to mitigate platform-specific noise sources, such as photon loss in optics or thermal fluctuations in superconductors.

Neutral atom array platforms have recently emerged as promising candidates for quantum protocols due to their inherent advantages, including excellent scalability, high gate fidelity, long coherence times, and flexible configurations~\cite{Graham_2022}. These platforms have been shown to be programmable for demonstrating quantum error correction~\cite{Xu2024, Bluvstein_2022}, which is closely related to entanglement purification. Experimental realizations of single-species neutral atom arrays have made significant progress, demonstrating reliable entanglement generation via the Rydberg blockade mechanism~\cite{Lukin2001Blockade,Tong_2004, Wilk2010PRL, Evered2023}. However, further optimization and generalization of these protocols are necessary to fully exploit their potential in quantum networking.

Dual-species neutral atom platforms, utilizing two species of Rydberg atoms (e.g., Rb and Cs), offer additional advantages, such as independent control of distinct atomic species and stronger Rydberg blockade through resonant dipole-dipole interactions~\cite{Singh2022PRX, anand2024dualspecies,Beterov2015PRA}. These features enable more efficient applications to various promising tasks proposed for single-species neutral atom platforms, including solving combinatorial optimization problems such as the maximum independent set~\cite{Ebadi_2022} and quadratic unconstrained binary optimization challenges~\cite{Nguyen_2023}. Moreover, dual-species Rydberg atom arrays naturally support native multi-qubit gates~\cite{Evered2023, Quera_2024}, providing significant flexibility for quantum circuit designs adaptable to various qubit connectivities~\cite{Labuhn_2016, Singh2022PRX, Sheng_2022}.

Rydberg atom arrays typically have slower gate operations compared to most quantum computing platforms (except for trapped-ion platforms) and mid-circuit atom losses from traps. However, by leveraging the unique advantages of dual-species systems, we may create a compact quantum circuit based on global control, which is feasible and favorable in most experiments. For single-species platforms, it has been demonstrated that the constraint of global control still allows flexibility in quantum simulation~\cite{hu2025}. Therefore, we expect that a more efficient quantum circuit for practical tasks can be implemented when we have additional species of atoms.

In this work, we focus on implementing EPPs within dual-species Rydberg atom arrays. We design protocols using a low-overhead set of convenient operations to ensure efficient atom manipulation. Within this constrained architecture, we propose a circuit compilation scheme for EPPs based on arbitrary stabilizer codes, generalizing the original approach by Bennett \textit{et al.}~\cite{bennett_1996_PRL}. Through detailed numerical simulations of noisy circuits, we demonstrate the robustness of the generalized circuit in terms of purification performance. Additionally, we discuss distillation rates and circuit optimizations. Our study provides clear guidance for practical implementations of entanglement purification on dual-species Rydberg platforms, paving the way for robust and scalable quantum networking and fault-tolerant quantum computing under experimentally realistic conditions.

This paper is organized as follows. Section~\ref{sec:background} reviews the concept of entanglement purification and its relationship with quantum error correction codes, while introducing the necessary components of dual-species atom arrays and the convenient operation set. Section~\ref{sec:2EPP} presents the two-way EPP as a generalization of the original recurrence method. Section~\ref{sec:1EPP} demonstrates that a similar circuit pattern applies to general stabilizer codes, validated through noisy circuit simulations. Finally, Section~\ref{sec:Discussion} discusses the universality of our framework and explores circuit optimization strategies.

\section{Background}\label{sec:background}

\subsection{Entanglement Purification}

Entanglement purification is a key technique in quantum information theory designed to counteract the degradation of entanglement caused by noise and imperfections in quantum channels. In quantum communication and computation, entangled states are essential resources that enable protocols such as quantum teleportation, superdense coding, and quantum cryptography. However, during transmission or storage, entangled particles are susceptible to environmental disturbances, which reduce their fidelity and yield mixed states with diminished utility.

The core concept of entanglement purification involves starting with a large ensemble of weakly entangled bipartite states and extracting a smaller number of high-fidelity entangled pairs through LOCC. Protocols that succeed probabilistically are known as stochastic LOCC protocols. The constraint of LOCC is particularly valuable in practical scenarios where quantum gates between distant parties are not feasible, as it enhances the quality of entanglement using pre-shared quantum resources and cheap classical communications.

A seminal contribution in this field was made by Bennett \textit{et al.}~\cite{bennett_1996_PRL}, which introduced a simple stochastic LOCC protocol that converts two noisy entangled pairs into one pair with higher fidelity. This EPP is referred to as the $2 \to 1$ recurrence method.  In the low-noise regime, the recurrence method reduces the infidelity by a factor of approximately $2/3$. Additionally, they proposed a one-way deterministic $n \to k$ purification scheme, known as the hashing protocol, which asymptotically distills a finite fraction of nearly perfect Bell states from the noisy inputs, if the number of input states goes to infinity. Remarkably, the combination of the above protocols gives a finite asymptotic yield whenever the input Werner state is entangled.

While the hashing protocol traditionally resembles a classical data structure technique, it can be fundamentally understood as a random stabilizer code, a type of quantum error-correcting code (QECC).
The hashing protocol identifies errors by a primitive random circuit. A major reason preventing the practical use of this protocol is the NP-hardness of decoding, that is, inferring the input error from the classical bit readout~\cite{Iyer_Poulin_2015}. Replacing the random stabilizer code with a carefully conceived QECC can solve this problem if it is equipped with efficient decoders.
Actually, from a higher point of view, EPPs are closely related to quantum error-correcting codes, as both of them aim to remove noise from valuable quantum coherence. 
The former concentrates the entanglement from many noisy copies of entangled pairs to a subset of these copies, while the latter recovers an unknown logical quantum state from a larger noisy quantum system with redundancy. 
Significantly, given a circuit of a specific QECC, such as a qubit stabilizer code, one can derive an entanglement purification protocol, which may use one-way or two-way classical communication to either correct or detect errors. 
This work will mainly focus on the case of qubit stabilizer codes, as it fits the atomic qubits in the atom array.

\subsection{Stabilizer Codes in QECC}
Stabilizer codes are a well-studied family of quantum error-correcting codes due to their versatility and similarity to classical error-correcting codes~\cite{Steane_1996, Calderbank_1996, Gottesman_1997}. 
Here, we briefly introduce the concept of stabilizer codes for qubits, as they are the only relevant codes in this work.

Qubit stabilizer codes are a fundamental class of quantum error-correcting codes used to protect quantum information in quantum computing systems. These codes encode logical qubits into a larger number of physical qubits, creating a protected subspace $\LL$ defined by a set of commuting Pauli operators called stabilizer generators. Given the Pauli matrices $X$, $Y$, and $Z$ acting on the $i$-th qubit, the $n$-qubit Pauli group is defined as $\mathcal{P}_n=\langle \{X,Z\}\rangle^{\otimes n}$.
A set of independent stabilizer generators $\{\hat{S}_i\}_{i=1}^{r_S} \subset \mathcal{P}_n$ consists of mutually commuting Pauli operators, $[\hat{S}_i,\hat{S}_j]=0$, such that, for $a_\ell\in\Fbb_2$, $\prod_{\ell=1}^{r_S}\hat{S}_\ell^{a_\ell}=I$ implies $a_\ell=0$ for every $\ell$. Stabilizers generate the stabilizer group by $\mathcal{S} = \langle \{\hat{S}_i\}_{i=1}^{r_S} \rangle \subset \mathcal{P}_n$, which is required to fulfill $-I\notin \mathcal{S}$. With the above definitions, $\mathcal{S}$ determines the logical subspace $\LL = \mathbb{C}^{2^k}$ with $k = n-r_S$, which consists of all logical quantum states $\ket{\psi}$ that satisfy $\hat{S}_i \ket{\psi} = +\ket{\psi}$.
Operators acting only within $\LL$ are known as logical operators. Let $N_{\mathcal{P}_n}(\mathcal{S})$ be the normalizer of $\mathcal{S}$ in $\mathcal{P}_n$; then the set of non-identity logical Pauli operators is $N_{\mathcal{P}_n}(\mathcal{S})\setminus \mathcal{S}$.

In classical parity-check codes, parity checks are used to infer errors in the data. The stabilizer operator $\hat{S}_i$ plays the role of the parity check in the quantum setting, and the logical subspace $\LL$ is the quantum analog of the codespace. 
In the qubit case, the stabilizer $\hat{S}_i$ is an observable, whose measurement will not disturb $\LL$. The stabilizer measurement outcome $s_i \in \{0, 1\}$ is known as the \textit{syndrome}, which is used to infer the Pauli error acting on the code under a given noise model. A well-designed $\mathcal{S}$ allows one to distinguish the dominant errors acting on the code and thereby counteract their effects to restore the quantum state.

Although the Hilbert space of $n$ qubits has an exponentially large dimensionality, large stabilizer codes can be studied on a classical computer due to the Gottesman-Knill theorem~\cite{Aaronson2004}. The insight here is that the stabilizer formalism is equivalent to (classical) additive self-orthogonal codes over $\Fbb_4$~\cite{calderbank1997}.
As in classical additive codes, the capability of detecting errors is given by the code distance $d$, which is equal to the minimum weight of a non-trivial logical Pauli operator in the stabilizer code. Here, the weight $w$ of a Pauli operator means it acts on $w$ qubits with non-identity operations.
We say a stabilizer code is an $[[n, k, d]]$ code if it encodes $k$ logical qubits in an $n$-qubit system and has a code distance $d$.
Any Pauli error causing non-zero syndromes is \textit{detectable}. Nevertheless, detecting an error does not certify that the details of the error are fully learnable and hence \textit{correctable}. 
An $[[n, k, d]]$ code guarantees the correction of any Pauli error of weight less than or equal to $t = \lfloor (d-1)/2 \rfloor$. So a distance $3$ code allows us to correct any single-qubit Pauli error.
We should also highlight that the parameter $t$ only characterizes the worst case. Typically, some correctable errors may have weights greater than $t$. This is similar to decoding a classical linear block code, where a good decoder is required to identify such large weight errors~\cite{mackay2003information}. 

Due to their deep connection with classical coding theory, significant efforts have been invested in studying qubit stabilizer codes in recent decades. 
Researchers are searching for quantum codes with better coding rates and code distances, such as through the exploration of quantum low-density parity-check codes~\cite{Breuckmann_Eberhardt_PRX_2021, DHLV_QLDPC_2022, PKCode_2022, QTannerCode_2022}.
Since quantum circuits are notoriously noisy, ensuring a QECC corrects more errors than it introduces is essential~\cite{gottesman2009introductionquantumerrorcorrection}. Such a discussion, falling into the category of fault-tolerant QECC, is beyond the scope of this paper.

\begin{figure}[t]
    \centering
    \includegraphics[width=0.99\linewidth]{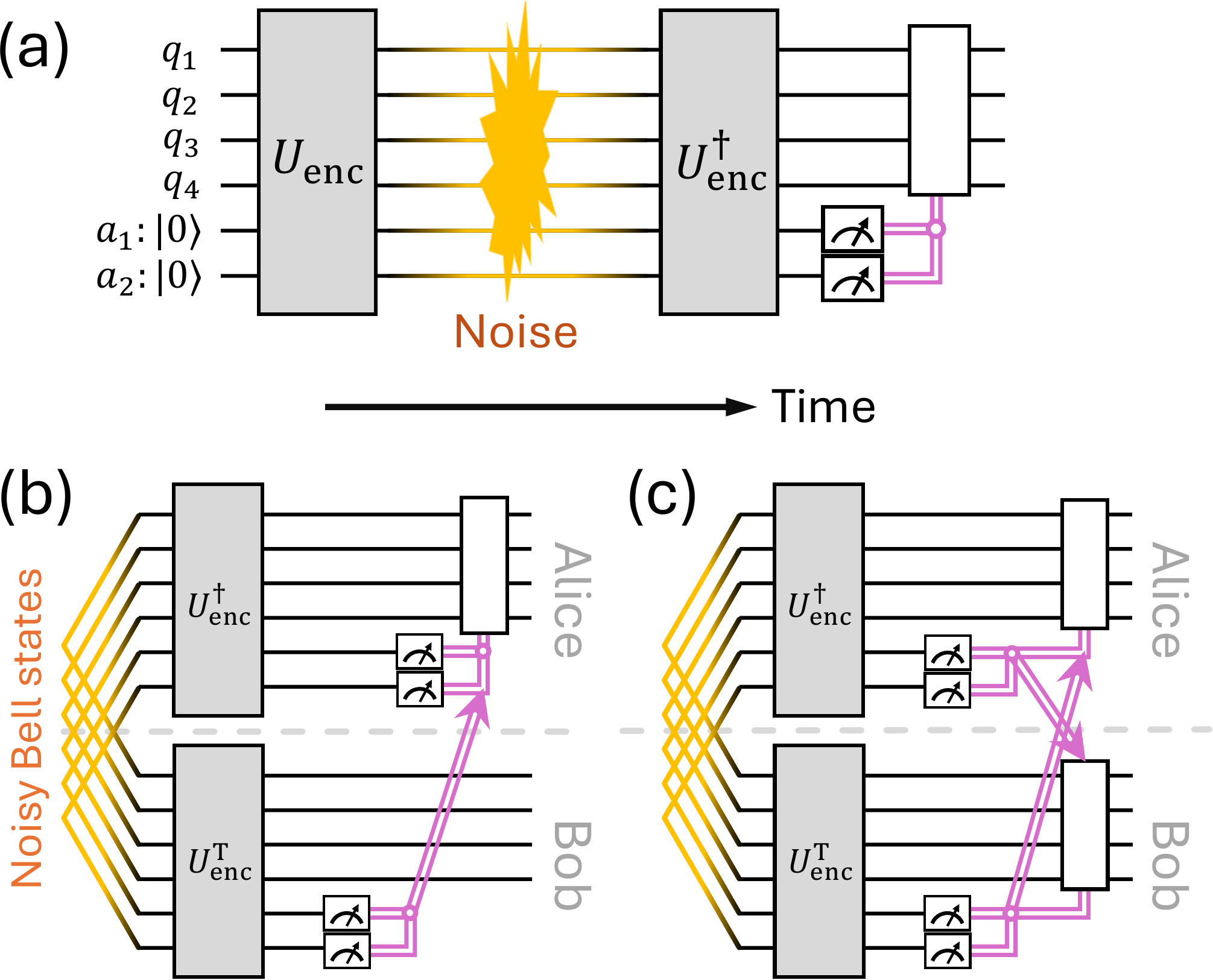}
    \caption{
    This figure illustrates the correspondence between the circuits of QECC and EPP. (a) A typical quantum circuit depicting the encoding and decoding processes for a QECC. (b, c) Circuits for one-way and two-way EPP, respectively. The white box without a label denotes local quantum operations conditioned on classical communication (indicated by purple lines and arrows). The circuits in (b) and (c) are derived by folding the circuit in (a), with minor modifications. The yellow V-shaped polyline represents a noisy initial Bell state $\Phi_+$.  
    }
    \label{fig:qecandepp}
\end{figure}

\subsection{Mapping QECC to EPP}

QECC and EPPs are closely related, and QECC can be used to construct EPPs \cite{bennett_1996_PRA, Dur_2007}. In this section, we discuss how quantum error correction corresponds to one-way and two-way EPP \cite{bennett_1996_PRA, Hostens_2004, Dur_2007, Zang_2024}. Briefly speaking, a QECC circuit that corrects errors can usually be transformed into a one-way EPP, whereas a QECC circuit that detects errors typically corresponds to a two-way EPP. 

We demonstrate this by ignoring any high-level aspects of quantum circuit fault tolerance and briefly reviewing the process of quantum error correction using an $[[n,k,d]]$ qubit stabilizer code. Typically, the encoding circuit of concern here maps the $k$-qubit input (represented by $q_i$ in Fig.~\ref{fig:qecandepp}) to an $n$-qubit output via a unitary operator $U_{\text{enc}}$, along with $n-k$ ancilla qubits (represented by $a_i$ in Fig.~\ref{fig:qecandepp}). For simplicity, all ancilla qubits are initialized in the state $\ket{0}$. After corruption by noise, a decoding unitary $U^{\dagger}_{\text{enc}}$ is applied to the system to reverse the encoding. This $U^{\dagger}_{\text{enc}}$ effectively transforms each stabilizer operator $\hat{S}_i$ into Pauli $Z$ operators acting on the ancilla qubits. By measuring these qubits in the computational basis ($Z$ basis), one obtains the syndrome set $\{s_i\}$ and thereby learns information about the error. 

Once the error is identified, it can either be corrected by reversing its effects or the current state can be discarded, followed by reinitialization of a clean state without error. Typically, the former approach is more valuable in the context of scalable quantum computation, whereas the latter is commonly employed in smaller protocols where the probability of no error occurring is reasonably high. 

The purpose of an EPP, specifically for a bipartite system, is to concentrate the entanglement from a large, noisy entangled state into a smaller entangled state, such that the resulting state is closer to a pure entangled state without noise. This process is particularly useful when the available gate sets are sufficiently noisy that they hinder the direct creation of distant entangled qubits. In fact, EPP is closely related to QEC. A QECC scheme can be transformed into an EPP by simply ``folding'' the encoding-decoding circuit, as shown in Fig.~\ref{fig:qecandepp}(b). This circuit folding can also be understood through the ``ricochet'' property of Bell state $\ket{\Phi_+}:=\frac{1}{\sqrt{2}}(\ket{00}+\ket{11})$. Let $\ket{\Omega}:=\ket{\Phi_+}^{\otimes n}$, and let $\Lambda$ be any linear map whose input and output are square operators acting on $n$ qubits. The ``ricochet'' trick then simply expresses the following relation:
\begin{equation}\label{eq:ricochet}
    \II\otimes \Lambda (\ket{\Omega}\!\bra{\Omega}) = \Lambda^\T  \otimes \II (\ket{\Omega}\!\bra{\Omega}),
\end{equation}
where $\II$ is the identity map. $\Lambda^\T $ is the linear map with transposed Kraus operators defined by $\langle A,  \Lambda \left(B\right)\rangle = \langle \Lambda^\T(A),  B\rangle$, where the inner product is defined by $\langle A, B \rangle:= \tr(A^\T B)$, and $A^\T$ is the transposition of the matrix $A$. 

In EPPs for Bell states, an initial ensemble of noisy entangled qubit pairs is shared between two distant parties, Alice and Bob. Each party applies unitary operations ($U_{\text{enc}}^\dagger$ and $U_{\text{enc}}^{\T}$, respectively) in their local laboratories and exchanges measurement outcomes via classical channels. Upon receiving these results, they compute syndromes $s_i = m_{A,i} \oplus m_{B,i}$ (where $\oplus$ denotes the XOR operation), analogous to quantum error correction (QEC) schemes. If the syndrome indicates a correctable error, they retain the output and apply local corrections to complete the protocol. Otherwise, they discard the states. When successful, the protocol extracts $k$ higher-fidelity output entangled pairs from $n$ input pairs.

The circuits in the gray boxes of Fig.~\ref{fig:qecandepp}(b,c) illustrate the purification of 4 higher-fidelity pairs from 6 noisy input pairs using one-way and two-way schemes, respectively. Alice and Bob apply the depicted unitaries and measure their lower two qubits in the $Z$ basis, yielding outcomes $m_{A,i}, m_{B,i} \in \{0,1\}$. An error is typically detected if $m_{A,i} \oplus m_{B,i} \neq 0$ for any $i$. They then infer the error from the syndrome and, if correctable, perform local operations on the remaining qubits to yield purified states. Otherwise, via two-way communication, they deem the protocol unsuccessful and restart.

\subsection{The dual-species atom model}
\begin{figure}[t]
    \centering
    \includegraphics[scale = 0.5]{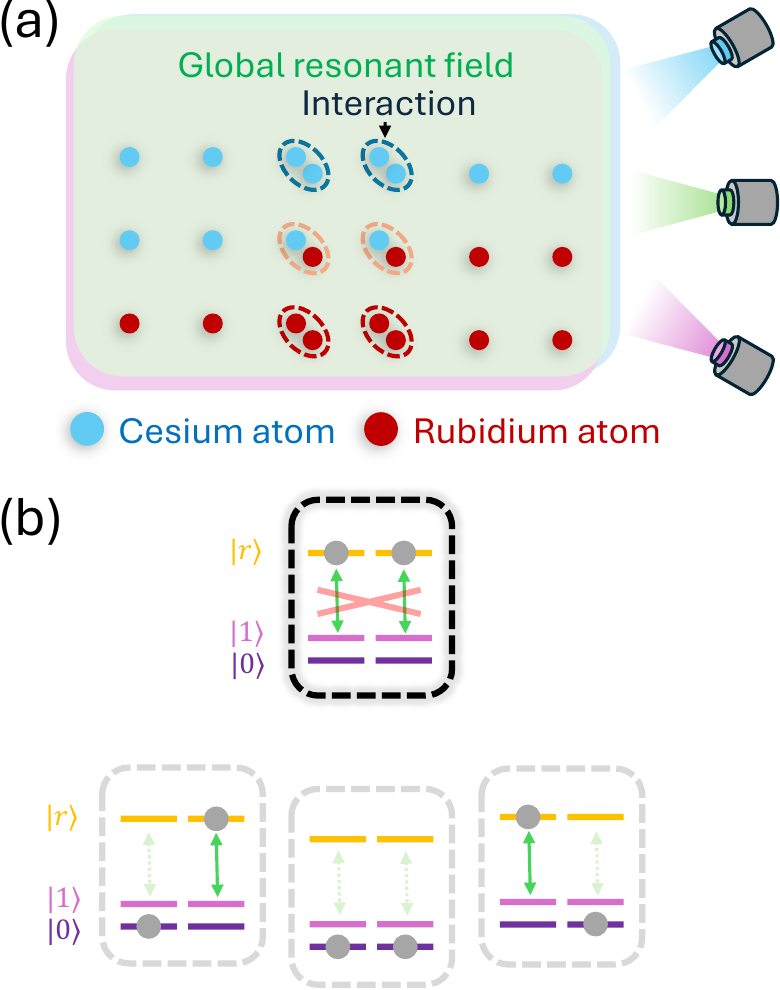}
    \caption{Panel (a) shows an array of Rydberg atoms comprising two species. The positions of the atoms can be manipulated using optical tweezers. All atoms stay in the same working zone, so that the internal qubit state is controlled via a global laser field. (b) The interaction between two qubits is mediated by the Rydberg blockade mechanism. Here, $\ket{0}$ and $\ket{1}$ represent two nearly degenerate ground states (e.g., hyperfine clock states), while $\ket{r}$ is a highly excited Rydberg state optically coupled to $\ket{1}$ by an external laser drive. Under Rydberg blockade, both atoms cannot be simultaneously excited to the $\ket{rr}$ state if they are initially prepared in $\ket{11}$. This scheme is also applicable to interactions between atoms of different species, where the energy spectra are distinct.}
    \label{Fig:interacting_atoms}
\end{figure}

While various quantum circuits with appealing performance have been proposed theoretically, their implementation on practical platforms can be particularly challenging or inconvenient, depending on the specific features of the devices.

This paper focuses on platforms utilizing two species of neutral atoms. However, we emphasize that our approach is also applicable to mathematically equivalent models, such as those involving two distinct Hilbert subspaces of a single atomic species.

A popular choice for these atomic species is rubidium (Rb) and cesium (Cs)~\cite{anand2024dualspecies}. In Fig.~\ref{Fig:interacting_atoms}, they are displayed as red dots and blue dots, respectively.

In this tweezer array, each qubit is encoded in the Hilbert subspace spanned by two states, denoted as $\ket{0}_{i,\alpha}$ and $\ket{1}_{i,\alpha}$, where the subscript $i$ indicates the $i$-th atom, and $\alpha\in\{\mathrm{Rb},\mathrm{Cs}\}$ refers to the atomic species. The corresponding Pauli operators are denoted as $I_{i,\alpha}$, $X_{i,\alpha}$, $Y_{i,\alpha}$, and $Z_{i,\alpha}$.
An additional highly excited state $\ket{r}_{i,\alpha}$ is utilized to manipulate the states of the atom; this state couples only to $\ket{1}_{i,\alpha}$ via a global resonant laser field.
In our notation, we may omit the subscripts without loss of clarity when it does not cause ambiguity.
A global coherent laser field is tuned to couple $\ket{1}_{i,\alpha}$ with the highly excited state $\ket{r}_{i,\alpha}$.

Since Rb and Cs atoms have \textit{distinct} spectra, properly designed global pulse sequences enable the performance of single-qubit unitaries and mid-circuit readout (by capturing fluorescence at a specific wavelength with a camera) on qubits of one atomic species without interfering with the other species~\cite{Singh_2023}.

The set of global single-qubit rotations is denoted as
\begin{equation}
    \UU_{\alpha}:=\bigg\{\bigotimes_i U_\alpha: \forall\, U_{\alpha} \in SU(2)\bigg\}.
\end{equation}
The set of Pauli $Z$ measurements, that is, collapsing all $\mathrm{Rb}/\mathrm{Cs}$ atoms into the computational basis, is denoted as
\begin{equation}
    \measset_{\alpha}:=\bigg\{\bigotimes_i \proj{x_i}_{i,\alpha}: x_i\in\{0,1\}\bigg\}.
\end{equation}
Besides species-wise global single-qubit gates and measurements, interactions between qubits are essential for performing nontrivial tasks. Various experiments have demonstrated the flexibility of relocating these atoms using optical tweezers~\cite{Fang_SciAdv_2025,Guttridge_PRL_2025}. The set of atomic relocation operations enabled by optical tweezers is denoted as $\mathcal{T}$.

By bringing two atoms close together within a certain interaction radius, van der Waals or dipole-dipole interactions can be harnessed to induce the Rydberg blockade. Specifically, the effective total Hamiltonian is 
\begin{equation}
    \HH = \sum_{\alpha} \HH^{\text{c}}_{\alpha} + \sum_{\alpha}\sum_{i<j}\HH^{\text{b}}_{(i,\alpha)(j,\alpha)}+ \sum_{i,j}\HH^{\text{b}}_{(i,\Cs)(j,\Rb)}\;,
\end{equation}
where the control term $\HH^{\text{c}}_\alpha$ and the Rydberg blockade term $\HH^{\text{b}}_{(i,\alpha)(j,\beta)}$ are defined as:
\begin{align*}
    &\HH^{\text{c}}_\alpha =  \sum_i \Omega_\alpha(t)\left(\ket{r}\!\bra{1}_{i,\alpha} + \text{h.c.}\right) + \Delta_\alpha(t) \proj{r}_{i,\alpha},\\
    &\HH^{\text{b}}_{(i,\alpha)(j,\beta)}=\lambda^{\text{b}}_{i,j,\alpha,\beta} \ket{r,r}\!\bra{r,r}_{(i,\alpha)(j,\beta)}.
\end{align*}

When two atoms are in close proximity, a strong coupling strength $\lambda^{\text{b}}_{i,j,\alpha,\beta}$ combines with a proper control pulse $\Omega_\alpha(t)$ and $\Delta_\alpha(t)$ to create the following phase differences in the computational basis of the two interacting qubits:
\begin{equation}
    \ket{00},\,e^{i\phi_1}\ket{01},\,e^{i\phi_1}\ket{10},\,e^{i\phi_2}\ket{11},
\end{equation}
whereas two non-interacting qubits acquire
\begin{equation}
    \ket{00},\,e^{i\phi_1}\ket{01},\,e^{i\phi_1}\ket{10},\,e^{2i\phi_1}\ket{11}.
\end{equation}

For both same-species and different-species cases, $\phi_1 = \phi_2 = \pi$ can be achieved using specialized pulse sequences, such that the desired $\CZgate$ gates are implemented in a manner robust to perturbations in the coupling strength~\cite{Saffman2010,Saffman2020,Lukin2001Blockade,anand2024dualspecies}. Here, we need not worry about additional $Z$ rotations on any atoms, as $\phi_1 = \pi$ ensures that the entire quantum state simply shifts to a different Pauli frame, which can be corrected via classical information postprocessing.

For convenience, the above controlled-Z gate set is denoted as
\begin{equation}
    \begin{aligned}
        \mathcal{CZ}:=\{\CZgate_{(i,\alpha),(j,\beta)}:{}&i,j\in\mathbb{N},\,\alpha,\beta\in\{\mathrm{Rb},\mathrm{Cs}\},\\[-2pt]
        &(i,\alpha)\ne(j,\beta)\}.
    \end{aligned}
\end{equation}
We then define the dual-species atom convenient operation set (DACOS) as
\begin{equation}
    \text{DACOS}:=\mathcal{T}\cup\mathcal{CZ}\cup \bigcup_{\alpha\in\{\mathrm{Rb},\mathrm{Cs}\}} (\UU_{\alpha} \cup \measset_{\alpha}).
\end{equation}

Our major contribution is to show that, for specific useful tasks such as EPP using stabilizer codes, straightforward circuit compilation under DACOS is possible without ancillary atoms, local control lasers, or intricate maneuvers of atoms. This simpler operation sequence facilitates the demonstration of EPP advantages in near-term experiments on dual-species neutral atom platforms.

\section{Two-way EPP with $d=2$ codes}\label{sec:2EPP}
\begin{figure}[t]
    \centering
    \includegraphics[width=0.9\linewidth]{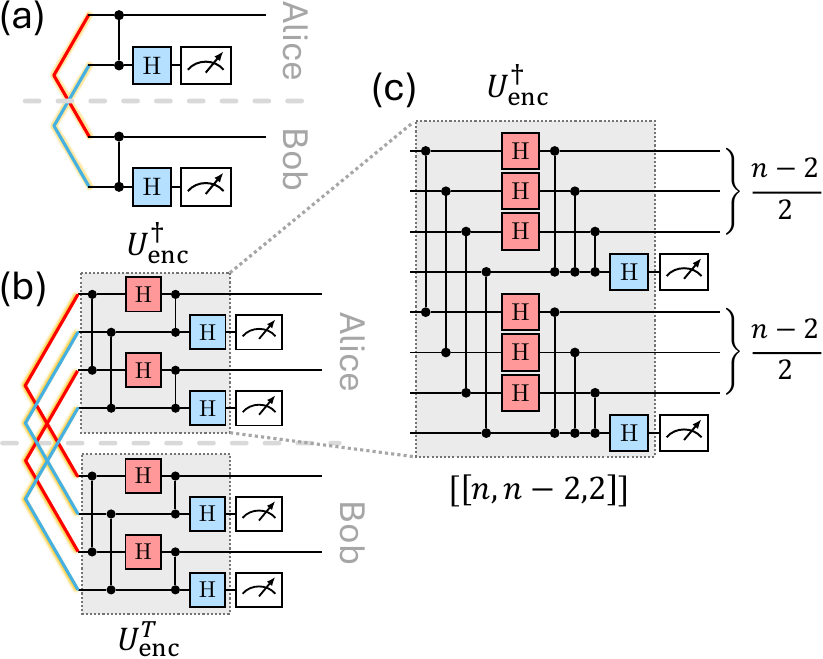}
    \caption{This figure uses different colors to distinguish qubits of different species of atoms (red for Rb and blue for Cs). (a) The original $2\to 1$ 2-EPP circuit~\cite{bennett_1996_PRL}. (b) The $[[4,2,2]]$ 2-EPP that purifies 2 pairs of entangled states from 4 pairs of noisy entangled inputs. Using an LC equivalent $[[n,n-2,2]]$ QECC, panel (c) generalizes the unitary in panel (a), which purifies $n-2$ pairs of entangled Rb atoms with a constant number of Cs atoms.} 
    \label{Fig:EPPnn22}
\end{figure}
The first purification protocol~\cite{bennett_1996_PRL} is a two-way entanglement purification protocol (2-EPP), in which Alice must communicate to Bob whether to retain the state after receiving his measurement outcome (see Fig.~\ref{Fig:EPPnn22}(a)). 
This is a $2 \to 1$ protocol that purifies two noisy Bell pairs into a single Bell pair with higher fidelity upon success.
Although this scheme cannot be directly mapped to a QECC, it can be generalized to an $n \to k$ protocol using any QECC that encodes $k$ logical qubits in $n$ physical qubits, thereby achieving a higher yield. 
In this section, we present a possible generalization based on $[[n, n-2, 2]]$ codes with $n\ge 4$ being an even number. 
This QECC has been introduced as the iceberg code in~\cite{Self2024}. When $n=4$, this approach is equivalent to the Leung-Shor method~\cite{Leung_Shor_2008}.
This section mainly demonstrates the efficient implementation of EPP on dual-species neutral atom arrays using the DACOS. 
Notably, the qubits to be retained and those to be measured belong to different atomic species, enabling efficient measurement via fluorescence imaging and facilitating subsequent protocol iterations.

As illustrated in Fig.~\ref{Fig:EPPnn22}(b,c), the iceberg-code-based 2-EPP is an $n \to n-2$ protocol. 
The V-shaped colored polylines represent noisy Bell states $\ket{\Phi_+}$ encoded with Rb (red) and Cs (blue) atoms, which are pre-distributed to Alice and Bob. In practice, Alice and Bob may represent two distant quantum computers based on dual-species neutral atoms, restricted to local quantum operations and classical communication over distance. 
The protocol enables the purification of one atomic species using the other, leveraging the DACOS and simple spatial arrangements.
The same purification cycle can be multiplexed and implemented recursively, supplemented by a sufficient number of noisy Bell states, to converge to high-fidelity Bell states as the fixed point of the protocol. 

We now describe the protocol assuming an ideal gate set. As shown in Fig.~\ref{Fig:EPPnn22}(b), the simplest version corresponds to $n=4$. Let us label the qubits from top to bottom as 1 to 4. Alice and Bob both measure their qubits 2 and 4 in the computational basis (i.e., the $Z$ basis). If Alice's and Bob's measurement outcomes match for both qubits, the protocol succeeds, and the Bell states involving qubits 1 and 3 are retained. Otherwise, all qubits are discarded, and the protocol fails. To simplify the discussion, we consider the noisy input quantum state as four copies of the following qubit isotropic state~\cite{footnote1}:
\begin{equation}\label{eq:IsotropicState}
    \begin{aligned}
        \hat{\rho}(p) &:= (1-p)\Phi_+ + \frac{p}{4} I \\
        &=
        (\II\otimes \EE^{(1)}_p) (\Phi_+) = (\EE^{(1)}_p\otimes \II) (\Phi_+) \\
        &= \EE^{(1)}_{\ptilde}\otimes \EE^{(1)}_{\ptilde} (\Phi_+),
    \end{aligned}
\end{equation}
where $I$ is the identity operator, $\II$ denotes the identity channel, $1-\ptilde = \sqrt{1-p}$, and $\Phi_+ := \ket{\Phi_+}\bra{\Phi_+}$. In the above equation, the $m$-qubit depolarizing channel $\EE^{(m)}_p$ is defined as
\begin{equation}
    \EE^{(m)}_p(\hat{\rho}) := (1-p)\hat{\rho} + p I/2^{m},
\end{equation}
which satisfies $\EE^{(m)}_{1-p}\circ\EE^{(m)}_{1-q} = \EE^{(m)}_{1-pq}$.
In the single-qubit case ($m=1$), $\EE_p^{(1)}$ can be represented by the following set of Kraus operators 
\begin{equation}
    \left\{\sqrt{1-3p/4} I,\,\sqrt{p/4}X,\,\sqrt{p/4}Y,\,\sqrt{p/4}Z\right\}.
\end{equation}
Owing to the ``ricochet'' property and $\EE^{(m)}_p = \EE^{(m)\T}_p$, as captured in the second line of Eq.~\eqref{eq:IsotropicState}, the qubit isotropic state $\hat{\rho}(p)$ can be interpreted as an ensemble of states $I\otimes \hat{E}(\Phi_+)$ (or equivalently $\hat{E}\otimes I(\Phi_+)$), where the error $\hat{E}$ is $X$, $Y$, or $Z$ with probability $p/4$ each, or $\hat{E}=I$ with probability $1-3p/4$. 
In laboratory settings, we emphasize that the actual noise is probably biased, and atom loss errors are likely involved~\cite{baranes2025,perrin2025quantumerrorcorrectionresilient}. However, we only model the depolarization noise as such for analytical simplicity.
We note that our compilation framework is fully general and applies to any stabilizer code, so it naturally accommodates codes optimized for biased noise such as dephasing-dominated noise where $Z$ errors are more likely than $X$ or $Y$ errors. CSS codes with asymmetric $X$ and $Z$ distances are a natural fit in this regime. The circuit structure and DACOS compilation procedure remain unchanged. Only the choice of underlying stabilizer code would differ. Furthermore, atom loss from optical traps can be modeled as an erasure error with known location~\cite{WuStern2019}, for which stabilizer codes can correct up to the code distance with a typically higher threshold than the depolarizing case. Our ancilla-free circuit design is particularly advantageous in this context, as fewer atoms in the array means a lower probability of experiencing atom loss during the protocol.
Throughout this work, we model circuit-level noise by applying a single-qubit depolarizing channel $\EE_q^{(1)}$ after each single-qubit gate and a two-qubit depolarizing channel $\EE_q^{(2)}$ after each CZ gate, where $q$ parameterizes the strength of these noise channels. The parameter $q$ thus quantifies the circuit noise level in all subsequent numerical simulations.

\begin{figure}[t]
    \centering
    \includegraphics[width=0.99\linewidth]{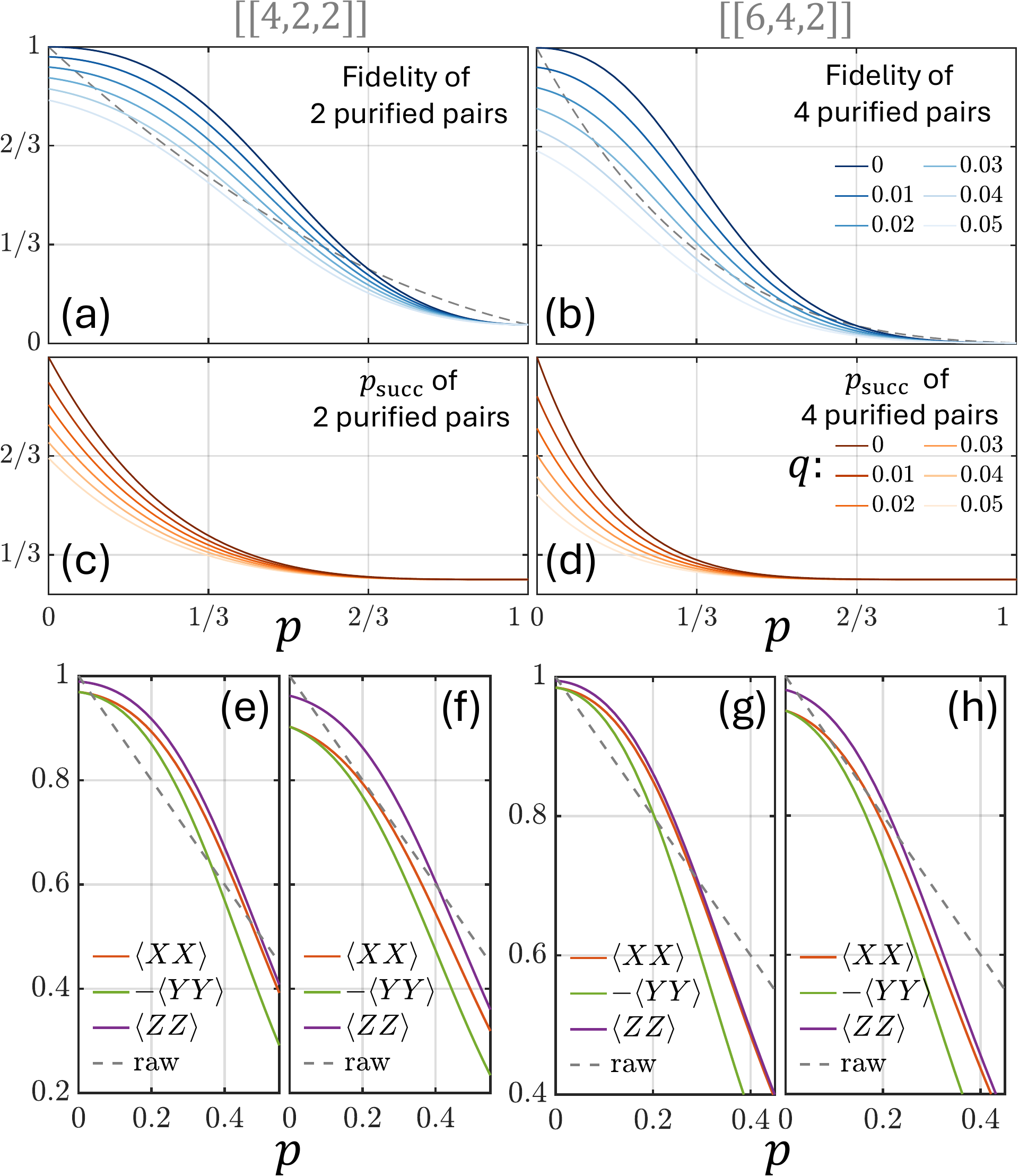}
    \caption{The figure illustrates the expectation values of the fidelity $F(\psi_0, \rho_{\text{out}})$, the success probabilities, and the values of non-zero correlators for the $[[4,2,2]]$ 2-EPP (panels (a,c,e,f)) and $[[6,4,2]]$ 2-EPP (panels (b,d,g,h)). Each panel considers different levels of noise characterized by the parameter $q$. For panels (e,f,g,h), the values of $q$ are $0.01$, $0.03$, $0.005$, and $0.015$.
    }
    \label{fig:curves_1}
\end{figure}
The EPP with $n=4$ will take $\hat{\rho}(p)^{\otimes 4}$ as the input, and perform the unitary operator as displayed in Fig.~\ref{Fig:EPPnn22}(b):
\begin{equation}
    U_{\text{enc}} = \CZgate_{13}\CZgate_{24}\Hgate_{1}\Hgate_{3}
    \CZgate_{12}\CZgate_{34}\Hgate_{2}\Hgate_{4},
\end{equation}
which transforms the $Z$ operators acting on qubits $2$ and $4$ into stabilizer operators:
\begin{equation}\label{eq:422stab}
    \begin{aligned}
        Z_{2}&\to \widetilde{S}_X := U_{\text{enc}}Z_{2}U_{\text{enc}}^\dagger = X_1X_2Z_3Z_4, \\
        Z_{4}&\to \widetilde{S}_Z :=  U_{\text{enc}}Z_{4}U_{\text{enc}}^\dagger = Z_1Z_2X_3X_4. \\
    \end{aligned}
\end{equation}
By the local Clifford gate $H_3H_4$, the above operators $\widetilde{S}_X$ and $\widetilde{S}_Z$ are equivalent to the stabilizers of the $[[4,2,2]]$ code:
\begin{equation}\label{eq:stab_iceberg}
    S_X:=\prod_{i=1}^n X_i,\quad S_Z:=\prod_{i=1}^n Z_i,
\end{equation}
where $n = 4$. This equivalence is known as local Clifford (LC) equivalence in the literature. Since it does not change the error correction features of our concern, we sometimes ignore the term ``LC equivalent'' in this work. Eq.~\eqref{eq:422stab} indicates the measurements of $Z_2$ and $Z_4$ at the end of the circuit are equivalent to the measurement of stabilizers $\widetilde{S}_X$ and $\widetilde{S}_Z$ directly on the input qubits. Recall that the input qubits are in qubit isotropic states, which can be understood as the ensemble of $4$ possible Bell basis states. Alice and Bob will always have consistent measurement outcomes if the input state happens to be the error-free case $\ket{\Phi^+}^{\otimes 4}$.
When one of the input Bell states is $I\otimes \hat{E}\ket{\Phi_+}$ with $\hat{E}\in\{X,Y,Z\}$, then, $\hat{E}$ alters the measurement consistency in Alice's or Bob's hands, since $\hat{E}$ anticommutes with either $\widetilde{S}_X$ or $\widetilde{S}_Z$. The inconsistent readouts between Alice and Bob indicate the existence of the impure input. Thus, without knowing the details of $\hat{E}$, the best strategy is to discard all qubits, and consider the protocol as failed.

Such a postselection does not guarantee obtaining error-free copies of $\ket{\Phi_+}$, since errors of weight greater than or equal to $2$, such as $\hat{E} = X_1X_2$ and $\hat{E} = X_2Y_3X_4$, are undetectable due to the commutation with $\widetilde{S}_X$ and $\widetilde{S}_Z$. In our input state ensemble $\hat{\rho}(p)^{\otimes 4}$, the total probability of such an undetectable error is $\left(1+3\left(1-p\right)^{4}\right)/4$. The derivation of this result and its generalization is presented in Appendix~\ref{app:iceberg_fid}.
Note that the only trivially acting Pauli operators acting on the logical space imposed by the noise channel are those Pauli operators from the stabilizer group $\mathcal{S}$. These harmless errors have a probability $(1-3p/4)^4+3(p/4)^{4}$ in total.
With the above discussion, the overlap between the output state $\rho_{\text{out}}$ and the ideal target state $\ket{\psi_0}:=\ket{\Phi_+}^{\otimes 2}$ is given by
\begin{equation}\label{eq:fid_422}
    \begin{aligned}
        \bra{\psi_{0}}\rho_{\text{out}}\ket{\psi_{0}} = \frac{\left(1-\frac{3p}{4}\right)^4 + 3\left(\frac{p}{4}\right)^4}{\frac{1}{4}+\frac{3}{4}\left(1-p\right)^{4}},\\
    \end{aligned}
\end{equation}
which is the Uhlmann fidelity between a pure state $\psi$ and a mixed state $\rho$: $F(\psi,\rho) := \bra{\psi}\rho\ket{\psi}$. Thus, Eq.~\eqref{eq:fid_422} gives an error infidelity suppression to higher order $F(\psi_{0},\rho_{\text{out}}) = 1 - \mathcal{O}(p^2)$, whose performance is displayed as the solid line with $q=0$ in Fig.~\ref{fig:curves_1}(a). We indicate that the output state $\rho_{\text{out}}\ne \rho_{\text{out},1}\otimes \rho_{\text{out},2}$ is generally not a product state of two entangled pairs.

Due to the symmetry of $\widetilde{S}_X$ and $\widetilde{S}_Z$, the circuit can be implemented straightforwardly using DACOS. Here, we present numerical simulations that account for noise in the gate set. Specifically, we model the EPP process with an additional single-qubit depolarizing channel $\EE_q^{(1)}$ after each single-qubit gate and a two-qubit depolarizing channel $\EE_q^{(2)}$ after each CZ gate.
As shown in Fig.~\ref{fig:curves_1}(a,e), the $[[4,2,2]]$ 2-EPP provides benefits when $q$ is small. For example, at $q=0.04$, it still yields higher fidelity than the unpurified case when $p \approx 1/3$. However, sufficiently large circuit-level noise reduces the overall success probability and bipartite correlations, as illustrated in Fig.~\ref{fig:curves_1}.

The $4 \to 2$ protocol generalizes readily to any $n \to n-2$ protocol based on the $[[n,n-2,2]]$ code, whose stabilizers are given by Eq.~\eqref{eq:stab_iceberg} for even $n \geq 4$. Similarly, using the encoding unitary $U_{\text{enc}}$ in Fig.~\ref{Fig:EPPnn22}(c), we derive stabilizers $\widetilde{S}_X$ and $\widetilde{S}_Z$ from $Z_1$ and $Z_2$. These are locally Clifford equivalent to $S_X$ and $S_Z$ in Eq.~\eqref{eq:stab_iceberg}.
In Fig.~\ref{fig:curves_1}(b,d,g,h), we show results for the output state fidelity, success probability, and correlators of the $[[6,4,2]]$ scheme under noisy gates, using a setup similar to that for the $[[4,2,2]]$ scheme. Results for higher $n$ at $q=0$ appear in Fig.~\ref{fig:app}(a), with analytical expressions provided in Appendix~\ref{app:iceberg_fid}.

\begin{figure}[t]
    \centering
    \includegraphics[scale=0.325]{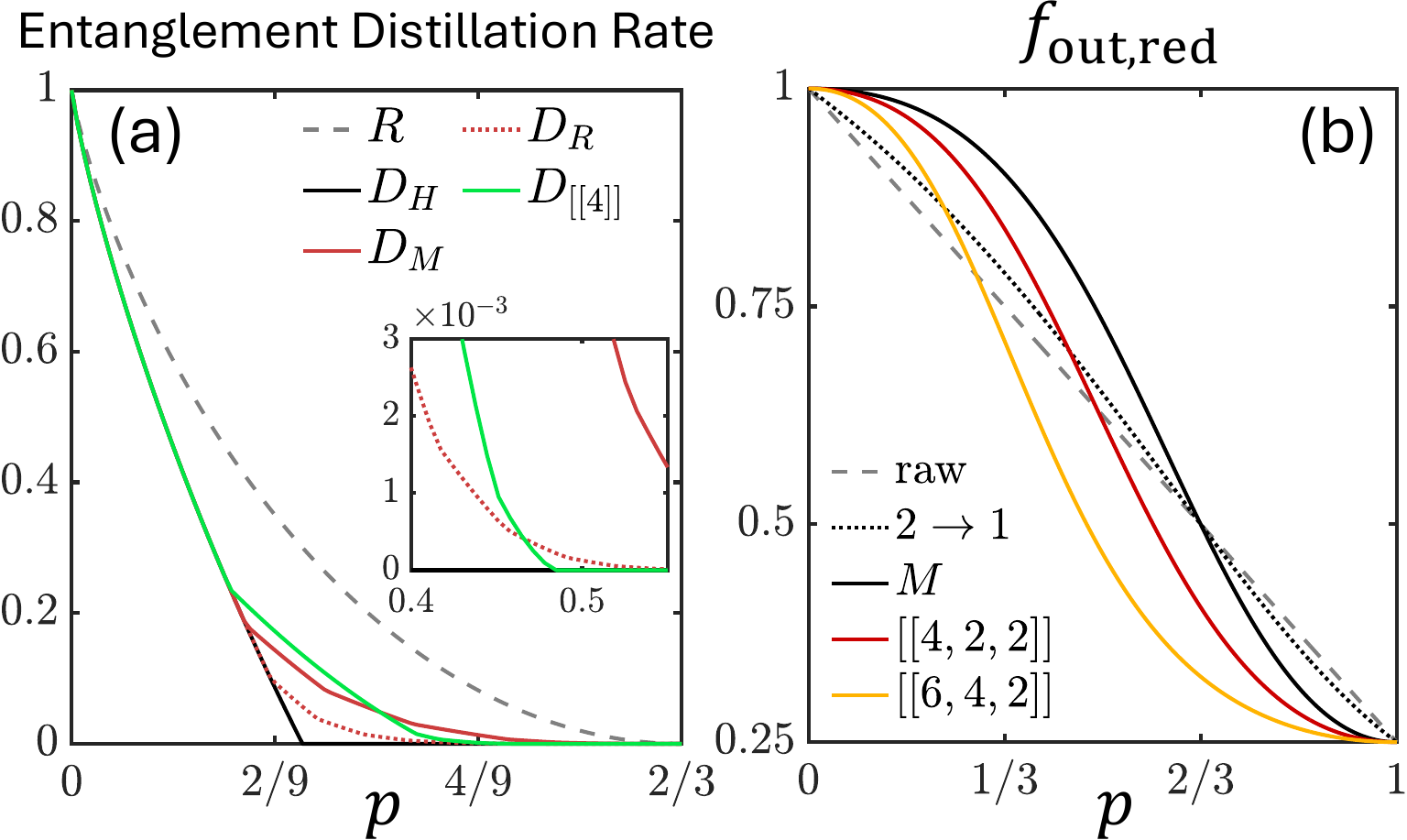}
    \caption{Panel (a) displays the asymptotic rate of distilled entanglement under the input $\lim_{N\to\infty}\hat{\rho}(p)^{\otimes N}$.
    The black solid line represents the celebrated hashing bound regarding $\EE^{(1)}_p$. The dashed black line represents the Rains bound $R(p)$ for $\hat{\rho}(p)$, an upper bound for the distillation rate. $D_R$ and $D_M$ are the rates given by the recurrence-hashing method and the Macchiavello method~\cite{bennett_1996_PRL}. The green solid line labels the rates $D_{[[4]]}$ based on $[[4,2,2]]$. The inset shows $D_{[[4]]}$ vanishes near $p\approx 0.51$.  The colored solid lines in panel (b) display the relationship between $p$ and the output fidelity of a single pair $f_{\text{out},\text{red}}$ under different 2-EPPs. }
    \label{fig:europe}
\end{figure}

Observe that when $q=0$, the iceberg code always yields a block fidelity improvement for $p < 2/3$, which can lead to a finite asymptotic distillation rate, similar to the $2 \to 1$ recurrence protocol in Ref.~\cite{bennett_1996_PRA}.
In this work, the asymptotic distillation rate for a given LOCC protocol $\Lambda$ is defined as the largest real number $D\ge 0$ such that
\begin{equation}
    \lim_{N\to\infty} \|\Lambda(\hat{\rho}(p)^{\otimes N}) - \Phi_+^{\otimes \lfloor D N\rfloor}\|_1 = 0,
\end{equation}
where $\|A\|_1 = \tr\sqrt{A^\dagger A}$ is the trace norm of the operator $A$. When discussing this asymptotic rate, we always ignore the circuit-level noise.

We compare the asymptotic distillation rate of our iceberg-code-based 2-EPP with that of the recurrence $2 \to 1$ protocol. The distillation rates of the original recurrence scheme and the improved scheme suggested by Macchiavello~\cite{bennett_1996_PRL} are denoted as $D_R$ and $D_M$, respectively.
As a simple extension, we replace the $2 \to 1$ subroutine with our $[[n,n-2,2]]$ 2-EPP. The noisy states improved by the 2-EPP iteration will be rearranged and sent to the hashing protocol for perfect state distillation, which has a yield $D_{[[n]]}$. The concrete technical details are presented in Appendix~\ref{app:ED}.
Numerically, we find that $D_{[[4]]}$ outperforms $D_R$ and $D_M$ at $p\lesssim 0.34$. During the numerical optimization, we also find that $D_{[[n]]}$ with $n=4$ gives the best performance. Usually, $D_{[[n]]}$ with a large $n$ is not better than $\hat{\rho}(p)$'s hashing bound $D_H(p) = \max\{0, 1 - h_2(\frac{3p}{4})-\frac{3p}{4}\log_2 3\}$, where $h_2(x):=-x\log_2 x - (1-x)\log_2 (1-x)$. The decreased yield of large $n$ is not only due to the decreasing success probability, but also due to the decreased average fidelity $f_i:=\bra{\Phi_+}\rho_{\text{out}, i}\ket{\Phi_+}$ (see Fig.~\ref{fig:europe}(b)), where $\rho_{\text{out},i}$ is the reduced state of the $i$-th pair of the output state.

In Fig.~\ref{fig:europe}(a), the Rains bound is an upper bound for the LOCC distillable entanglement, which is given by $R(p) = \max\{0,1-h_2(1-\frac{3p}{4})\}$ in our case~\cite{Rains_2001}.
We emphasize that these asymptotic distillation procedures are neither theoretically optimal nor experimentally practical. The demonstration of higher rates simply highlights the advantages of incorporating diverse quantum error-correcting codes into EPPs. Better protocols can be found in Ref.~\cite{AJOSS2025}.

In Fig.~\ref{fig:europe}(b), we display the output fidelity of a single qubit pair for different protocols, when the circuit-level noise is absent. Due to the output state $\rho_{\text{out}}$ of interest having identical reduced states $\rho_{\text{out},\text{red}}=\tr_{\text{other pair}}\rho_{\text{out}}$ for each bipartite qubit pair, we use $f_{\text{out},\text{red}}(p) = \bra{\Phi_+}\rho_{\text{out},\text{red}}\ket{\Phi_+}$ to characterize this performance. The dashed line and dotted line represent the input state fidelity and the output state fidelity for the $2\to 1$ recurrence method. The black solid line represents the Macchiavello method with two rounds, which collects the successful outputs of two copies of the $2\to 1$ EPP, then combines them with another $2\to 1$ EPP after applying a bilateral Hadamard operation (also see details in~\cite{Jiang2007}). The Macchiavello method of two rounds can also be viewed as a $[[4,1,2]]$ code-based 2-EPP. The red and yellow lines display the value of $f_{\text{out},\text{red}}(p)$ for the $[[4,2,2]]$ and $[[6,4,2]]$ 2-EPPs.

\section{The general EPP with stabilizer codes}\label{sec:1EPP}
\begin{figure}[t]
    \centering
    \includegraphics[scale = 0.30]{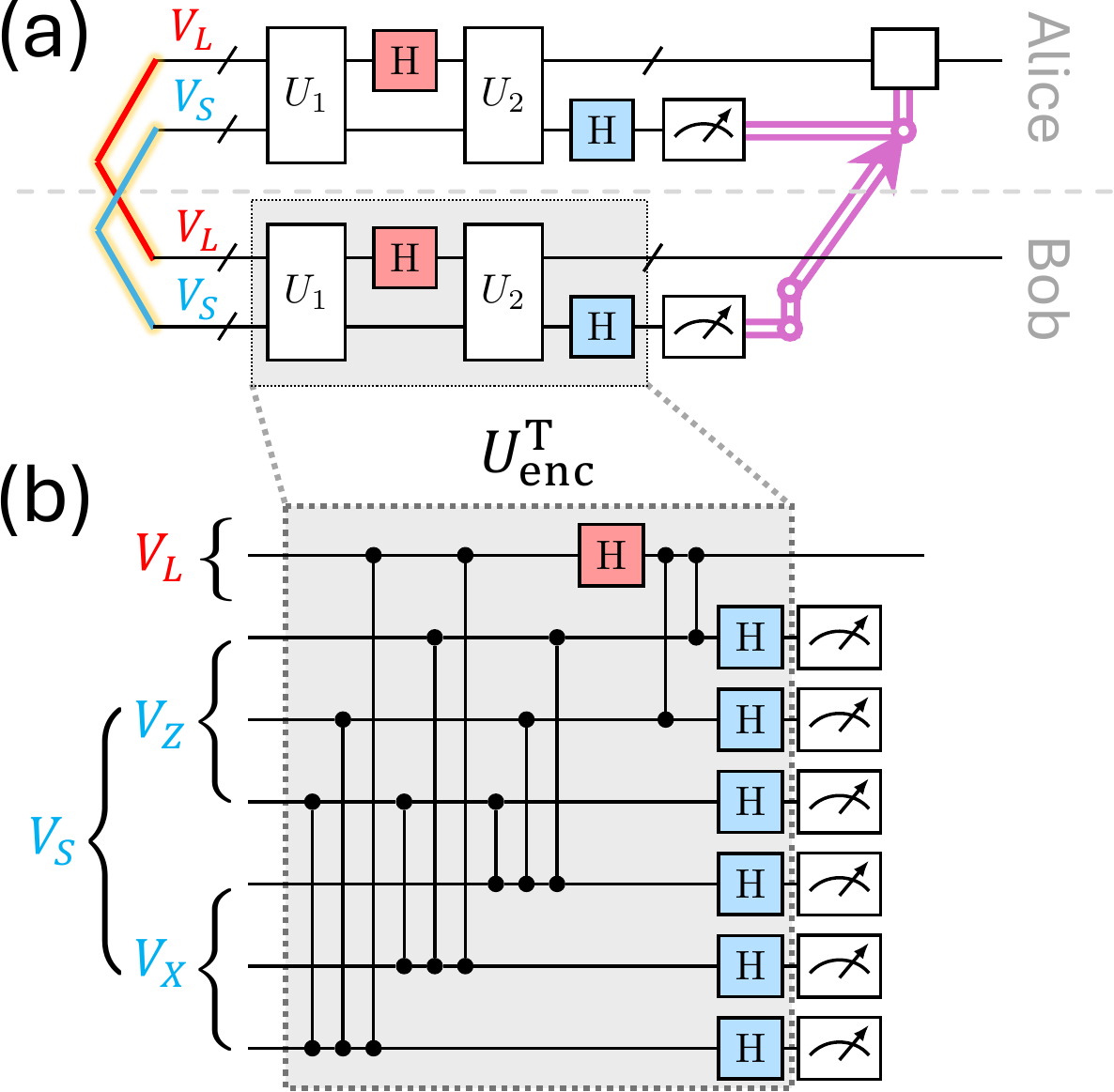}
    \caption{(a) General circuit implementation for EPP based on any stabilizer code, shown here for one round of one-way classical communication (generalizable to two-way). $V_L$ denotes the set of noisy Bell pairs to be purified, while $V_S$ is the set of qubits measured in the computational basis at the end. Here, $V_L$ and $V_S$ are realized using two distinct atomic species (red and blue). Hadamard gates (red and blue boxes) are applied globally to all qubits of the respective species. $U_1$ and $U_2$ consist of CZ gate sequences. (b) An example circuit for the LC equivalent $[[7,1,3]]$ CSS code, where each vertical line connecting black dots represents a $\CZgate$ gate. }
    \label{fig:stab_circuit}
\end{figure}

As presented in the previous section, the 2-EPP based on the $[[n,n-2,2]]$ code can be conveniently implemented on a dual-species neutral atom platform. So a natural question is whether EPPs based on codes with larger distances can be implemented in a similar manner. Here, we show that the encoding unitary $U_{\text{enc}}$ for any qubit stabilizer code can be compiled analogously. As illustrated in Fig.~\ref{fig:stab_circuit}(a), given a qubit stabilizer code, our EPP is applied to two sets of atoms, $V_S$ and $V_L$, each containing atoms of only a single species (e.g., Rb or Cs). We require that $|V_L| = k$, the number of logical qubits in the stabilizer code, and $|V_S| = n - k$, the number of independent stabilizer operators ($\hat{S}_i$).

The encoding unitary is a real matrix satisfying $U_{\text{enc}}^\T = U_{\text{enc}}^\dagger$. Specifically, this unitary comprises two blocks of Hadamard gates and two blocks of CZ sequences. The Hadamard gate blocks act separately on all atoms of each species. Moreover, the two blocks of CZ sequences are not unique, providing flexibility for further optimization. These features can be conveniently programmed and implemented on a dual-species neutral atom array platform.

\subsection{Circuit Compilation}\label{Sec:stab_code_epp}
As one can see in Fig.~\ref{fig:stab_circuit}(a), different $U_{\text{enc}}$ differ mainly in the CZ blocks $U_1$ and $U_2$.
We demonstrate how to obtain $U_1$ and $U_2$ by decomposing $U_{\text{enc}}$.
Given the complete set of stabilizers of an $[[n,k,d]]$ stabilizer code, we first choose $r_S = n-k$ \textit{linearly independent} stabilizers, which still generate the stabilizer group $\mathcal{S}$.
The tableau matrix representation (or the binary symplectic format) for these stabilizer operators can be written as
\begin{equation}
    \Tb_0=
    \left[
    \arraycolsep=1.5pt \def\arraystretch{0.9}
    \begin{array}{c|c}
        \Tb_X & \Tb_Z\\
    \end{array}\right],
\end{equation}
where $\Tb_0\in \Fbb_2^{r_S\times 2n}$ must be a full rank matrix, and $\Tb_X$ has the same size as $\Tb_Z$, since these stabilizers are independent. The Clifford unitary transformation of stabilizers by $U$ is represented by right multiplying $\Tb_0$ by a matrix $\Cb(U)\in \text{Sp}(2n,\Fbb_2)$, where $\text{Sp}(2n,\Fbb_2)$ is the symplectic group acting on vector space $\Fbb_2^{2n}$.
If we denote $V_S = \{1,2,\cdots,r_S\}$, $V_L = \{r_S+1,r_S+2,\cdots, n\}$, then our target is obtaining
\begin{equation}\label{eq:Tf}
    \Tb_{f} = \left[
    \arraycolsep=1.5pt \def\arraystretch{0.9}
    \begin{array}{c|cc}
         0&I_{r_S}&0\\
    \end{array}\right]
\end{equation}
from $\Tb$ with the DACOS, where $I_r$ is an $r\times r$ identity matrix. We show that the Clifford unitary transforming $\Tb_0 \to \Tb_f$ can be decomposed into the following steps.
\begin{enumerate}[label=(\roman*)]
    \item Let $I_m$ be the $m\times m$ identity matrix. By proper row transformation and permuting columns of $\Tb_X$ and $\Tb_Z$ in the same way, the $\Tb$ matrix can always be transformed into the standard form~\cite{Gottesman_1997,Shi2025PRXQ}:
    \begin{equation}\label{eq:T1}
        \Tb\mapsto
        \Tb_1
        =\left[
        \arraycolsep=1.5pt \def\arraystretch{0.9}
        \begin{array}{ccc|ccc}
            I_{r_X} &J_1&J_2&L_1&0&L_2 \\
            0&0&0&K_1&I_{r_Z}&K_2 
        \end{array}\right] \\
    \end{equation}
    in which  $J_1 \in\Fbb_2^{r_X\times r_Z}$, $J_2,L_2 \in\Fbb_2^{r_X\times k}$, $K_1 \in\Fbb_2^{r_Z\times r_X}$, $K_2 \in\Fbb_2^{r_Z\times k}$ and $r_X + r_Z = r_S$. 
    For convenience, let us denote two disjoint sets $V_X = \{1,\cdots,r_X\}$ and $V_Z = \{r_X+1,\cdots, r_S\}$, so we have $V_S = V_X\cup V_Z$. 
    Here, any linear transformation of rows keeps the represented stabilizer group invariant, and the column permutation means swapping qubits, which can be done by the atom relocation operation $\hat{T}\in\mathcal{T}$. Since stabilizers are commutative, the $J_i$, $K_i$ and $L_i$ submatrices satisfy:
    \begin{align}
    & L_1 + L_2J_2^\T = (L_1 + L_2J_2^\T)^\T,  \label{eq:commutation_cond_1}\\
    & K_1 + J_1^\T + K_2 J_2^\T = 0 .\label{eq:commutation_cond_2}
    \end{align}
    
    \item Performing the unitary $H_{(1)}:=\bigotimes_{i\in V_Z\cup V_L}\Hgate_i$ gives:
    \begin{equation}
        \Tb_1
        \mapsto \Tb_2:=
        \left[
        \arraycolsep=1.5pt \def\arraystretch{0.9}
        \begin{array}{ccc|ccc}
            I_{r_X} &0&L_2&L_1&J_1&J_2 \\
            0&I_{r_Z}&K_2&K_1&0&0 
        \end{array}
        \right].
    \end{equation}
    Note that $H_{(1)}$ acts on two species of atoms, which is inconvenient, but we will show that it can be removed in our final circuit.
    \item Given two disjoint sets $V_1$ and $V_2$, we denote the product of controlled-Z gates with biadjacency matrix $B\in\Fbb_2^{|V_1|\times |V_2|}$ as
    \begin{equation}
        CZ(V_1, V_2;B):= \prod_{i\in V_1, j\in V_2} \CZgate_{ij}^{B_{ij}}.
    \end{equation} 
    Considering the unitary gate
    \begin{equation}\label{eq:U1}
        U_1:=CZ(V_X, V_Z\cup V_L;[J_1,J_2]),
    \end{equation}
    which transforms the columns of $\Tb_2$ by the following $\Cb(U_1)$
    \begin{equation}
        \Cb(U_1)
        =
        \left[
        \arraycolsep=1.5pt \def\arraystretch{0.9}
        \begin{array}{ccc|ccc}
            I_{r_X}&0&0&0&J_1&J_2\\
            0&I_{r_Z}&0&J_1^\T &0&0\\
            0&0&I_{k}&J_2^\T &0&0\\
            \hline
            0&0&0&I_{r_X}&0&0\\
            0&0&0&0&I_{r_Z}&0\\
            0&0&0&0&0&I_{k}\\
        \end{array}
        \right].
    \end{equation}
    Thus, 
    \begin{equation}
        \begin{aligned}
            \Tb_2
            &\mapsto
            \Tb_3:=
            \Tb_2
            \Cb(U_1)\\
            &=
            \left[
            \begin{array}{ccc|ccc}
                I_{r_X} &0&L_2&\Gamma&0&0 \\
                0&I_{r_Z}&K_2&0&0&0 
            \end{array}
            \right]
            ,
        \end{aligned}
    \end{equation}
    where $\Gamma:=L_1+L_2 J_2^\T$ is a symmetric matrix due to Eq.~\eqref{eq:commutation_cond_1}. And we have used Eq.~\eqref{eq:commutation_cond_2} to eliminate $K_1+J_1^\T +K_2J_2^\T$. 
    \item Note that the symmetric $\Gamma$ may have a nonzero diagonal. Let $\gamma\in\Fbb_2^{r_X}$ be the diagonal of $\Gamma$. We perform the unitary $P_{(1)}:=\bigotimes_{i\in V_X} \Pgate_i^{\gamma_i}$ with phase gates, such that $\Gamma$ is replaced by a symmetric matrix
    \begin{equation}
        \Gamma_0 = \Gamma + \operatorname{diag}(\gamma) \in \Fbb_2^{r_X\times r_X},
    \end{equation}
    with zero diagonal elements, which defines an adjacency matrix of a simple graph. Again, we have to be cautious about $P_{(1)}$, since it acts on only a subset of one species.
    \item Performing the unitary $H_{(2)} := \bigotimes_{i\in V_L}\Hgate_i$ yields 
    \begin{equation}
        \Tb_3
        \mapsto
        \Tb_4:=
        \left[
        \begin{array}{ccc|ccc}
            I_{r_X} &0&0&\Gamma_0&0&L_2 \\
            0&I_{r_Z}&0&0&0&K_2
        \end{array}
        \right].
    \end{equation}
    \item Considering the unitary gate
    \begin{equation}\label{eq:U2}
        U_2 = CZ(V_S, V_L; \begin{bmatrix}
            L_2 \\
            K_2
        \end{bmatrix})\prod_{\substack{i,j\in V_X\\ i<j}} \CZgate_{ij}^{(\Gamma_0)_{ij}} 
    \end{equation}
    which transforms the columns of $\Tb_4$ by
    \begin{equation}
        \Cb(U_2)
        =
        \left[
        \arraycolsep=1.5pt \def\arraystretch{0.9}
        \begin{array}{ccc|ccc}
            I_{r_X}&0&0&\Gamma_0&0&L_2\\
            0&I_{r_Z}&0&0&0&K_2\\
            0&0&I_{k}&L_2^\T&K_2^\T &0\\
            \hline
            0&0&0&I_{r_X}&0&0\\
            0&0&0&0&I_{r_Z}&0\\
            0&0&0&0&0&I_{k}\\
        \end{array}
        \right].
    \end{equation}
    Thus,
    \begin{equation}
    \begin{aligned}
        \Tb_4
        \mapsto
        \Tb_5&:=\Tb_4\Cb(U_2) \\&=
        \left[
        \begin{array}{ccc|ccc}
            I_{r_X} &0&0&0&0&0 \\
            0&I_{r_Z}&0&0&0&0
        \end{array}
        \right].
    \end{aligned}
    \end{equation} 
    \item Finally, performing the unitary $H_{(3)} = \bigotimes_{i\in V_S} \Hgate_i$ gives $\Tb_5\mapsto \Tb_f$ in Eq.~\eqref{eq:Tf}.
\end{enumerate}

\begin{figure}[t]
    \centering
    \includegraphics[scale = 0.28]{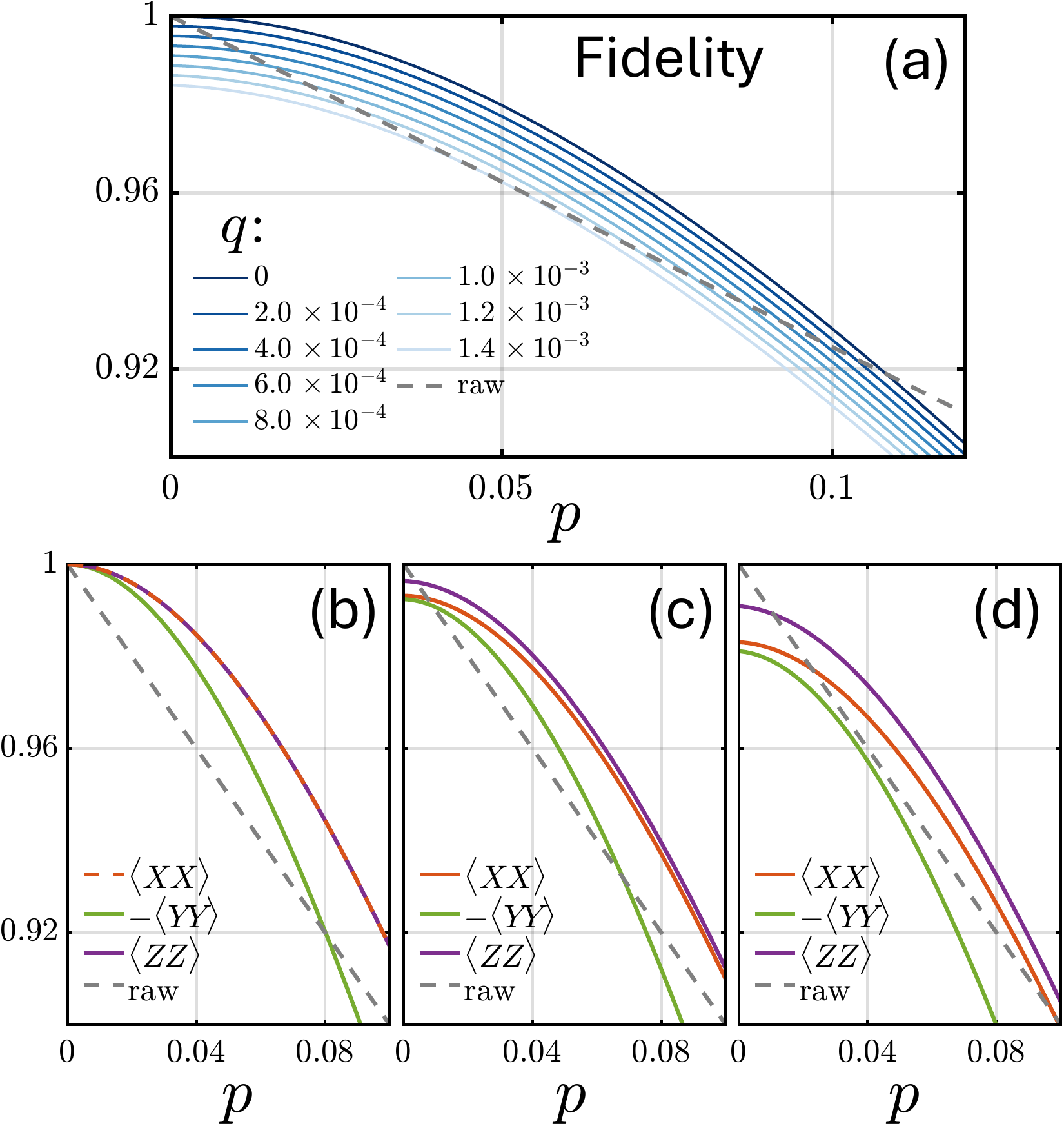}
    \caption{This figure shows the performance of the LC equivalent $[[7,1,3]]$ 1-EPP implemented in the noisy circuit of Fig.~\ref{fig:stab_circuit}(b), using the componentwise decoder in Table~\ref{tab:713decode}. (a) Output fidelity as a function of the input noise parameter $p$ for various circuit noise levels $q$. The gray dashed line represents the fidelity of the input state $\hat{\rho}(p)$. (b--d) Nonzero two-point correlators of the first pair of output entangled states under circuit noise levels $q=0$, $4\times 10^{-4}$, and $10^{-3}$, respectively. The gray dashed lines indicate the corresponding correlators for the input state $\hat{\rho}(p)$.}
    \label{fig:713curves}
\end{figure}

The total unitary operator of the above process from (i) to (vii) is given by $H_{(3)} U_2 H_{(2)}P_{(1)}U_1 H_{(1)}\hat{T}$. However, by commuting $P_{(1)}$ with $U_1$, and omitting $P_{(1)}H_{(1)}\hat{T}$, the total unitary operator can be reduced to
\begin{equation}\label{eq:Uenc_stab}
    U_{\text{enc}}^\dagger = H_{(3)} U_2 H_{(2)}U_1,
\end{equation}
which is compatible with DACOS.
Here, $\hat{T}$ is the initialization of the atom configuration, and $P_{(1)}H_{(1)}$ yields an LC equivalent stabilizer code. We can get rid of these gates by assuming this deformed stabilizer code works as well as its original counterpart.

Now we have achieved the goal of implementing Fig.~\ref{fig:stab_circuit}(a) -- in the decomposition of Eq.~\eqref{eq:Uenc_stab}. It is straightforward to see how the unitary part of the EPP can be realized with the DACOS. $H_{(2)}$ and $H_{(3)}$ are Hadamard gates acting on all atoms of the same species, separately. $U_1$ and $U_2$ are sequences of CZ gates entangling different qubits. 
For the feedback operations (pink arrows in Fig.~\ref{fig:stab_circuit}), note that all purified entangled atoms belong to the same species. When a unitary operation is needed on a specific atom, it cannot be done with a global laser field. However, in stabilizer codes, the unitary feedback operations are typically Pauli operations, which simply turn $\ket{\Phi_+}$ to other Bell states $\ket{\Phi_-}$ or $\ket{\Psi_\pm}$. Such a Pauli operation can be omitted before further distribution of bipartite entangled qubits since Alice can infer the sequence of Bell states she purifies by the syndrome measurement.

\subsection{1-EPP Examples}
\paragraph{Steane code} An example of the LC equivalent $[[7,1,3]]$ Steane code is presented in Fig.~\ref{fig:stab_circuit}(b), which displays the details of $U_{\text{enc}}$. In this case, we have 
\begin{equation*}
    \Tb_0 = \left[\begin{array}{c|c}
        H_X & 0\\
        0& H_Z
    \end{array}\right],
\end{equation*}
with $H_X = H_Z = H_{[7,4]}$, where $H_{[7,4]}$ is the conventional check matrix of the $[7,4]$ Hamming code:
\begin{equation}
    H_{[7,4]} = \left[ 
	\arraycolsep=1.5pt \def\arraystretch{0.9} 
		\begin{array}{ccccccc} 
			 1 & 1 & 0 & 1 & 1 & 0 & 0 \\ 
			 1 & 0 & 1 & 1 & 0 & 1 & 0 \\ 
			 0 & 1 & 1 & 1 & 0 & 0 & 1 \\ 
	\end{array} 
\right].
\end{equation}
Thus, $H_X$ and $H_Z$ can be transformed into
\begin{align}
    R_XH_XP &= \left[ 
	\arraycolsep=1.5pt \def\arraystretch{0.9} 
		\begin{array}{ccccccc} 
			 1 & 0 & 0 & 1 & 0 & 1 & 1 \\ 
			 0 & 1 & 0 & 1 & 1 & 0 & 1 \\ 
			 0 & 0 & 1 & 1 & 1 & 1 & 0 \\ 
	\end{array} 
    \right] \\
    R_ZH_ZP &= \left[ 
	\arraycolsep=1.5pt \def\arraystretch{0.9} 
		\begin{array}{ccccccc} 
			 1 & 1 & 1 & 1 & 0 & 0 & 0 \\ 
			 1 & 0 & 1 & 0 & 1 & 0 & 1 \\ 
			 0 & 1 & 1 & 0 & 0 & 1 & 1 \\ 
	\end{array} 
    \right] 
\end{align}
with some row transformations $R_X$, $R_Z$ and column transformation $P$. Compared to Eq.~\eqref{eq:T1}, we have
\begin{align}
    \left[J_1,J_2\right] &= \left[ 
	\arraycolsep=1.5pt \def\arraystretch{0.9} 
		\begin{array}{cccc} 
			  1 & 0 & 1 & 1 \\ 
			  1 & 1 & 0 & 1 \\ 
			  1 & 1 & 1 & 0 \\ 
	\end{array} 
    \right],
    K_2 = \left[ 
	\arraycolsep=1.5pt \def\arraystretch{0.9} 
    \begin{array}{c} 
           0 \\ 
           1 \\ 
           1 \\ 
	\end{array} 
    \right],
\end{align}
and $L_i = 0$. The matrices $[J_1,J_2]$ and $K_2$ determine $U_1$ and $U_2$, respectively, as displayed in Fig.~\ref{fig:stab_circuit}(b). After Bob sends his measurement outcome to Alice, Alice computes the syndrome $s=m_A\oplus m_B$, independently decodes its bit-flip and phase-flip components, and applies the resulting Pauli feedback operation to finish the 1-EPP. The correspondence between the feedback unitary $U_7$ and the measured syndrome is provided in Table~\ref{tab:713decode}.
The performance of this LC-equivalent $[[7,1,3]]$ 1-EPP is shown in Fig.~\ref{fig:713curves}. Panel (a) displays the output fidelity as a function of the input noise parameter $p$ for various circuit noise levels $q$, while panels (b--d) show the nonzero two-point correlators of the first pair of output entangled states.

\begin{table}[ht]
\centering
\setlength{\tabcolsep}{6pt}
\begin{tabular}[t]{|cc||cc|}
\hline
$(s_1,s_2,s_3,s_4,s_5,s_6)$ & $U_{7}$ & $(s_1,s_2,s_3,s_4,s_5,s_6)$ & $U_{7}$ \\
\hline
\hline
$0\,\,0\,\,0\,\,0\,\,1\,\,1$ & $X$ & $1\,\,0\,\,1\,\,0\,\,0\,\,0$ & $Z$ \\
$0\,\,0\,\,0\,\,1\,\,0\,\,1$ & $X$ & $1\,\,0\,\,1\,\,0\,\,0\,\,1$ & $Z$ \\
$0\,\,0\,\,0\,\,1\,\,1\,\,0$ & $X$ & $1\,\,0\,\,1\,\,0\,\,1\,\,0$ & $Z$ \\
$0\,\,0\,\,1\,\,0\,\,1\,\,1$ & $X$ & $1\,\,0\,\,1\,\,0\,\,1\,\,1$ & $Y$ \\
$0\,\,0\,\,1\,\,1\,\,0\,\,1$ & $X$ & $1\,\,0\,\,1\,\,1\,\,0\,\,0$ & $Z$ \\
$0\,\,0\,\,1\,\,1\,\,1\,\,0$ & $X$ & $1\,\,0\,\,1\,\,1\,\,0\,\,1$ & $Y$ \\
$0\,\,1\,\,0\,\,0\,\,1\,\,1$ & $X$ & $1\,\,0\,\,1\,\,1\,\,1\,\,0$ & $Y$ \\
$0\,\,1\,\,0\,\,1\,\,0\,\,1$ & $X$ & $1\,\,0\,\,1\,\,1\,\,1\,\,1$ & $Z$ \\
$0\,\,1\,\,0\,\,1\,\,1\,\,0$ & $X$ & $1\,\,1\,\,0\,\,0\,\,0\,\,0$ & $Z$ \\
$0\,\,1\,\,1\,\,0\,\,0\,\,0$ & $Z$ & $1\,\,1\,\,0\,\,0\,\,0\,\,1$ & $Z$ \\
$0\,\,1\,\,1\,\,0\,\,0\,\,1$ & $Z$ & $1\,\,1\,\,0\,\,0\,\,1\,\,0$ & $Z$ \\
$0\,\,1\,\,1\,\,0\,\,1\,\,0$ & $Z$ & $1\,\,1\,\,0\,\,0\,\,1\,\,1$ & $Y$ \\
$0\,\,1\,\,1\,\,0\,\,1\,\,1$ & $Y$ & $1\,\,1\,\,0\,\,1\,\,0\,\,0$ & $Z$ \\
$0\,\,1\,\,1\,\,1\,\,0\,\,0$ & $Z$ & $1\,\,1\,\,0\,\,1\,\,0\,\,1$ & $Y$ \\
$0\,\,1\,\,1\,\,1\,\,0\,\,1$ & $Y$ & $1\,\,1\,\,0\,\,1\,\,1\,\,0$ & $Y$ \\
$0\,\,1\,\,1\,\,1\,\,1\,\,0$ & $Y$ & $1\,\,1\,\,0\,\,1\,\,1\,\,1$ & $Z$ \\
$0\,\,1\,\,1\,\,1\,\,1\,\,1$ & $Z$ & $1\,\,1\,\,1\,\,0\,\,1\,\,1$ & $X$ \\
$1\,\,0\,\,0\,\,0\,\,1\,\,1$ & $X$ & $1\,\,1\,\,1\,\,1\,\,0\,\,1$ & $X$ \\
$1\,\,0\,\,0\,\,1\,\,0\,\,1$ & $X$ & $1\,\,1\,\,1\,\,1\,\,1\,\,0$ & $X$ \\
$1\,\,0\,\,0\,\,1\,\,1\,\,0$ & $X$ & Otherwise & $I$ \\
\hline
\hline
\end{tabular}
\caption{Decoder table for the LC-equivalent $[[7,1,3]]$ code whose encoding unitary is shown in Fig.~\ref{fig:stab_circuit}(b). The two three-bit syndrome components are decoded independently in the CSS frame by selecting minimum-weight bit- and phase-flip representatives, whose residual Pauli actions determine $U_7$. This corrects every single-qubit error and selects a minimum-weight representative for each of the $64-(1+3\times 7)=42$ remaining syndromes. Under i.i.d.\ depolarizing input errors, each such syndrome has three equiprobable weight-two representatives with distinct feedback operations.}
\label{tab:713decode}
\end{table}%

\begin{figure}[t]
    \centering
    \includegraphics[scale = 0.28]{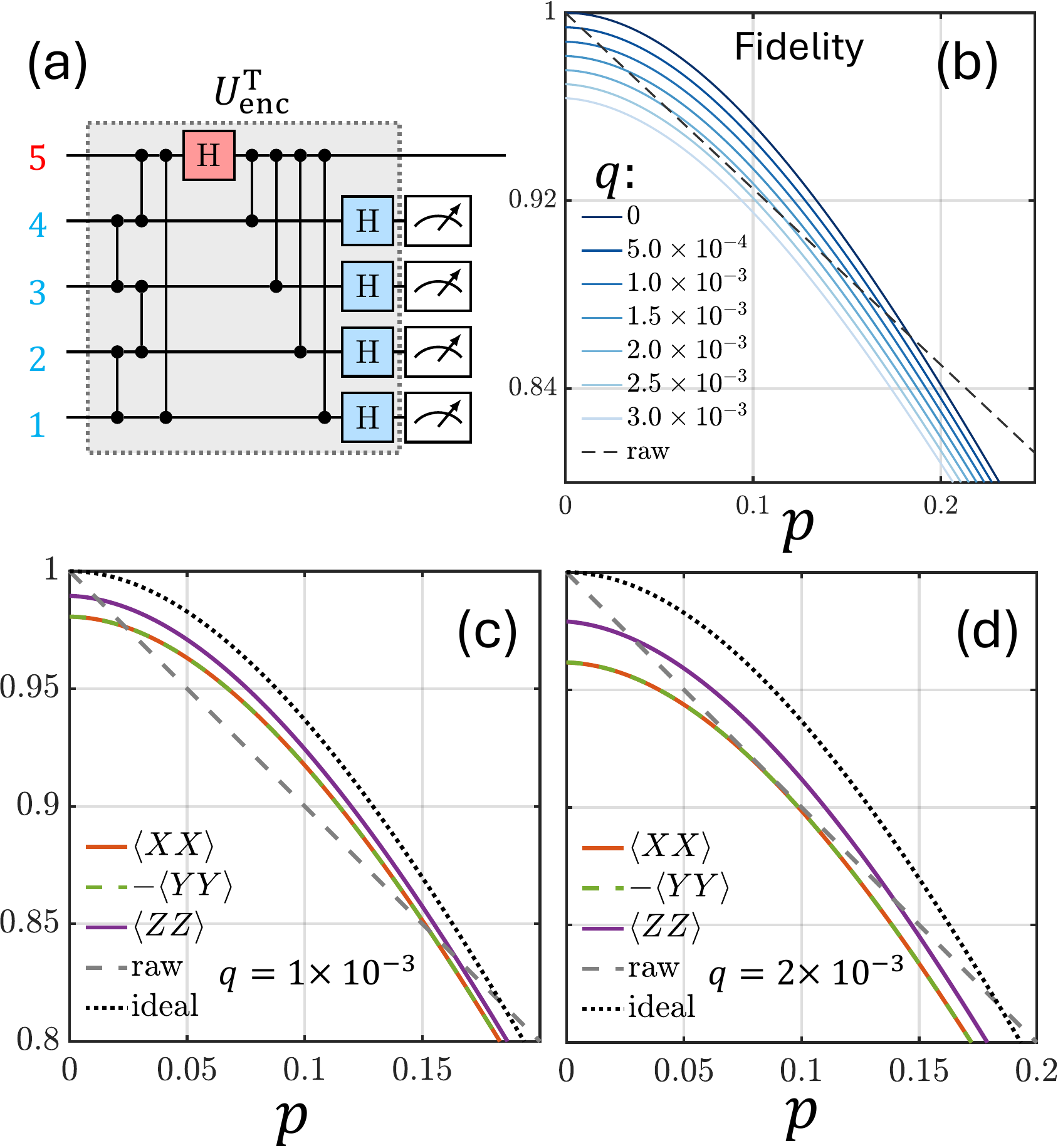}
    \caption{This figure shows the implementation and performance of a 1-EPP based on the $[[5,1,3]]$ code. (a) A highly symmetric decoding circuit for $U_{\text{enc}}^\T$. An equivalent implementation is displayed in Fig.~\ref{fig:CZ_coloring}(c). (b) Output fidelity versus the input noise parameter $p$ for various circuit noise levels $q$. The gray dashed line represents the fidelity of the input state $\hat{\rho}(p)$. (c,d) Two-point correlators of the output state for specific circuit noise levels $q$. For $q=0$, the curves coincide with the black dotted lines. The gray dashed lines indicate the correlators of the input state $\hat{\rho}(p)$.}
    \label{fig:513curves}
\end{figure}

\begin{table}[ht]
\centering
\setlength{\tabcolsep}{8pt}
\begin{tabular}[t]{|cc||cc|}
\hline
$(s_1,s_2,s_3,s_4)$ & $U_{5}$ & $(s_1,s_2,s_3,s_4)$ & $U_{5}$ \\
\hline
\hline
$0\,\,0\,\,1\,\,1$ & $Y$ & $1\,\,0\,\,1\,\,1$ & $Y$ \\   
$0\,\,1\,\,0\,\,1$ & $Z$ & $1\,\,1\,\,0\,\,0$ & $Y$ \\   
$0\,\,1\,\,1\,\,0$ & $Y$ & $1\,\,1\,\,0\,\,1$ & $Y$ \\   
$0\,\,1\,\,1\,\,1$ & $Z$ & $1\,\,1\,\,1\,\,0$ & $Z$ \\   
$1\,\,0\,\,0\,\,1$ & $Z$ & $1\,\,1\,\,1\,\,1$ & $X$ \\   
$1\,\,0\,\,1\,\,0$ & $Z$ & Otherwise & $I$ \\ 
\hline
\hline
\end{tabular}
\caption{Decoder table for the $[[5,1,3]]$ code, with the encoding unitary shown in Fig.~\ref{fig:513curves}(a). As the $[[5,1,3]]$ code is perfect, every nontrivial syndrome corresponds to a unique weight-1 Pauli error, so this assignment realizes the minimum-weight decoder.}
\label{tab:513decoder}
\end{table}

\paragraph{The five-qubit perfect code} Another good example for demonstration is the $[[5,1,3]]$ perfect code, which presents a symmetric circuit in our framework of DACOS. The unitary $U_{\text{enc}}$ displayed in Fig.~\ref{fig:513curves}(a) gives the following transformation:
\begin{equation}
    \begin{aligned}
        Z_1 &\to \widetilde{S}_1=Y_1Z_2Z_4Y_5 ,\\
        Z_2 &\to \widetilde{S}_2=X_2Z_3Z_4X_5 ,\\
        Z_3 &\to \widetilde{S}_3=Z_1Z_2X_3X_5 ,\\
        Z_4 &\to \widetilde{S}_4=Z_1Z_3Y_4Y_5 .\\
    \end{aligned}
\end{equation}
Observe that $\{\widetilde{S}_2, \widetilde{S}_1 \widetilde{S}_4, \widetilde{S}_2 \widetilde{S}_4, \widetilde{S}_3\}$ is exactly the set of conventional stabilizer generators for the $[[5,1,3]]$ code with cyclic symmetry~\cite{gottesman2009introductionquantumerrorcorrection}. 
To see how the circuit in Fig.~\ref{fig:513curves}(a) is obtained, we start from the original $[[5,1,3]]$ stabilizers and follow the aforementioned procedure, and find the standard form: 
\begin{equation}
    \Tb_X = \left[
    \arraycolsep=1.5pt \def\arraystretch{0.9}
    \begin{array}{ccccc}
        1	&0	&0	&0	&1 \\
        0	&1	&0	&0	&1 \\
        0	&0	&1	&0	&1 \\
        0	&0	&0	&1	&1
    \end{array}\right],\,
    \Tb_Z = \left[
    \arraycolsep=1.5pt \def\arraystretch{0.9}
    \begin{array}{ccccc}
        1	&1	&0	&1	&1 \\
        0	&0	&1	&1	&0 \\ 
        1	&1	&0	&0	&0 \\
        1	&0	&1	&1	&1
    \end{array}\right].
\end{equation}
Thus, there are $r_X = 4$, $r_Z = 0$ (i.e., $V_Z\equiv \emptyset$), $V_L = \{5\}$, $H_{(1)}=P_{(1)} = I$, and
\begin{equation}\label{eq:JKL513}
\begin{aligned}
    \left[J_1\,J_2\right] & = 
    \left[\arraycolsep=1.5pt \def\arraystretch{0.9}
    \begin{array}{c}
    1\\0\\0\\1
    \end{array}\right],\, 
    \left[\arraycolsep=1.5pt \def\arraystretch{0.9}
    \begin{array}{c}
        L_2\\ K_2
    \end{array}\right]
    = \left[
    \arraycolsep=1.5pt \def\arraystretch{0.9}
    \begin{array}{c}
    1\\1\\1\\1
    \end{array}\right],\,
    \Gamma_0= \left[
    \arraycolsep=1.5pt \def\arraystretch{0.9}
    \begin{array}{cccc}
        0&1&0&0\\
        1&0&1&0\\
        0&1&0&1\\
        0&0&1&0
    \end{array}\right].
\end{aligned}    
\end{equation}
The above binary matrices yield 9 CZ gates, which are together rearranged into the symmetric layout in Fig.~\ref{fig:513curves}(a).

This quantum code with distance $3$ can be used for one-way EPP, as any single-qubit Pauli error can be detected and corrected. For convenience, we present the corresponding table for the feedback unitary conditioned on the observed syndrome (Table~\ref{tab:513decoder}). With this decoder, Fig.~\ref{fig:513curves} demonstrates the performance of this $[[5,1,3]]$ 1-EPP, including fidelity and correlators at different levels of noise, which suggests that the EPP provides benefit when $q\le 2\times 10^{-3}$. When $q=0$, the protocol ideally gives a fidelity $1 - \frac{45}{8}p^2+\frac{75}{8}p^3-\frac{45}{8}p^4 + \frac{9}{8}p^5$~\cite{RW2005PRL}, shown as the darkest curve in Fig.~\ref{fig:513curves}(b). In this work, we briefly compare the performance of the $[[4,2,2]]$, $[[5,1,3]]$, and $[[7,1,3]]$ approaches in Table~\ref{tab:compare}.

\begin{table*}[t]
\centering
\setlength{\tabcolsep}{8pt}
\begin{tabular}{c c c c c c}
\hline
Code  & $\CZgate$ layers  & Fidelity gain & $p_s$ & Type & $\Tpre$ (DACOS\,/\,SS-opt) \\
\hline
\hline
$[[4,2,2]]$  & 2  & $97\%\to 99.75\% $ & $0.88$ & 2-EPP & $\lesssim 10\,\mu$s\,/\,$\sim 1$\,ms \\
$[[5,1,3]]$  & 6  & $97\%\to 98.52\% $ & $1$ & 1-EPP & $\lesssim 10\,\mu$s\,/\,$\sim 1.5$\,ms \\
$[[7,1,3]]$  & 4  & $97\%\to 98.06\% $ & $1$ & 1-EPP & $\lesssim 10\,\mu$s\,/\,$\sim 1.5$\,ms \\
\hline
\hline
\end{tabular}
\caption{Comparisons between EPPs constructed by different codes. The fidelity and success probability are evaluated at $p=0.04, q=0.0005$. The fidelity refers to the fidelity of the reduced state of one qubit pair (if $\rho_{\text{out}}$ is a multi-qubit-pair state). The $\CZgate$ layer number is based on the discussion in Sec.~\ref{sec:coloring}. The last column estimates the per-lab per-round \emph{pre-measurement} runtime $\Tpre$ via Eq.~\eqref{eq:runtime}: the DACOS entry counts only gate pulses in a single working zone with $N_{\mathrm{move}}=0$ inter-zone moves, while the SS-opt entry assumes an optimally compiled single-species zoned implementation~\cite{Lin2024,Jang2025} with $\tLS\approx 300\,\mu$s~\cite{Bluvstein2024,Jang2025}. The measurement contribution $\TM=t_{M}\sim 7$\,ms~\cite{Singh_2023} adds identically to both columns and is omitted here as a common bottleneck across the neutral-atom platform. See the $[[4,2,2]]$ worked example in Sec.~\ref{sec:dual_advantage}.}
\label{tab:compare}
\end{table*}

\section{Discussion}\label{sec:Discussion}

\subsection{Universality of DACOS}
The dual-species atomic platform equipped with DACOS offers great flexibility for various protocols. Recent work~\cite{Cesa2023PRL} has demonstrated that the universality of quantum circuits can be achieved by representing the world line of each qubit as a one-dimensional atom array, where superatoms are required to implement the protocol effectively. However, employing superatoms demands careful maneuvering when trapping multiple atoms in close proximity, presenting an additional experimental challenge.

When focusing on entanglement purification protocols, our approach minimizes the use of ancillary structures, as reducing the size of the atom array can potentially mitigate the impact of atom loss. We demonstrate that, without any ancillary atoms, the encoding unitary of arbitrary (LC-equivalent) stabilizer codes can be efficiently compiled under a global laser field. This set of arbitrary stabilizer codes includes pure stabilizer states, which constitute a special case with zero logical qubits.

Furthermore, DACOS is sufficient to implement an arbitrary non-Clifford quantum unitary using a small number of additional ancillary atoms, provided that swap gates are compiled with $\mathcal{CZ}$ and Hadamard gates. In this scenario, ancillary atoms of one species are employed to inject and extract quantum information into and from the data atoms of the other species, with details provided in Appendix~\ref{app:univ_circuit}. A potential drawback is that a lengthy operation sequence may be required, which could degrade quantum state coherence due to accumulated errors, thereby contradicting our intent for efficiency.

In summary, based on DACOS, we have found an efficient realization of stabilizer-code-based entanglement purification protocols on dual-species Rydberg atom arrays without ancillary atoms or a lengthy operation sequence. The framework accommodates both one-way and two-way protocols, generalizes to arbitrary stabilizer codes, and optimizes for shallow-depth $\mathcal{CZ}$ sequences. By aligning algorithmic design with the strengths of current Rb–Cs hardware, this work offers a practical pathway to high-yield, high-fidelity entanglement distribution in near-term neutral-atom quantum networks.

\subsection{Advantage of dual-species architecture}\label{sec:dual_advantage}
The advantage of the dual-species architecture for EPP goes beyond merely reducing atom movements. We highlight the key structural benefits compared to a single-species implementation:
\begin{enumerate}[label=(\roman*)]
    \item \textit{Species-selective global control without local addressing.} In our DACOS framework, single-qubit rotations are applied globally to all atoms of a given species via a single laser pulse. This eliminates the need for local addressing beams entirely. In a single-species array, implementing the same EPP circuit would require either local addressing beams, which introduce crosstalk errors, or shuttling atoms between spatially separated processor and storage zones, which adds decoherence.
    \item \textit{Non-destructive species-selective measurement.} The retained logical qubits and the measured syndrome qubits belong to different atomic species. Measurement of one species via fluorescence imaging at a species-specific wavelength does not disturb the other species, enabling efficient mid-circuit measurement without physical separation of atoms.
    \item \textit{Ancilla-free, shallow circuits.} As reported in Table~\ref{tab:compare}, the DACOS compilation produces circuits with only 2--6 CZ layers and no ancillary atoms. This compact structure exploits the fact that Hadamard gates on different species commute with each other and can be applied simultaneously, which is not possible under single-species global control.
    \item \textit{Interspecies CZ gates.} The interspecies Rydberg interaction, such as the Rb--Cs resonant dipole-dipole coupling, can provide a stronger blockade than the single-species van der Waals interaction at the same interatomic distance, potentially enabling higher two-qubit gate fidelity or relaxing the proximity requirement.
\end{enumerate}

To make these advantages quantitative, we compare with a transport-aware runtime model. For both dual-species and zoned single-species realizations, we split the duration $T$ of one EPP round into a pre-measurement contribution $\Tpre$ for transport and gate pulses, and a measurement contribution $\TM$ for fluorescence imaging of the syndrome qubits,
\begin{equation}\label{eq:runtime}
\begin{aligned}
    T &= \Tpre + \TM, \\
    \Tpre &:= N_{\mathrm{move}}\, \tLS + N_{\mathrm{CZ}}\, \tCZ + N_{1q}\, \toneq,\quad \\
    \TM &:= N_{M}\, t_{M},
\end{aligned}
\end{equation}
where $\tLS$ denotes the inter-zone Load/Store time, $\tCZ$ and $\toneq$ denote two-qubit and single-qubit gate durations, $N_{M}$ is the number of measurement steps per round, and $t_{M}$ is the per-measurement readout time. Here $N_{\mathrm{move}}$ counts inter-zone Load/Store cycles only. Intra-zone AOD-tweezer transits within a working zone are a one-time setup and reset cost estimated below, and are not included in $\Tpre$. Current neutral-atom experiments report $\tCZ \approx 380$\,ns~\cite{Evered2023,Jang2025} and $\toneq \approx 625$\,ns~\cite{Evered2023,Jang2025}, whereas a typical Load/Store cycle takes $\tLS\approx 0.2$--$0.5$\,ms with median $\sim 530\,\mu$s~\cite{Bluvstein2024,Jang2025}, dominated by two $\sim 150\,\mu$s AOD--SLM trap handoffs per cycle~\cite{Jang2025}. Hence $\tLS/\tCZ\sim 10^{3}$, and inter-zone movements consume $78$--$90\%$ of the pre-measurement runtime even in compiled programs~\cite{Jang2025}, so movement overhead is the dominant cost \emph{within} $\Tpre$. The measurement time $t_{M}\sim 7$\,ms for dual-species mid-circuit fluorescence readout~\cite{Singh_2023} is, in contrast, a bottleneck shared across the neutral-atom platform: $t_{M} \gg \tLS \gg \tCZ$ in any current neutral-atom EPP, regardless of the choice between single- and dual-species architectures. We therefore claim a runtime advantage only for $\Tpre$. $\TM$ enters identically in both architectures and could be reduced to the $\mu$s regime by recently proposed schemes for fast Rydberg-assisted readout~\cite{Norcia2024}.

The dual-species implementation reduces movement overhead by replacing selective local operations and selective readout with species-selective global operations, directly reducing $N_{\mathrm{move}}$. On top of this hardware-level gain, our DACOS compilation further reduces effective depth through shallow CZ layering. Starting from the stabilizer parity-check matrix, our method directly constructs DACOS-compatible circuits and optimizes the CZ schedule via multigraph edge coloring as discussed in Sec.~\ref{sec:coloring}, yielding shallow CZ depths of 2, 4, and 6 layers for the representative codes in Table~\ref{tab:compare}, and avoiding unnecessary intermediate rearrangements. In contrast, a naive gate-by-gate mapping of the foundational protocol~\cite{bennett_1996_PRL} does not enforce this structure and therefore leads to larger movement overhead and deeper effective schedules. These reductions translate into better practical performance under finite circuit noise $q$, as reflected in higher output fidelity and/or success probability in the same operating regime (Table~\ref{tab:compare}).

As a concrete worked example, consider one round of the $[[4,2,2]]$ 2-EPP at Alice's lab, shown in Fig.~\ref{Fig:EPPnn22}(b): four atoms, two $\CZgate$ layers, and a single selective $Z$-basis readout of two atoms at the end of the round, so $N_{M}=1$. In the DACOS implementation, all qubit operations stay within a single working zone. Single-qubit gates and fluorescence readout act selectively on one species at a time and require no inter-zone shuttling, so $N_{\mathrm{move}} = 0$ and $\Tpre^{(\mathrm{DACOS})} \lesssim 5\,\mu$s per round, dominated by gate pulses. We further assume that atoms occupy fixed Rb--Cs lattice sites with built-in adjacency for interspecies Rydberg blockade, as in current dual-species array experiments~\cite{Singh2022PRX,anand2024dualspecies}, so successive CZ layers are realized by switching the species-selective laser drive rather than by repositioning atoms. The intra-zone AOD-tweezer transit, typically $\sim 20\,\mu$s per lattice edge~\cite{Jang2025}, is incurred only for one-time setup and reset, not per CZ layer, and is therefore negligible compared with the inter-zone Load/Store cost in zoned single-species architectures. For comparison, we consider the optimally compiled single-species zoned baseline, denoted SS-opt~\cite{Lin2024,Jang2025}, where selective single-qubit control and selective $Z$-basis readout still require shuttling between storage, entanglement, and readout zones. A representative budget is $N_{\mathrm{move}}^{(\mathrm{SS\text{-}opt})} \sim 4$ inter-zone moves per lab per round, giving $\Tpre^{(\mathrm{SS\text{-}opt})} \sim 1$\,ms, more than two orders of magnitude longer than $\Tpre^{(\mathrm{DACOS})}$. The measurement contribution $\TM=t_{M}\sim 7$\,ms~\cite{Singh_2023} adds equally to both implementations, and is a bottleneck shared across the neutral-atom platform rather than an architectural difference. The corresponding cumulative transport-induced infidelity, taking the experimentally reported per-move transfer fidelity $\sim 99.9\%$~\cite{Lin2024} and parallel-rearrangement cycle survival $\sim 99.6\%$~\cite{Knottnerus2025}, accumulates to $\gtrsim 1\%$ per round in the single-species case versus $0$ in DACOS, comparable to the per-round fidelity gain of the protocol from $97\%$ to $99.75\%$ reported in Table~\ref{tab:compare}.

Looking ahead, if the recently proposed schemes for fast Rydberg-assisted measurement on dual-species arrays~\cite{Norcia2024} mature, $t_{M}$ could drop into the $\mu$s regime. This would eliminate the common measurement bottleneck and restore the DACOS runtime advantage of several orders of magnitude in $\Tpre$ as the dominant cost of the full EPP round.

\subsection{Alternative EPP schemes}
The dual-species setup is naturally suited for stabilizer measurements using ancillary qubits, where data qubits are assigned to atoms of one species, and the qubits used for extracting syndromes are assigned to atoms of the other species. There are two major ways of implementing EPPs based on ancilla qubits. 

Suppose we have two distant labs. The first approach is distributing noisy entanglement to the atoms of the \textit{same} species in each lab. Then the relatively pure ancilla qubits are prepared locally on the other species of atoms, which ensures reliable stabilizer measurements. This approach ends up with purified entangled qubits on the \textit{logical level}, as in the scheme shown in~\cite{Ataides2025}. 

The second approach requires a good QECC allowing successful decoding with noisy ancilla qubits (as noisy as the data qubits). Any good quantum low-density parity-check code with a single-shot decoder~\cite{Gu2024} will be an efficient candidate, as only one round of stabilizer measurement is needed. In this case, we consider distributing noisy entanglement to two \textit{different} species of atoms in each lab. Each lab uses the atoms of one species as the data qubits and the other ones for the stabilizer measurements. If such a good QECC operates above the noise threshold, we end up with entangled qubits on the \textit{physical level} by consuming noisy entangled ancillae.

\subsection{Related EPP literature}\label{sec:related_epp}
Our approach to EPP, built from stabilizer codes, connects to several lines of work. Aschauer~\cite{Aschauer2005thesis} established the theoretical foundations for constructing EPPs from quantum error-correcting codes, but did not address compilation onto a specific hardware platform. Our contribution is to provide an explicit hardware-aware compilation scheme. We show that the encoding circuit of any stabilizer code can be decomposed into the structured DACOS-compatible form (Eq.~\eqref{eq:Uenc_stab}), which is directly implementable on dual-species atom arrays without ancillary atoms, local addressing, or complex atom rearrangements.

Jansen \textit{et al.}~\cite{Jansen2022} provide a comprehensive catalogue of two-way EPPs for small numbers of input pairs using symmetry reduction techniques. These protocols are derived without reference to a specific hardware platform. Translating a generic EPP from this catalogue into a dual-species setup would require case-by-case circuit decomposition, since the unitary operations do not, in general, have the structured H-CZ-H-CZ-H form that DACOS provides. Our framework takes a complementary approach: by starting from stabilizer codes, we obtain EPPs that are \emph{by construction} compilable into DACOS-compatible circuits with guaranteed shallow CZ depth. Protocols from~\cite{Jansen2022} that happen to be equivalent to stabilizer-code-based EPPs can be compiled through our framework.

The software tool of Krastanov \textit{et al.}~\cite{Krastanov2019} searches over EPP circuits to optimize fidelity and success probability, treating the circuit as an abstract sequence of gates. One could in principle use such a tool to identify high-performing stabilizer codes or near-stabilizer protocols, and then apply our DACOS compilation framework to map the result onto a dual-species atom array. The CZ layer count in our framework is determined by the chromatic index of the CZ connectivity graph (Sec.~\ref{sec:coloring}), which depends on the specific stabilizer code. The trade-off between code distance (error-correcting capability) and CZ depth (circuit noise accumulation) is summarized in Table~\ref{tab:compare}.

Finally, recent work by Miguel-Ramiro \textit{et al.}~\cite{MiguelRamiro2025} proposes non-Clifford EPPs that achieve performance beyond what Clifford EPPs (built from stabilizer codes) can offer. Regarding DACOS compatibility, as discussed above, DACOS is sufficient to implement arbitrary non-Clifford unitaries with a small number of additional ancillary atoms (see Appendix~\ref{app:univ_circuit}). Therefore, the non-Clifford EPPs of~\cite{MiguelRamiro2025} can in principle be compiled for dual-species atom arrays, but at the cost of introducing ancillary atoms and potentially lengthier operation sequences. For near-term implementations that prioritize shallow circuits and minimal ancilla overhead, we focus on Clifford EPPs based on stabilizer codes. As hardware improves and coherence times extend, non-Clifford EPPs may become practical on dual-species platforms.

\subsection{Optimization of CZ sequences}\label{sec:coloring}
In real neutral atom platforms that employ Rydberg blockade for CZ gates, CZ gate infidelity represents a primary source of error. Careful calibrations are essential to achieve high gate fidelity. To maintain tractability under Rydberg blockade, we restrict pairings to one-to-one atom interactions. In this framework, any two CZ gates can be implemented simultaneously if they operate on disjoint sets of atoms. We assume that switching laser fields is far more convenient and desirable than relocating atom coordinates. 
So a layer of parallelizable CZ gates simply consists of pairing atoms, inducing Rydberg blockade by a laser pulse sequence, and subsequent atom separation.

\begin{figure}[t]
    \centering
    \includegraphics[scale = 0.55]{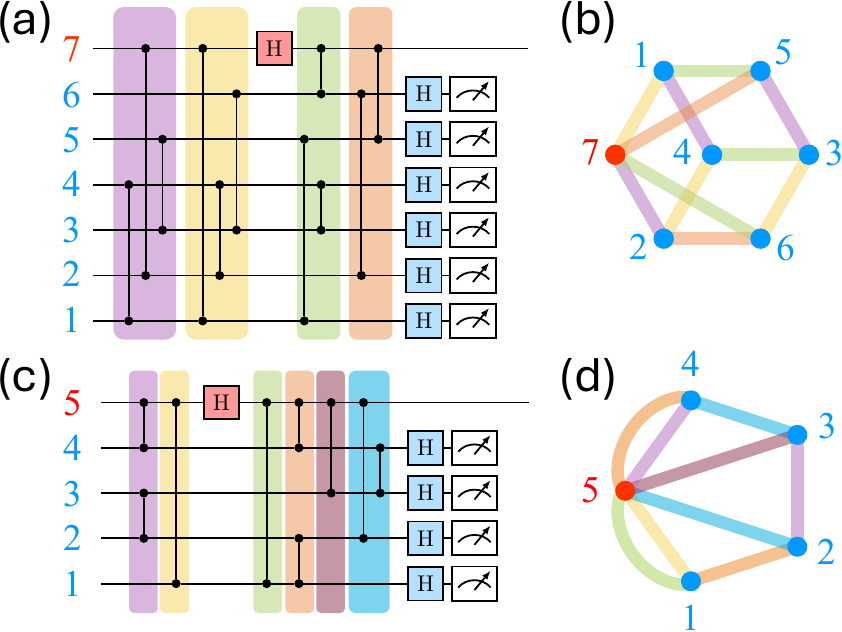}
    \caption{Panels (a) and (c) show equivalent implementations of the circuits in Fig.~\ref{fig:stab_circuit}(b) and Fig.~\ref{fig:513curves}(a), respectively.  Panels (b) and (d) are the multigraphs corresponding to the CZ sequences in (a) and (c).}
    \label{fig:CZ_coloring}
\end{figure}

For $U_{\text{enc}}$, we anticipate a small number of CZ layers, resulting in a shallow overall circuit. Given a fixed stabilizer code, the flexibility of $\hat{T}$ and row transformation can lead to different outcomes of $J_i$, $K_i$, and $L_i$, which determine the specific form of $U_1$ and $U_2$. 
The total connectivity of $U_1$ and $U_2$ is represented by a loop-less undirected \textit{multigraph} $G=(V,E)$, where $V$ and $E$ are sets of vertices and edges. Moreover, the global maximum edge multiplicity $\mu(G)$ fulfills $\mu(G)\le 2$ (due to the two $U_i$'s). We can represent the multigraph $G$ by the following \textit{integer-valued} adjacency matrix:
\begin{equation}
    A(G) = \begin{bmatrix}
        \Gamma_0&J_1&J_2+L_2&\\
        J_1^\T&0&K_2&\\
        J_2^\T+ L_2^\T&K_2^\T&0
    \end{bmatrix} \in \mathbb{Z}^{n\times n}.
\end{equation}
Specifically, when $L_2 = 0$, $G$ reduces to a simple graph ($\mu(G) = 1$).
Rearranging the CZ sequence into $\ell$ layers of parallel CZ gates equates to determining the chromatic index $\chi'(G)$ of $G$, that is $\ell = \chi'(G)$. From Favrholdt, Stiebitz and Toft~\cite{stiebitz2012graph}, we have
\begin{equation}
    \delta(G) \le \ell  \le \min_{v\in V}\{\delta(G)+\mu(G-v),\lfloor 3\delta(G)/2\rfloor\}, 
\end{equation}
where $\delta(G)$ is the maximum degree of $G$ among all vertices, and $G-v$ is a subgraph of $G$, obtained by removing vertex $v$ and its incident edges from $G$. 
In Fig.~\ref{fig:CZ_coloring}, graph edges and CZ layers share the same colors to illustrate the correspondences.
Particularly, Fig.~\ref{fig:CZ_coloring}(b) depicts the multigraph $G$ for the circuit in Fig.~\ref{fig:CZ_coloring}(a), which is reduced to a simple graph due to $L_2 = 0$. The chromatic index $4$ indicates that $\ell = 4$ of such disjoint CZ gate layers are needed. However, Fig.~\ref{fig:CZ_coloring}(d) is a multigraph with $\mu(G) = 2$, which corresponds to the circuit in Fig.~\ref{fig:CZ_coloring}(c). The chromatic index $6$ suggests that $\ell = 6$ layers of disjoint CZ sequences are needed.
Usually, determining the chromatic index of $G$ is NP-complete. Together with the flexibility of $A(G)$, the circuit optimization is overall a challenging problem. We verified that $4$ layers and $6$ layers are optimal in Fig.~\ref{fig:CZ_coloring} after exhaustive searches.

\section{Acknowledgements}
We thank Allen Zang for useful discussions regarding the distillation rate.
This material is based upon work supported by the U.S. Department of Energy, Office of Science, Advanced Scientific Computing Research (ASCR) program under contract number DE-AC02-06CH11357 as part of the InterQnet quantum networking project. B.L. and L.J. also acknowledge support from the ARO (W911NF-23-1-0077), ARO MURI (W911NF-21-1-0325), AFOSR MURI (FA9550-19-1-0399, FA9550-21-1-0209, FA9550-23-1-0338), DARPA (HR0011-24-9-0359, HR0011-24-9-0361), NSF (OMA-1936118, ERC-1941583, OMA-2137642, OSI-2326767, CCF-2312755), NTT Research, Packard Foundation (2020-71479), and the Marshall and Arlene Bennett Family Research Program. R.W. acknowledges support from the NSF GRFP (Grant No. 2140001).
T.H. and R.A. acknowledge support from the Marshall and Arlene Bennett Family Research Program, the Peter and Patricia Gruber Award, and the Air Force Office of Scientific Research under award number FA9550-22-1-0391. 
R.A. was further generously supported by the Koshland Research Fund and is the Daniel E. Koshland Career Development Chair.

\appendix
\section{Universality of Performing Quantum Circuits}\label{app:univ_circuit}

To demonstrate universality, we assume that all quantum gates at the physical level are implemented perfectly.
Suppose we wish to perform an arbitrary Clifford unitary operation $U_C$ on the quantum state supported on Rb atoms labeled $1 \le i \le N$. This can be achieved using an additional ancillary Cs atom labeled $i=0$. Specifically, various works have discussed the decomposition of $U_C$ into the fundamental Clifford gate set $\{\Hgate_i, \Pgate_i, \CXgate_{ij}\}$~\cite{Aaronson2004,Bravyi_2021TIT}, where $\Hgate_i = \frac{1}{\sqrt{2}} \left(X_i + Z_i \right)$, $\Pgate_i = \ket{0}\!\bra{0}_i + i \ket{1}\!\bra{1}_i$, and $\CXgate_{ij} = \Hgate_j \CZgate_{ij} \Hgate_j$.

Let $\Hgate_{i,\alpha}$ represent the Hadamard gate acting on the $i$-th qubit of atom species $\alpha$, and $\Hgate_{\alpha}:=\otimes_i \Hgate_{i,\alpha}$ be the global Hadamard gate. Implementing a single-qubit unitary $U$ on a specific Rb atom $i$ can then be accomplished via the following sequence:
\begin{equation}\label{eq:Ui_Rb}
    U_{i,\Rb} = \mathrm{SWAP}_{(i,\Rb),(0,\Cs)} \, U_{0,\Cs} \, \mathrm{SWAP}_{(i,\Rb),(0,\Cs)},
\end{equation}
where the $\mathrm{SWAP}$ gate is realized by the unitary sequence
\begin{equation}\label{eq:SWAP0i}
    \begin{aligned}
        &\mathrm{SWAP}_{(i,\Rb),(0,\Cs)} = \Hgate_{\Rb} \CZgate_{(i,\Rb),(0,\Cs)} \\
        &\qquad\times(\Hgate_{\Rb} \Hgate_{0,\Cs} \CZgate_{(i,\Rb),(0,\Cs)})^2 \Hgate_{\Rb}.
    \end{aligned}
\end{equation}
Since any $U_{i,\Rb}$ is accessible, it is straightforward to generate an arbitrary unitary together with the entangling gate $\CZgate_{ij, \Rb}$ within the set of qubits of the Rb atoms. As for implementing computational basis measurement on the $i$-th Rb atom, the SWAP trick is applied to transfer the quantum state of a particular Rb atom to the ancillary Cs atom. Then the measurement $\mathcal{M}_{\Cs}$ allows one to measure the desired qubit.

The above approach may not be efficient for near-term devices. For instance, when implementing the SWAP trick, the globally implemented $\Hgate_{\Rb}$ may introduce unnecessary noise on idling atoms.

In contrast, our work aims for ancilla-free efficient protocols. For certain specific tasks on noisy intermediate-scale devices, a well-calibrated DACOS subset $\mathcal{T}\cup\{\Hgate_{\Rb}, \Hgate_{\Cs}\}\cup \mathcal{CZ}$ is sufficient for a large family of EPPs under compact circuits.

\section{Fidelity Calculations for $[[n,n-2,2]]$ 2-EPP}\label{app:iceberg_fid}

\begin{figure}[t]
    \centering
    \includegraphics[scale=0.3]{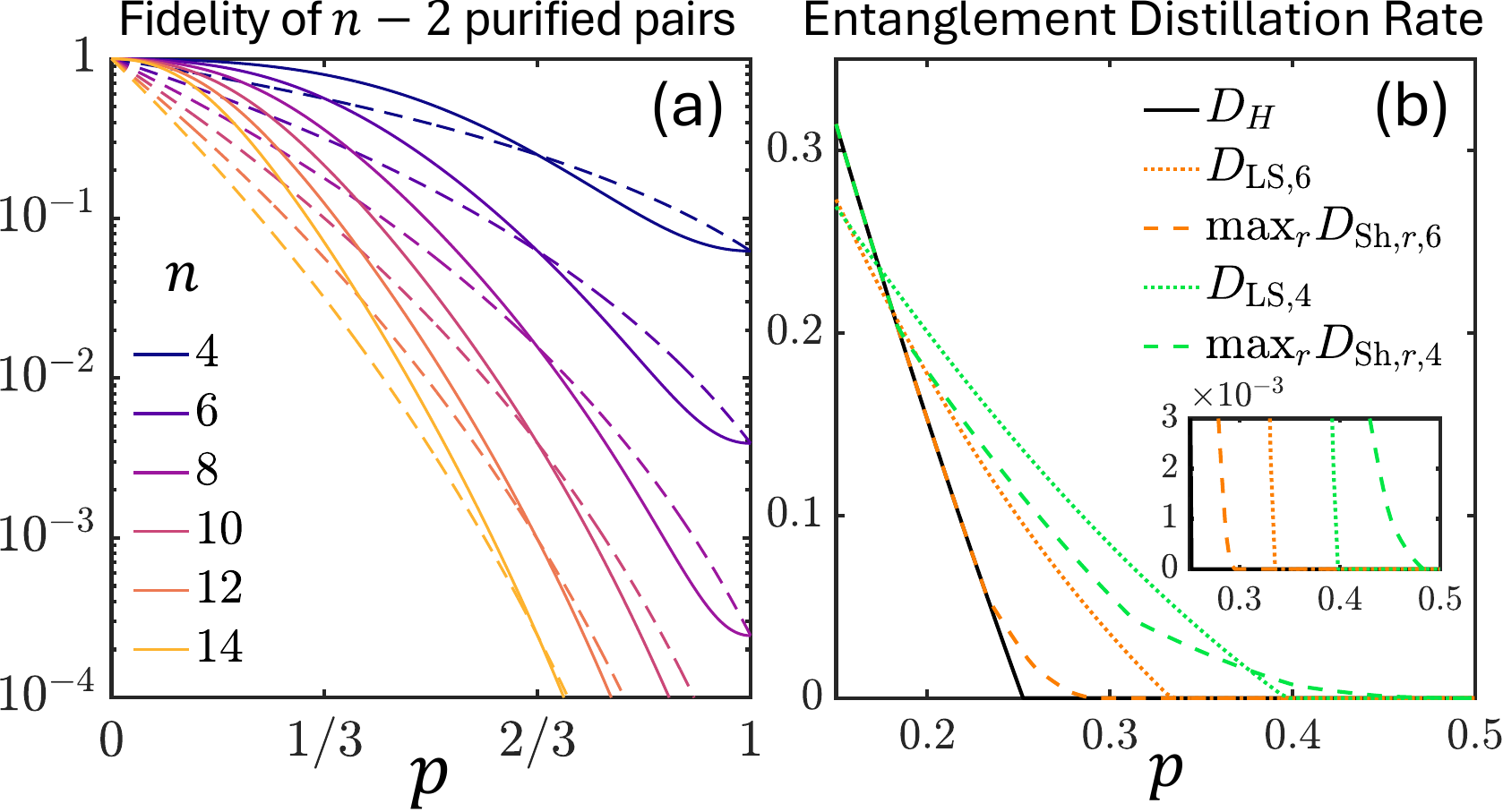}
    \caption{The solid lines in panel (a) display the fidelity Eq.~\eqref{eq:iceberg_fid} for different values of $n$ and $p$. The dashed lines are the values of $(1-3p/4)^{n-2}$ of the corresponding $n$, which is the fidelity of the $\hat{\rho}(p)^{\otimes(n-2)}$. Panel (b) displays the values of $D_{\text{LS},n}(p)$ and $\max_r D_{\text{Sh},r,n}$ for $n = 4$ and $6$. $D_H$ is the hashing bound for the isotropic state $\hat{\rho}(p)$.}
    \label{fig:app}
\end{figure}

In general, if an $[[n,n-2,2]]$ 2-EPP ($n\ge 4$) can be performed perfectly against a noisy quantum channel $\EE_p^{(1)\otimes n}$, the fidelity with respect to the target state $\ket{\psi_0} := \ket{\Phi_+}^{\otimes (n-2)}$ is given by
\begin{equation}\label{eq:iceberg_fid}
    \begin{aligned}
        F(\psi_0, \rho_{\text{out}}) &=\bra{\psi_0}\rho_{\text{out}}\ket{\psi_0} \\
        &=  \frac{\left(1 - \frac{3p}{4}\right)^{n} + 3 \left(\frac{p}{4}\right)^{n}}{\frac{1}{4} + \frac{3}{4} (1-p)^{n}}\\
        &=1-\frac{3n(n-1)}{32} p^2 + \mathcal{O}(p^3).
    \end{aligned}
\end{equation}
We display the above result in Fig.~\ref{fig:app}. One can explicitly verify that $F(\psi_0,\rho_{\text{out}}) = (1-3p/4)^{n-2}$ at $p=2/3$, and conclude that the \textit{block} fidelity is improved if $0<p<2/3$ for all $n\ge 4$. Such a fidelity improvement implies that the $[[n,n-2,2]]$ 2-EPP scheme may be treated as a generalization of the recurrence method. The latter can distill the qubit isotropic state whenever it is distillable~\cite{bennett_1996_PRL,Concurrence1997PRL}.

The derivation of Eq.~\eqref{eq:iceberg_fid} proceeds as follows. Assume that the Pauli operators $X$, $Y$, and $Z$ occur independently on each qubit with probability $p/4$. The probability of no logical error corresponds to the error-free case or the same type of Pauli operator acting on all qubits simultaneously, namely
\begin{equation}\label{eq:p_good}
    \left(1 - \frac{3p}{4}\right)^{n} + 3 \left(\frac{p}{4}\right)^{n}.
\end{equation}
The denominator of $F(\psi_0, \rho_{\text{out}})$ represents the success probability of the protocol. The protocol succeeds if all stabilizers of the $[[n,n-2,2]]$ code commute with the Pauli operator $\hat{E}$ imposed by $\EE^{(1)\otimes n}_p$. Let $\hat{E}$ consist of $w_x$ Pauli $X$'s, $w_y$ Pauli $Y$'s, and $w_z$ Pauli $Z$'s. For commutativity, it must hold that $w_x \equiv w_y \equiv w_z \pmod{2}$, ensuring that $\hat{E}$ commutes with all stabilizers and is thus undetectable. There are
\begin{equation}
    n_w = \binom{n}{w} \sum_{\substack{w_x,w_y,w_z\ge 0\\w_x+w_y+w_z=w}} \frac{w!}{w_x! w_y! w_z!} \delta^{(2)}_{w_x,w_y} \delta^{(2)}_{w_x,w_z}
\end{equation}
different configurations of such $\hat{E}$, where $w := w_x + w_y + w_z$ is the weight of $\hat{E}$, and $\delta^{(2)}_{a,b}=1$ if $a\equiv b\pmod 2$ and $0$ otherwise. These parity delta functions can be expressed as
\begin{equation}
    \begin{aligned}
        &\quad \delta^{(2)}_{w_x,w_y} \delta^{(2)}_{w_x,w_z} \\
        &= \frac{1}{4} \left( 1 + (-1)^{w_x + w_y} + (-1)^{w_y + w_z} + (-1)^{w_z + w_x} \right) \\
        &= \frac{1}{4} \left( 1 + (-1)^{w - w_z} + (-1)^{w - w_x} + (-1)^{w - w_y} \right),
    \end{aligned}
\end{equation}
which yields
\begin{equation}
    n_w = \frac{1}{4} \binom{n}{w} \left( 3^w + (-1)^w 3\right).
\end{equation}
Given that the probability for each $X$, $Y$, or $Z$ error is $p/4$, the total probability of undetectable $\hat{E}$ is
\begin{equation}\label{eq:p_undetected}
    \sum_{w=0}^{n} n_w \left(1 - \frac{3p}{4}\right)^{n - w} \left( \frac{p}{4} \right)^w = \frac{1 + 3 (1-p)^{n}}{4}.
\end{equation}
Combining Eq.~\eqref{eq:p_good} and Eq.~\eqref{eq:p_undetected} thus gives Eq.~\eqref{eq:iceberg_fid}.

\section{Asymptotic Distillation Rate Calculation}\label{app:ED}
This section presents the calculation of the distillation rate, leading to the results shown in Fig.~\ref{fig:europe}. The original recurrence-hashing method comprises two major steps~\cite{bennett_1996_PRL}: (i) converting the input $\hat{\rho}(p)^{\otimes N}$ to $\hat{\rho}(\ptilde)^{\otimes \Ntilde}$ using the $2\to 1$ EPP; (ii) applying the hashing protocol to $\hat{\rho}(\ptilde)^{\otimes \Ntilde}$, which yields $\hat{\rho}(p_f)^{\otimes N_f}$. When $N$ is sufficiently large, $N_f/N$ converges to a finite ratio while $p_f$ converges to $0$. The ratio $\lim_{N\to \infty} N_f/N$ is the desired asymptotic entanglement distillation rate. 

When the $2\to 1$ subroutine is replaced with the $[[n,n-2,2]]$ code 2-EPP, the entire distillation protocol can be more complicated due to the correlated multi-pair output. 
The correlated Pauli errors reduce the effectiveness of direct iteration of the $[[n,n-2,2]]$ code.
So, a straightforward solution is to apply the hashing protocol to identify the typical Pauli error. When $n=4$, it is known as the Leung-Shor method~\cite{Leung_Shor_2008}. We can generalize it to the $n\ge 4$ case with any even $n$, and call it the $\text{LS}(n)$ protocol. For the normalized joint output state $\rho_{\text{out}}(p)$ of a successful round, its distillation rate is
\[
    \begin{aligned}
        D_{\text{LS},n}(p)
        &=\frac{p_s(p)}{n}\max\left\{0,n-2-S\!\left(\rho_{\text{out}}(p)\right)\right\},\\
        p_s(p)&=\frac{1+3(1-p)^n}{4},
    \end{aligned}
\]
where $p_s(p)$ is the round success probability and $S(\rho):=-\tr(\rho\log_2\rho)$ is the von Neumann entropy.

However, by slightly scaling up $N$ and careful shuffling (see Appendix~\ref {app:shuffling}), we can always gather the \textit{uncorrelated pairs} for the next level of $[[n,n-2,2]]$ 2-EPP and the hashing protocol, without any wasted qubits. Due to the symmetry of $[[n,n-2,2]]$, when there is no circuit-level noise, the $n-2$ entangled pairs are in general correlated, but they share an identical reduced state. Thus, we can concatenate the $[[n,n-2,2]]$ 2-EPP with identical uncorrelated inputs for $r$ rounds, and finish the protocol with $O(1)$ qubit hashing protocols. We denote this scheme with shuffling as the $\text{Sh}(r,n)$ protocol. Its distillation rate is denoted as $D_{\text{Sh},r,n}(p)$ $(r\ge 0)$. Note that, when $r=0$, $D_{\text{Sh},0,n}(p)$ represents the hashing bound of $\hat{\rho}(p)$. When $r=1$, $D_{\text{Sh},1,n}(p)$ is generally worse than $D_{\text{LS},n}(p)$ due to the usage of uncorrelated outcomes.
However, $\max_rD_{\text{Sh},r,n}(p)$ may still be greater than $D_{\text{LS},n}(p)$ at some $p$ as shown in Fig.~\ref{fig:app}(b).

Based on the $[[n,n-2,2]]$ 2-EPP, there are many other approaches for asymptotic entanglement distillation. Due to the optimization complexity, we only consider the best distillation rate defined by:
\begin{equation}
    D_{[[n]]}(p) := \max\left\{D_{\text{LS},n}(p),\,\max_{r\ge 0}D_{\text{Sh},r,n}(p)\right\}.
\end{equation}
In Fig.~\ref{fig:app}(b), we display the values of $D_{\text{LS},n}(p)$ and $\max_r D_{\text{Sh},r,n}$ for $n = 4$ and $6$. The performance with a greater $n$ is worse. Optimizing the result of $n=4$ gives $D_{[[4]]}$, as presented in Fig.~\ref{fig:europe}(a). For the numerical curves in Figs.~\ref{fig:europe}(a) and~\ref{fig:app}(b), the shuffled-rate maximum is evaluated over $0\le r\le 4$, while the threshold $p\approx0.51$ quoted in Fig.~\ref{fig:europe}(a) refers to the unrestricted-round limit.

\subsection{The shuffling subroutine}~\label{app:shuffling}
Now we elaborate the shuffling details in the $\text{Sh}(r,n)$ protocol. Unlike the $2\to 1$ scheme, the output state of the $[[n,n-2,2]]$ 2-EPP $\hat{\rho}_{\text{out}}$ consists of $n-2\ge 2$ pairs of qubits ($n\ge 4$). Moreover, $\hat{\rho}_{\text{out}}$ \textit{cannot} be decoupled as a product of its reduced states:
\begin{equation}
    \hat{\rho}_{\text{out}} \not\to \rho_{\text{red},1}\otimes \rho_{\text{red},2}\otimes\cdots \otimes \rho_{\text{red},{n-2}}
\end{equation}
by any \textit{linear} map, where $\rho_{\text{red},i}$ denotes the reduced state of the $i$-th qubit pair of $\hat{\rho}_{\text{out}}$. Since a map that outputs the product of its input's marginals is necessarily nonlinear.
When implementing the $[[n,n-2,2]]$ 2-EPP and the hashing protocol, the errors can be effectively removed if the input state is assumed to be a product state of independent copies. The existence of input correlation may degrade the outcome performance.

Here, we complement the details of enforcing independent inputs by scaling up the entire distillation scheme with a careful arrangement.
Suppose there are $r \ge 0$ rounds of $[[n,n-2,2]]$ 2-EPP concatenation, with the total input state given by $\hat{\rho}(p)^{\otimes N}$, where $N = n^r M$ is a large number with scalable $M$ and constant $n$ and $r$. We assign each qubit pair in $\hat{\rho}(p)^{\otimes N}$ a unique serial number:
\begin{equation}
    \mathbf{x}:=(x_0;x_1,x_2,\cdots,x_r),
\end{equation}
where $x_0 \in \{1,2,\cdots, M\}$ and $x_i \in \{1,2,\cdots, n\}$ for $1 \le i \le r$.
When $r = 0$, no EPP is performed, and all input states are sent directly to the hashing protocol.
In the $i$-th round ($i \ge 1$) of protocol concatenation, $n$ qubit pairs with serial numbers $\mathbf{x}$ that differ \textit{only} in $x_i$ are fed into the same 2-EPP instance.
If a 2-EPP instance succeeds, we remove the qubit pairs with $x_i \in \{n-1,n\}$ due to the entangled state consumption. Otherwise, in the failed 2-EPP instance, all involved qubit pairs are removed from our scheme. 

Under this arrangement, the input states are always uncorrelated in each 2-EPP instance, since any two serial numbers $\mathbf{x}$'s that differ only in $x_i$ never interact before the $i$-th round of 2-EPP. After the $r$ rounds of concatenated 2-EPPs, $\Ntilde$ qubit pairs remain, each with the same Bell diagonal reduced state $\omega$ and a serial number:
\begin{equation}
    \widetilde{\mathbf{x}}:=(x_0;\xtilde_1,\xtilde_2,\cdots,\xtilde_r),
\end{equation}
where $\xtilde_i \in \{1,2,\cdots, n-2\}$ for $1 \le i \le r$. Since we consider only successful protocols, some serial numbers corresponding to discarded qubits are ignored, such that the serial number set of remaining qubit pairs $\{\widetilde{\mathbf{x}}\}$ satisfies $|\{\widetilde{\mathbf{x}}\}| = \Ntilde \le (n-2)^rM$.
The remaining pairs do not, in general, form one global product state because outputs with the same successful EPP ancestry can remain correlated. Instead, for each fixed suffix $\mathbf{a}:=(\xtilde_1,\xtilde_2,\ldots,\xtilde_r)$, let $\widetilde{M}_{\mathbf{a}}$ be the number of surviving pairs obtained by varying $x_0$. The pairs in this class have distinct EPP ancestries and form the product state $\omega^{\otimes \widetilde{M}_{\mathbf{a}}}$. Moreover,
\begin{equation}
    \sum_{\mathbf{a}}\widetilde{M}_{\mathbf{a}}=\Ntilde,
    \qquad
    \frac{\widetilde{M}_{\mathbf{a}}}{M}\longrightarrow \prod_{i=1}^r p_{s,i}
\end{equation}
with high probability as $M\to\infty$, where $p_{s,i}$ is the success probability of each single 2-EPP instance in the $i$-th round of concatenation. Since there are $(n-2)^r$ fixed-suffix classes and $N=n^rM$, the yield fraction satisfies
\begin{equation}
    \frac{\Ntilde}{N}\longrightarrow \prod_{i=1}^r \frac{n-2}{n}p_{s,i}.
\end{equation}

The next step is to perform qubit hashing protocols. For each fixed suffix $\mathbf{a}$, qubit pairs with serial numbers $\widetilde{\mathbf{x}}$ that vary only in $x_0$ are fed into the same hashing protocol subroutine. In other words, $(n-2)^r=\mathcal{O}(1)$ separate hashing protocols are implemented.
Each of these hashing protocols takes the uncorrelated input $\omega^{\otimes \widetilde{M}_{\mathbf{a}}}$, giving the overall target asymptotic distillation rate
\begin{equation}
    D_{\mathrm{Sh},r,n}(p)=D_H(\omega)\prod_{i=1}^r \frac{n-2}{n}p_{s,i},
\end{equation}
where $p_{s,i}$ and $\omega$ depend on $p$, $r$, and $n$. The hashing bound $D_H(\omega)$ for the Bell diagonal state $\omega$ is given by
\begin{equation}
    D_H(\omega):= \max\left\{0, 1 + \tr(\omega\log_2\omega)\right\}.
\end{equation}

\vfill

\small

\noindent\framebox{\parbox{0.95\linewidth}{
The submitted manuscript has been created by UChicago Argonne, LLC, Operator of 
Argonne National Laboratory (``Argonne''). Argonne, a U.S.\ Department of 
Energy Office of Science laboratory, is operated under Contract No.\ 
DE-AC02-06CH11357. 
The U.S.\ Government retains for itself, and others acting on its behalf, a 
paid-up nonexclusive, irrevocable worldwide license in said article to 
reproduce, prepare derivative works, distribute copies to the public, and 
perform publicly and display publicly, by or on behalf of the Government.  The 
Department of Energy will provide public access to these results of federally 
sponsored research in accordance with the DOE Public Access Plan. 
http://energy.gov/downloads/doe-public-access-plan.}}\\


\begin{thebibliography}{85}%
\makeatletter
\providecommand \@ifxundefined [1]{%
 \@ifx{#1\undefined}
}%
\providecommand \@ifnum [1]{%
 \ifnum #1\expandafter \@firstoftwo
 \else \expandafter \@secondoftwo
 \fi
}%
\providecommand \@ifx [1]{%
 \ifx #1\expandafter \@firstoftwo
 \else \expandafter \@secondoftwo
 \fi
}%
\providecommand \natexlab [1]{#1}%
\providecommand \enquote  [1]{``#1''}%
\providecommand \bibnamefont  [1]{#1}%
\providecommand \bibfnamefont [1]{#1}%
\providecommand \citenamefont [1]{#1}%
\providecommand \href@noop [0]{\@secondoftwo}%
\providecommand \href [0]{\begingroup \@sanitize@url \@href}%
\providecommand \@href[1]{\@@startlink{#1}\@@href}%
\providecommand \@@href[1]{\endgroup#1\@@endlink}%
\providecommand \@sanitize@url [0]{\catcode `\\12\catcode `\$12\catcode
  `\&12\catcode `\#12\catcode `\^12\catcode `\_12\catcode `\%12\relax}%
\providecommand \@@startlink[1]{}%
\providecommand \@@endlink[0]{}%
\providecommand \url  [0]{\begingroup\@sanitize@url \@url }%
\providecommand \@url [1]{\endgroup\@href {#1}{\urlprefix }}%
\providecommand \urlprefix  [0]{URL }%
\providecommand \Eprint [0]{\href }%
\providecommand \doibase [0]{https://doi.org/}%
\providecommand \selectlanguage [0]{\@gobble}%
\providecommand \bibinfo  [0]{\@secondoftwo}%
\providecommand \bibfield  [0]{\@secondoftwo}%
\providecommand \translation [1]{[#1]}%
\providecommand \BibitemOpen [0]{}%
\providecommand \bibitemStop [0]{}%
\providecommand \bibitemNoStop [0]{.\EOS\space}%
\providecommand \EOS [0]{\spacefactor3000\relax}%
\providecommand \BibitemShut  [1]{\csname bibitem#1\endcsname}%
\let\auto@bib@innerbib\@empty
%</preamble>
\bibitem [{\citenamefont {Bruzewicz}\ \emph {et~al.}(2019)\citenamefont
  {Bruzewicz}, \citenamefont {Chiaverini}, \citenamefont {McConnell},\ and\
  \citenamefont {Sage}}]{Bruzewicz_2019}%
  \BibitemOpen
  \bibfield  {author} {\bibinfo {author} {\bibfnamefont {C.~D.}\ \bibnamefont
  {Bruzewicz}}, \bibinfo {author} {\bibfnamefont {J.}~\bibnamefont
  {Chiaverini}}, \bibinfo {author} {\bibfnamefont {R.}~\bibnamefont
  {McConnell}},\ and\ \bibinfo {author} {\bibfnamefont {J.~M.}\ \bibnamefont
  {Sage}},\ }\bibfield  {title} {\bibinfo {title} {Trapped-ion quantum
  computing: Progress and challenges},\ }\href
  {https://doi.org/10.1063/1.5088164} {\bibfield  {journal} {\bibinfo
  {journal} {Applied Physics Reviews}\ }\textbf {\bibinfo {volume} {6}},\
  \bibinfo {pages} {021314} (\bibinfo {year} {2019})}\BibitemShut {NoStop}%
\bibitem [{\citenamefont {Iqbal}\ \emph {et~al.}(2024)\citenamefont {Iqbal},
  \citenamefont {Tantivasadakarn}, \citenamefont {Verresen}, \citenamefont
  {Campbell}, \citenamefont {Dreiling}, \citenamefont {Figgatt}, \citenamefont
  {Gaebler}, \citenamefont {Johansen}, \citenamefont {Mills}, \citenamefont
  {Moses}, \citenamefont {Pino}, \citenamefont {Ransford}, \citenamefont
  {Rowe}, \citenamefont {Siegfried}, \citenamefont {Stutz}, \citenamefont
  {Foss-Feig}, \citenamefont {Vishwanath},\ and\ \citenamefont
  {Dreyer}}]{Iqbal_2024}%
  \BibitemOpen
  \bibfield  {author} {\bibinfo {author} {\bibfnamefont {M.}~\bibnamefont
  {Iqbal}}, \bibinfo {author} {\bibfnamefont {N.}~\bibnamefont
  {Tantivasadakarn}}, \bibinfo {author} {\bibfnamefont {R.}~\bibnamefont
  {Verresen}}, \bibinfo {author} {\bibfnamefont {S.~L.}\ \bibnamefont
  {Campbell}}, \bibinfo {author} {\bibfnamefont {J.~M.}\ \bibnamefont
  {Dreiling}}, \bibinfo {author} {\bibfnamefont {C.}~\bibnamefont {Figgatt}},
  \bibinfo {author} {\bibfnamefont {J.~P.}\ \bibnamefont {Gaebler}}, \bibinfo
  {author} {\bibfnamefont {J.}~\bibnamefont {Johansen}}, \bibinfo {author}
  {\bibfnamefont {M.}~\bibnamefont {Mills}}, \bibinfo {author} {\bibfnamefont
  {S.~A.}\ \bibnamefont {Moses}}, \bibinfo {author} {\bibfnamefont {J.~M.}\
  \bibnamefont {Pino}}, \bibinfo {author} {\bibfnamefont {A.}~\bibnamefont
  {Ransford}}, \bibinfo {author} {\bibfnamefont {M.}~\bibnamefont {Rowe}},
  \bibinfo {author} {\bibfnamefont {P.}~\bibnamefont {Siegfried}}, \bibinfo
  {author} {\bibfnamefont {R.~P.}\ \bibnamefont {Stutz}}, \bibinfo {author}
  {\bibfnamefont {M.}~\bibnamefont {Foss-Feig}}, \bibinfo {author}
  {\bibfnamefont {A.}~\bibnamefont {Vishwanath}},\ and\ \bibinfo {author}
  {\bibfnamefont {H.}~\bibnamefont {Dreyer}},\ }\bibfield  {title} {\bibinfo
  {title} {Non-Abelian topological order and anyons on a trapped-ion
  processor},\ }\href {https://doi.org/10.1038/s41586-023-06934-4} {\bibfield
  {journal} {\bibinfo  {journal} {Nature}\ }\textbf {\bibinfo {volume} {626}},\
  \bibinfo {pages} {505} (\bibinfo {year} {2024})}\BibitemShut {NoStop}%
\bibitem [{\citenamefont {Siddiqi}(2021)}]{Siddiqi_2021}%
  \BibitemOpen
  \bibfield  {author} {\bibinfo {author} {\bibfnamefont {I.}~\bibnamefont
  {Siddiqi}},\ }\bibfield  {title} {\bibinfo {title} {Engineering
  high-coherence superconducting qubits},\ }\href
  {https://doi.org/10.1038/s41578-021-00370-4} {\bibfield  {journal} {\bibinfo
  {journal} {Nature Reviews Materials}\ }\textbf {\bibinfo {volume} {6}},\
  \bibinfo {pages} {875} (\bibinfo {year} {2021})}\BibitemShut {NoStop}%
\bibitem [{\citenamefont {Giustina}(2023)}]{Giustina_2023}%
  \BibitemOpen
  \bibfield  {author} {\bibinfo {author} {\bibfnamefont {M.}~\bibnamefont
  {Giustina}},\ }\bibfield  {title} {\bibinfo {title} {Superconducting qubits
  cover new distances},\ }\href {https://doi.org/10.1038/d41586-023-01488-x} {\bibfield  {journal} {\bibinfo  {journal} {Nature}\ }\textbf {\bibinfo {volume} {617}},\ \bibinfo {pages} {254} (\bibinfo {year} {2023})}\BibitemShut
  {NoStop}%
\bibitem [{\citenamefont {Henriet}\ \emph {et~al.}(2020)\citenamefont
  {Henriet}, \citenamefont {Beguin}, \citenamefont {Signoles}, \citenamefont
  {Lahaye}, \citenamefont {Browaeys}, \citenamefont {Reymond},\ and\
  \citenamefont {Jurczak}}]{Henriet_2020}%
  \BibitemOpen
  \bibfield  {author} {\bibinfo {author} {\bibfnamefont {L.}~\bibnamefont
  {Henriet}}, \bibinfo {author} {\bibfnamefont {L.}~\bibnamefont {Beguin}},
  \bibinfo {author} {\bibfnamefont {A.}~\bibnamefont {Signoles}}, \bibinfo
  {author} {\bibfnamefont {T.}~\bibnamefont {Lahaye}}, \bibinfo {author}
  {\bibfnamefont {A.}~\bibnamefont {Browaeys}}, \bibinfo {author}
  {\bibfnamefont {G.-O.}\ \bibnamefont {Reymond}},\ and\ \bibinfo {author}
  {\bibfnamefont {C.}~\bibnamefont {Jurczak}},\ }\bibfield  {title} {\bibinfo
  {title} {Quantum computing with neutral atoms},\ }\href
  {https://doi.org/10.22331/q-2020-09-21-327} {\bibfield  {journal} {\bibinfo
  {journal} {{Quantum}}\ }\textbf {\bibinfo {volume} {4}},\ \bibinfo {pages}
  {327} (\bibinfo {year} {2020})}\BibitemShut {NoStop}%
\bibitem [{\citenamefont {Graham}\ \emph {et~al.}(2022)\citenamefont {Graham},
  \citenamefont {Song}, \citenamefont {Scott}, \citenamefont {Poole},
  \citenamefont {Phuttitarn}, \citenamefont {Jooya}, \citenamefont {Eichler},
  \citenamefont {Jiang}, \citenamefont {Marra}, \citenamefont {Grinkemeyer},
  \citenamefont {Kwon}, \citenamefont {Ebert}, \citenamefont {Cherek},
  \citenamefont {Lichtman}, \citenamefont {Gillette}, \citenamefont {Gilbert},
  \citenamefont {Bowman}, \citenamefont {Ballance}, \citenamefont {Campbell},
  \citenamefont {Dahl}, \citenamefont {Crawford}, \citenamefont {Blunt},
  \citenamefont {Rogers}, \citenamefont {Noel},\ and\ \citenamefont
  {Saffman}}]{Graham_2022}%
  \BibitemOpen
  \bibfield  {author} {\bibinfo {author} {\bibfnamefont {T.~M.}\ \bibnamefont
  {Graham}}, \bibinfo {author} {\bibfnamefont {Y.}~\bibnamefont {Song}},
  \bibinfo {author} {\bibfnamefont {J.}~\bibnamefont {Scott}}, \bibinfo
  {author} {\bibfnamefont {C.}~\bibnamefont {Poole}}, \bibinfo {author}
  {\bibfnamefont {L.}~\bibnamefont {Phuttitarn}}, \bibinfo {author}
  {\bibfnamefont {K.}~\bibnamefont {Jooya}}, \bibinfo {author} {\bibfnamefont
  {P.}~\bibnamefont {Eichler}}, \bibinfo {author} {\bibfnamefont
  {X.}~\bibnamefont {Jiang}}, \bibinfo {author} {\bibfnamefont
  {A.}~\bibnamefont {Marra}}, \bibinfo {author} {\bibfnamefont
  {B.}~\bibnamefont {Grinkemeyer}}, \bibinfo {author} {\bibfnamefont
  {M.}~\bibnamefont {Kwon}}, \bibinfo {author} {\bibfnamefont {M.}~\bibnamefont
  {Ebert}}, \bibinfo {author} {\bibfnamefont {J.}~\bibnamefont {Cherek}},
  \bibinfo {author} {\bibfnamefont {M.~T.}\ \bibnamefont {Lichtman}}, \bibinfo
  {author} {\bibfnamefont {M.}~\bibnamefont {Gillette}}, \bibinfo {author}
  {\bibfnamefont {J.}~\bibnamefont {Gilbert}}, \bibinfo {author} {\bibfnamefont
  {D.}~\bibnamefont {Bowman}}, \bibinfo {author} {\bibfnamefont
  {T.}~\bibnamefont {Ballance}}, \bibinfo {author} {\bibfnamefont
  {C.}~\bibnamefont {Campbell}}, \bibinfo {author} {\bibfnamefont {E.~D.}\
  \bibnamefont {Dahl}}, \bibinfo {author} {\bibfnamefont {O.}~\bibnamefont
  {Crawford}}, \bibinfo {author} {\bibfnamefont {N.~S.}\ \bibnamefont {Blunt}},
  \bibinfo {author} {\bibfnamefont {B.}~\bibnamefont {Rogers}}, \bibinfo
  {author} {\bibfnamefont {T.}~\bibnamefont {Noel}},\ and\ \bibinfo {author}
  {\bibfnamefont {M.}~\bibnamefont {Saffman}},\ }\bibfield  {title} {\bibinfo
  {title} {Multi-qubit entanglement and algorithms on a neutral-atom quantum
  computer},\ }\href {https://doi.org/10.1038/s41586-022-04603-6} {\bibfield
  {journal} {\bibinfo  {journal} {Nature}\ }\textbf {\bibinfo {volume} {604}},\
  \bibinfo {pages} {457} (\bibinfo {year} {2022})}\BibitemShut {NoStop}%
\bibitem [{\citenamefont {Đorđević}\ \emph {et~al.}(2021)\citenamefont
  {Đorđević}, \citenamefont {Samutpraphoot}, \citenamefont {Ocola},
  \citenamefont {Bernien}, \citenamefont {Grinkemeyer}, \citenamefont
  {Dimitrova}, \citenamefont {Vuletić},\ and\ \citenamefont
  {Lukin}}]{Tamara2021Science}%
  \BibitemOpen
  \bibfield  {author} {\bibinfo {author} {\bibfnamefont {T.}~\bibnamefont
  {Đorđević}}, \bibinfo {author} {\bibfnamefont {P.}~\bibnamefont
  {Samutpraphoot}}, \bibinfo {author} {\bibfnamefont {P.~L.}\ \bibnamefont
  {Ocola}}, \bibinfo {author} {\bibfnamefont {H.}~\bibnamefont {Bernien}},
  \bibinfo {author} {\bibfnamefont {B.}~\bibnamefont {Grinkemeyer}}, \bibinfo
  {author} {\bibfnamefont {I.}~\bibnamefont {Dimitrova}}, \bibinfo {author}
  {\bibfnamefont {V.}~\bibnamefont {Vuletić}},\ and\ \bibinfo {author}
  {\bibfnamefont {M.~D.}\ \bibnamefont {Lukin}},\ }\bibfield  {title} {\bibinfo
  {title} {Entanglement transport and a nanophotonic interface for atoms in
  optical tweezers},\ }\href {https://doi.org/10.1126/science.abi9917}
  {\bibfield  {journal} {\bibinfo  {journal} {Science}\ }\textbf {\bibinfo
  {volume} {373}},\ \bibinfo {pages} {1511} (\bibinfo {year} {2021})} \BibitemShut
  {NoStop}%
\bibitem [{\citenamefont {Covey}\ \emph {et~al.}(2023)\citenamefont {Covey},
  \citenamefont {Weinfurter},\ and\ \citenamefont {Bernien}}]{Covey2023}%
  \BibitemOpen
  \bibfield  {author} {\bibinfo {author} {\bibfnamefont {J.~P.}\ \bibnamefont
  {Covey}}, \bibinfo {author} {\bibfnamefont {H.}~\bibnamefont {Weinfurter}},\
  and\ \bibinfo {author} {\bibfnamefont {H.}~\bibnamefont {Bernien}},\
  }\bibfield  {title} {\bibinfo {title} {Quantum networks with neutral atom
  processing nodes},\ }\href {https://doi.org/10.1038/s41534-023-00759-9}
  {\bibfield  {journal} {\bibinfo  {journal} {npj Quantum Information}\
  }\textbf {\bibinfo {volume} {9}},\ \bibinfo {pages} {90} (\bibinfo {year}
  {2023})}\BibitemShut {NoStop}%
\bibitem [{\citenamefont {Saha}\ \emph {et~al.}(2025)\citenamefont {Saha},
  \citenamefont {Shalaev}, \citenamefont {O'Reilly}, \citenamefont {Goetting},
  \citenamefont {Toh}, \citenamefont {Kalakuntla}, \citenamefont {Yu},\ and\
  \citenamefont {Monroe}}]{Saha2025}%
  \BibitemOpen
  \bibfield  {author} {\bibinfo {author} {\bibfnamefont {S.}~\bibnamefont
  {Saha}}, \bibinfo {author} {\bibfnamefont {M.}~\bibnamefont {Shalaev}},
  \bibinfo {author} {\bibfnamefont {J.}~\bibnamefont {O'Reilly}}, \bibinfo
  {author} {\bibfnamefont {I.}~\bibnamefont {Goetting}}, \bibinfo {author}
  {\bibfnamefont {G.}~\bibnamefont {Toh}}, \bibinfo {author} {\bibfnamefont
  {A.}~\bibnamefont {Kalakuntla}}, \bibinfo {author} {\bibfnamefont
  {Y.}~\bibnamefont {Yu}},\ and\ \bibinfo {author} {\bibfnamefont
  {C.}~\bibnamefont {Monroe}},\ }\bibfield  {title} {\bibinfo {title}
  {High-fidelity remote entanglement of trapped atoms mediated by time-bin
  photons},\ }\href {https://doi.org/10.1038/s41467-025-57557-4} {\bibfield
  {journal} {\bibinfo  {journal} {Nature Communications}\ }\textbf {\bibinfo
  {volume} {16}},\ \bibinfo {pages} {2533} (\bibinfo {year}
  {2025})}\BibitemShut {NoStop}%
\bibitem [{\citenamefont {Bennett}\ \emph
  {et~al.}(1996{\natexlab{a}})\citenamefont {Bennett}, \citenamefont
  {Brassard}, \citenamefont {Popescu}, \citenamefont {Schumacher},
  \citenamefont {Smolin},\ and\ \citenamefont {Wootters}}]{bennett_1996_PRL}%
  \BibitemOpen
  \bibfield  {author} {\bibinfo {author} {\bibfnamefont {C.~H.}\ \bibnamefont
  {Bennett}}, \bibinfo {author} {\bibfnamefont {G.}~\bibnamefont {Brassard}},
  \bibinfo {author} {\bibfnamefont {S.}~\bibnamefont {Popescu}}, \bibinfo
  {author} {\bibfnamefont {B.}~\bibnamefont {Schumacher}}, \bibinfo {author}
  {\bibfnamefont {J.~A.}\ \bibnamefont {Smolin}},\ and\ \bibinfo {author}
  {\bibfnamefont {W.~K.}\ \bibnamefont {Wootters}},\ }\bibfield  {title}
  {\bibinfo {title} {Purification of noisy entanglement and faithful
  teleportation via noisy channels},\ }\href
  {https://doi.org/10.1103/physrevlett.76.722} {\bibfield  {journal} {\bibinfo
  {journal} {Physical Review Letters}\ }\textbf {\bibinfo {volume} {76}},\
  \bibinfo {pages} {722–725} (\bibinfo {year}
  {1996}{\natexlab{a}})}\BibitemShut {NoStop}%
\bibitem [{\citenamefont {Horodecki}\ \emph {et~al.}(1998)\citenamefont
  {Horodecki}, \citenamefont {Horodecki},\ and\ \citenamefont
  {Horodecki}}]{Horodecki1998PRL}%
  \BibitemOpen
  \bibfield  {author} {\bibinfo {author} {\bibfnamefont {M.}~\bibnamefont
  {Horodecki}}, \bibinfo {author} {\bibfnamefont {P.}~\bibnamefont
  {Horodecki}},\ and\ \bibinfo {author} {\bibfnamefont {R.}~\bibnamefont
  {Horodecki}},\ }\bibfield  {title} {\bibinfo {title} {Mixed-state
  entanglement and distillation: Is there a ``bound'' entanglement in
  nature?},\ }\href {https://doi.org/10.1103/PhysRevLett.80.5239} {\bibfield
  {journal} {\bibinfo  {journal} {Phys. Rev. Lett.}\ }\textbf {\bibinfo
  {volume} {80}},\ \bibinfo {pages} {5239} (\bibinfo {year}
  {1998})}\BibitemShut {NoStop}%
\bibitem [{\citenamefont {Bennett}\ \emph
  {et~al.}(1996{\natexlab{b}})\citenamefont {Bennett}, \citenamefont
  {DiVincenzo}, \citenamefont {Smolin},\ and\ \citenamefont
  {Wootters}}]{bennett_1996_PRA}%
  \BibitemOpen
  \bibfield  {author} {\bibinfo {author} {\bibfnamefont {C.~H.}\ \bibnamefont
  {Bennett}}, \bibinfo {author} {\bibfnamefont {D.~P.}\ \bibnamefont
  {DiVincenzo}}, \bibinfo {author} {\bibfnamefont {J.~A.}\ \bibnamefont
  {Smolin}},\ and\ \bibinfo {author} {\bibfnamefont {W.~K.}\ \bibnamefont
  {Wootters}},\ }\bibfield  {title} {\bibinfo {title} {Mixed-state entanglement
  and quantum error correction},\ }\href
  {https://doi.org/10.1103/PhysRevA.54.3824} {\bibfield  {journal} {\bibinfo
  {journal} {Phys. Rev. A}\ }\textbf {\bibinfo {volume} {54}},\ \bibinfo
  {pages} {3824} (\bibinfo {year} {1996}{\natexlab{b}})}\BibitemShut {NoStop}%
\bibitem [{\citenamefont {Watrous}(2004)}]{Watrous2004PRL_distillation}%
  \BibitemOpen
  \bibfield  {author} {\bibinfo {author} {\bibfnamefont {J.}~\bibnamefont
  {Watrous}},\ }\bibfield  {title} {\bibinfo {title} {Many copies may be
  required for entanglement distillation},\ }\href
  {https://doi.org/10.1103/PhysRevLett.93.010502} {\bibfield  {journal}
  {\bibinfo  {journal} {Phys. Rev. Lett.}\ }\textbf {\bibinfo {volume} {93}},\
  \bibinfo {pages} {010502} (\bibinfo {year} {2004})}\BibitemShut {NoStop}%
\bibitem [{\citenamefont {Rozp\ifmmode~\mbox{\k{e}}\else \k{e}\fi{}dek}\ \emph
  {et~al.}(2018)\citenamefont {Rozp\ifmmode~\mbox{\k{e}}\else \k{e}\fi{}dek},
  \citenamefont {Schiet}, \citenamefont {Thinh}, \citenamefont {Elkouss},
  \citenamefont {Doherty},\ and\ \citenamefont {Wehner}}]{PhysRevA.97.062333}%
  \BibitemOpen
  \bibfield  {author} {\bibinfo {author} {\bibfnamefont {F.}~\bibnamefont
  {Rozp\ifmmode~\mbox{\k{e}}\else \k{e}\fi{}dek}}, \bibinfo {author}
  {\bibfnamefont {T.}~\bibnamefont {Schiet}}, \bibinfo {author} {\bibfnamefont
  {L.~P.}\ \bibnamefont {Thinh}}, \bibinfo {author} {\bibfnamefont
  {D.}~\bibnamefont {Elkouss}}, \bibinfo {author} {\bibfnamefont {A.~C.}\
  \bibnamefont {Doherty}},\ and\ \bibinfo {author} {\bibfnamefont
  {S.}~\bibnamefont {Wehner}},\ }\bibfield  {title} {\bibinfo {title}
  {Optimizing practical entanglement distillation},\ }\href
  {https://doi.org/10.1103/PhysRevA.97.062333} {\bibfield  {journal} {\bibinfo
  {journal} {Phys. Rev. A}\ }\textbf {\bibinfo {volume} {97}},\ \bibinfo
  {pages} {062333} (\bibinfo {year} {2018})}\BibitemShut {NoStop}%
\bibitem [{\citenamefont {Chen}\ and\ \citenamefont
  {\DJ{}okovi\ifmmode~\acute{c}\else \'{c}\fi{}}(2016)}]{DistillRank4_2016}%
  \BibitemOpen
  \bibfield  {author} {\bibinfo {author} {\bibfnamefont {L.}~\bibnamefont
  {Chen}}\ and\ \bibinfo {author} {\bibfnamefont {D.~\v{Z}.}\ \bibnamefont
  {\DJ{}okovi\ifmmode~\acute{c}\else \'{c}\fi{}}},\ }\bibfield  {title}
  {\bibinfo {title} {Distillability of non-positive-partial-transpose bipartite
  quantum states of rank four},\ }\href
  {https://doi.org/10.1103/PhysRevA.94.052318} {\bibfield  {journal} {\bibinfo
  {journal} {Phys. Rev. A}\ }\textbf {\bibinfo {volume} {94}},\ \bibinfo
  {pages} {052318} (\bibinfo {year} {2016})}\BibitemShut {NoStop}%
\bibitem [{\citenamefont {Chitambar}\ \emph {et~al.}(2020)\citenamefont
  {Chitambar}, \citenamefont {de~Vicente}, \citenamefont {Girard},\ and\
  \citenamefont {Gour}}]{Chitambar_2020}%
  \BibitemOpen
  \bibfield  {author} {\bibinfo {author} {\bibfnamefont {E.}~\bibnamefont
  {Chitambar}}, \bibinfo {author} {\bibfnamefont {J.~I.}\ \bibnamefont
  {de~Vicente}}, \bibinfo {author} {\bibfnamefont {M.~W.}\ \bibnamefont
  {Girard}},\ and\ \bibinfo {author} {\bibfnamefont {G.}~\bibnamefont {Gour}},\
  }\bibfield  {title} {\bibinfo {title} {Entanglement manipulation beyond local
  operations and classical communication},\ }\href
  {https://doi.org/10.1063/1.5124109} {\bibfield  {journal} {\bibinfo
  {journal} {Journal of Mathematical Physics}\ }\textbf {\bibinfo {volume}
  {61}},\ \bibinfo {pages} {042201} (\bibinfo {year} {2020})} \BibitemShut {NoStop}%
\bibitem [{\citenamefont {Lami}\ and\ \citenamefont {Regula}(2024)}]{Lami2024}%
  \BibitemOpen
  \bibfield  {author} {\bibinfo {author} {\bibfnamefont {L.}~\bibnamefont
  {Lami}}\ and\ \bibinfo {author} {\bibfnamefont {B.}~\bibnamefont {Regula}},\
  }\bibfield  {title} {\bibinfo {title} {Distillable entanglement under dually
  non-entangling operations},\ }\href
  {https://doi.org/10.1038/s41467-024-54201-5} {\bibfield  {journal} {\bibinfo
  {journal} {Nature Communications}\ }\textbf {\bibinfo {volume} {15}},\
  \bibinfo {pages} {10120} (\bibinfo {year} {2024})}\BibitemShut {NoStop}%
\bibitem [{\citenamefont {Ataides}\ \emph {et~al.}(2025)\citenamefont
  {Ataides}, \citenamefont {Zhou}, \citenamefont {Xu}, \citenamefont {Baranes},
  \citenamefont {Li}, \citenamefont {Lukin},\ and\ \citenamefont
  {Jiang}}]{Ataides2025}%
  \BibitemOpen
  \bibfield  {author} {\bibinfo {author} {\bibfnamefont {J.~P.~B.}\
  \bibnamefont {Ataides}}, \bibinfo {author} {\bibfnamefont {H.}~\bibnamefont
  {Zhou}}, \bibinfo {author} {\bibfnamefont {Q.}~\bibnamefont {Xu}}, \bibinfo
  {author} {\bibfnamefont {G.}~\bibnamefont {Baranes}}, \bibinfo {author}
  {\bibfnamefont {B.}~\bibnamefont {Li}}, \bibinfo {author} {\bibfnamefont
  {M.~D.}\ \bibnamefont {Lukin}},\ and\ \bibinfo {author} {\bibfnamefont
  {L.}~\bibnamefont {Jiang}},\ }\bibfield  {title} {\bibinfo {title} {Constant-overhead fault-tolerant Bell-pair
  distillation using high-rate codes},\ }\href {https://doi.org/10.1103/s39k-r2kq}
  {\bibfield  {journal} {\bibinfo  {journal} {Phys. Rev. Lett.}\ }\textbf
  {\bibinfo {volume} {135}},\ \bibinfo {pages} {130804} (\bibinfo {year}
  {2025})}\BibitemShut {NoStop}%
\bibitem [{\citenamefont {Pan}\ \emph {et~al.}(2003)\citenamefont {Pan},
  \citenamefont {Gasparoni}, \citenamefont {Ursin}, \citenamefont {Weihs},\
  and\ \citenamefont {Zeilinger}}]{Pan2003}%
  \BibitemOpen
  \bibfield  {author} {\bibinfo {author} {\bibfnamefont {J.-W.}\ \bibnamefont
  {Pan}}, \bibinfo {author} {\bibfnamefont {S.}~\bibnamefont {Gasparoni}},
  \bibinfo {author} {\bibfnamefont {R.}~\bibnamefont {Ursin}}, \bibinfo
  {author} {\bibfnamefont {G.}~\bibnamefont {Weihs}},\ and\ \bibinfo {author}
  {\bibfnamefont {A.}~\bibnamefont {Zeilinger}},\ }\bibfield  {title} {\bibinfo
  {title} {Experimental entanglement purification of arbitrary unknown
  states},\ }\href {https://doi.org/10.1038/nature01623} {\bibfield  {journal}
  {\bibinfo  {journal} {Nature}\ }\textbf {\bibinfo {volume} {423}},\ \bibinfo
  {pages} {417} (\bibinfo {year} {2003})}\BibitemShut {NoStop}%
\bibitem [{\citenamefont {Hu}\ \emph {et~al.}(2021)\citenamefont {Hu},
  \citenamefont {Huang}, \citenamefont {Sheng}, \citenamefont {Zhou},
  \citenamefont {Liu}, \citenamefont {Guo}, \citenamefont {Zhang},
  \citenamefont {Xing}, \citenamefont {Huang}, \citenamefont {Li},\ and\
  \citenamefont {Guo}}]{Hu2021PRL_photonicEPP}%
  \BibitemOpen
  \bibfield  {author} {\bibinfo {author} {\bibfnamefont {X.-M.}\ \bibnamefont
  {Hu}}, \bibinfo {author} {\bibfnamefont {C.-X.}\ \bibnamefont {Huang}},
  \bibinfo {author} {\bibfnamefont {Y.-B.}\ \bibnamefont {Sheng}}, \bibinfo
  {author} {\bibfnamefont {L.}~\bibnamefont {Zhou}}, \bibinfo {author}
  {\bibfnamefont {B.-H.}\ \bibnamefont {Liu}}, \bibinfo {author} {\bibfnamefont
  {Y.}~\bibnamefont {Guo}}, \bibinfo {author} {\bibfnamefont {C.}~\bibnamefont
  {Zhang}}, \bibinfo {author} {\bibfnamefont {W.-B.}\ \bibnamefont {Xing}},
  \bibinfo {author} {\bibfnamefont {Y.-F.}\ \bibnamefont {Huang}}, \bibinfo
  {author} {\bibfnamefont {C.-F.}\ \bibnamefont {Li}},\ and\ \bibinfo {author}
  {\bibfnamefont {G.-C.}\ \bibnamefont {Guo}},\ }\bibfield  {title} {\bibinfo
  {title} {Long-distance entanglement purification for quantum communication},\
  }\href {https://doi.org/10.1103/PhysRevLett.126.010503} {\bibfield  {journal}
  {\bibinfo  {journal} {Phys. Rev. Lett.}\ }\textbf {\bibinfo {volume} {126}},\
  \bibinfo {pages} {010503} (\bibinfo {year} {2021})}\BibitemShut {NoStop}%
\bibitem [{\citenamefont {Yu}\ \emph {et~al.}(2025)\citenamefont {Yu},
  \citenamefont {Zhou}, \citenamefont {Zahidy}, \citenamefont {Vigliar},
  \citenamefont {Rottwitt}, \citenamefont {Oxenl{\o}we},\ and\ \citenamefont
  {Ding}}]{yu2025_integratedphotonics_EPP}%
  \BibitemOpen
  \bibfield  {author} {\bibinfo {author} {\bibfnamefont {Y.}~\bibnamefont
  {Yu}}, \bibinfo {author} {\bibfnamefont {S.}~\bibnamefont {Zhou}}, \bibinfo
  {author} {\bibfnamefont {M.}~\bibnamefont {Zahidy}}, \bibinfo {author}
  {\bibfnamefont {C.}~\bibnamefont {Vigliar}}, \bibinfo {author} {\bibfnamefont
  {K.}~\bibnamefont {Rottwitt}}, \bibinfo {author} {\bibfnamefont {L.~K.}\
  \bibnamefont {Oxenl{\o}we}},\ and\ \bibinfo {author} {\bibfnamefont
  {Y.}~\bibnamefont {Ding}},\ }\href {https://arxiv.org/abs/2507.03604}
  {\bibinfo {title} {Entanglement purification by integrated silicon
  photonics}} (\bibinfo {year} {2025}),\ \Eprint
  {https://arxiv.org/abs/2507.03604} {arXiv:2507.03604 [quant-ph]} \BibitemShut
  {NoStop}%
\bibitem [{\citenamefont {Zhou}\ \emph {et~al.}(2025)\citenamefont {Zhou},
  \citenamefont {Huang}, \citenamefont {Sheng}, \citenamefont {Guo},
  \citenamefont {Hu}, \citenamefont {Huang}, \citenamefont {Li}, \citenamefont
  {Guo},\ and\ \citenamefont {Liu}}]{Zhou2025PRL}%
  \BibitemOpen
  \bibfield  {author} {\bibinfo {author} {\bibfnamefont {L.}~\bibnamefont
  {Zhou}}, \bibinfo {author} {\bibfnamefont {C.-X.}\ \bibnamefont {Huang}},
  \bibinfo {author} {\bibfnamefont {Y.-B.}\ \bibnamefont {Sheng}}, \bibinfo
  {author} {\bibfnamefont {Y.}~\bibnamefont {Guo}}, \bibinfo {author}
  {\bibfnamefont {X.-M.}\ \bibnamefont {Hu}}, \bibinfo {author} {\bibfnamefont
  {Y.-F.}\ \bibnamefont {Huang}}, \bibinfo {author} {\bibfnamefont {C.-F.}\
  \bibnamefont {Li}}, \bibinfo {author} {\bibfnamefont {G.-C.}\ \bibnamefont
  {Guo}},\ and\ \bibinfo {author} {\bibfnamefont {B.-H.}\ \bibnamefont {Liu}},\
  }\bibfield  {title} {\bibinfo {title} {Observation of residual entanglement
  in entanglement purification},\ }\href {https://doi.org/10.1103/kbw2-fdqn}
  {\bibfield  {journal} {\bibinfo  {journal} {Phys. Rev. Lett.}\ }\textbf
  {\bibinfo {volume} {135}},\ \bibinfo {pages} {050801} (\bibinfo {year}
  {2025})}\BibitemShut {NoStop}%
\bibitem [{\citenamefont {Yan}\ \emph {et~al.}(2022)\citenamefont {Yan},
  \citenamefont {Zhong}, \citenamefont {Chang}, \citenamefont {Bienfait},
  \citenamefont {Chou}, \citenamefont {Conner}, \citenamefont {Dumur},
  \citenamefont {Grebel}, \citenamefont {Povey},\ and\ \citenamefont
  {Cleland}}]{Yan2022PRL}%
  \BibitemOpen
  \bibfield  {author} {\bibinfo {author} {\bibfnamefont {H.}~\bibnamefont
  {Yan}}, \bibinfo {author} {\bibfnamefont {Y.}~\bibnamefont {Zhong}}, \bibinfo
  {author} {\bibfnamefont {H.-S.}\ \bibnamefont {Chang}}, \bibinfo {author}
  {\bibfnamefont {A.}~\bibnamefont {Bienfait}}, \bibinfo {author}
  {\bibfnamefont {M.-H.}\ \bibnamefont {Chou}}, \bibinfo {author}
  {\bibfnamefont {C.~R.}\ \bibnamefont {Conner}}, \bibinfo {author}
  {\bibfnamefont {\'{E}.}~\bibnamefont {Dumur}}, \bibinfo {author} {\bibfnamefont
  {J.}~\bibnamefont {Grebel}}, \bibinfo {author} {\bibfnamefont {R.~G.}\
  \bibnamefont {Povey}},\ and\ \bibinfo {author} {\bibfnamefont {A.~N.}\
  \bibnamefont {Cleland}},\ }\bibfield  {title} {\bibinfo {title} {Entanglement
  purification and protection in a superconducting quantum network},\ }\href
  {https://doi.org/10.1103/PhysRevLett.128.080504} {\bibfield  {journal}
  {\bibinfo  {journal} {Phys. Rev. Lett.}\ }\textbf {\bibinfo {volume} {128}},\
  \bibinfo {pages} {080504} (\bibinfo {year} {2022})}\BibitemShut {NoStop}%
\bibitem [{\citenamefont {Reichle}\ \emph {et~al.}(2006)\citenamefont
  {Reichle}, \citenamefont {Leibfried}, \citenamefont {Knill}, \citenamefont
  {Britton}, \citenamefont {Blakestad}, \citenamefont {Jost}, \citenamefont
  {Langer}, \citenamefont {Ozeri}, \citenamefont {Seidelin},\ and\
  \citenamefont {Wineland}}]{Reichle2006}%
  \BibitemOpen
  \bibfield  {author} {\bibinfo {author} {\bibfnamefont {R.}~\bibnamefont
  {Reichle}}, \bibinfo {author} {\bibfnamefont {D.}~\bibnamefont {Leibfried}},
  \bibinfo {author} {\bibfnamefont {E.}~\bibnamefont {Knill}}, \bibinfo
  {author} {\bibfnamefont {J.}~\bibnamefont {Britton}}, \bibinfo {author}
  {\bibfnamefont {R.~B.}\ \bibnamefont {Blakestad}}, \bibinfo {author}
  {\bibfnamefont {J.~D.}\ \bibnamefont {Jost}}, \bibinfo {author}
  {\bibfnamefont {C.}~\bibnamefont {Langer}}, \bibinfo {author} {\bibfnamefont
  {R.}~\bibnamefont {Ozeri}}, \bibinfo {author} {\bibfnamefont
  {S.}~\bibnamefont {Seidelin}},\ and\ \bibinfo {author} {\bibfnamefont
  {D.~J.}\ \bibnamefont {Wineland}},\ }\bibfield  {title} {\bibinfo {title}
  {Experimental purification of two-atom entanglement},\ }\href
  {https://doi.org/10.1038/nature05146} {\bibfield  {journal} {\bibinfo
  {journal} {Nature}\ }\textbf {\bibinfo {volume} {443}},\ \bibinfo {pages}
  {838} (\bibinfo {year} {2006})}\BibitemShut {NoStop}%
\bibitem [{\citenamefont {Kalb}\ \emph {et~al.}(2017)\citenamefont {Kalb},
  \citenamefont {Reiserer}, \citenamefont {Humphreys}, \citenamefont
  {Bakermans}, \citenamefont {Kamerling}, \citenamefont {Nickerson},
  \citenamefont {Benjamin}, \citenamefont {Twitchen}, \citenamefont {Markham},\
  and\ \citenamefont {Hanson}}]{Kalb_2017_science}%
  \BibitemOpen
  \bibfield  {author} {\bibinfo {author} {\bibfnamefont {N.}~\bibnamefont
  {Kalb}}, \bibinfo {author} {\bibfnamefont {A.~A.}\ \bibnamefont {Reiserer}},
  \bibinfo {author} {\bibfnamefont {P.~C.}\ \bibnamefont {Humphreys}}, \bibinfo
  {author} {\bibfnamefont {J.~J.~W.}\ \bibnamefont {Bakermans}}, \bibinfo
  {author} {\bibfnamefont {S.~J.}\ \bibnamefont {Kamerling}}, \bibinfo {author}
  {\bibfnamefont {N.~H.}\ \bibnamefont {Nickerson}}, \bibinfo {author}
  {\bibfnamefont {S.~C.}\ \bibnamefont {Benjamin}}, \bibinfo {author}
  {\bibfnamefont {D.~J.}\ \bibnamefont {Twitchen}}, \bibinfo {author}
  {\bibfnamefont {M.}~\bibnamefont {Markham}},\ and\ \bibinfo {author}
  {\bibfnamefont {R.}~\bibnamefont {Hanson}},\ }\bibfield  {title} {\bibinfo
  {title} {Entanglement distillation between solid-state quantum network
  nodes},\ }\href {https://doi.org/10.1126/science.aan0070} {\bibfield
  {journal} {\bibinfo  {journal} {Science}\ }\textbf {\bibinfo {volume}
  {356}},\ \bibinfo {pages} {928} (\bibinfo {year} {2017})} \BibitemShut
  {NoStop}%
\bibitem [{\citenamefont {Xu}\ \emph {et~al.}(2024)\citenamefont {Xu},
  \citenamefont {Bonilla~Ataides}, \citenamefont {Pattison}, \citenamefont
  {Raveendran}, \citenamefont {Bluvstein}, \citenamefont {Wurtz}, \citenamefont
  {Vasi{\'{c}}}, \citenamefont {Lukin}, \citenamefont {Jiang},\ and\
  \citenamefont {Zhou}}]{Xu2024}%
  \BibitemOpen
  \bibfield  {author} {\bibinfo {author} {\bibfnamefont {Q.}~\bibnamefont
  {Xu}}, \bibinfo {author} {\bibfnamefont {J.~P.}\ \bibnamefont
  {Bonilla~Ataides}}, \bibinfo {author} {\bibfnamefont {C.~A.}\ \bibnamefont
  {Pattison}}, \bibinfo {author} {\bibfnamefont {N.}~\bibnamefont
  {Raveendran}}, \bibinfo {author} {\bibfnamefont {D.}~\bibnamefont
  {Bluvstein}}, \bibinfo {author} {\bibfnamefont {J.}~\bibnamefont {Wurtz}},
  \bibinfo {author} {\bibfnamefont {B.}~\bibnamefont {Vasi{\'{c}}}}, \bibinfo
  {author} {\bibfnamefont {M.~D.}\ \bibnamefont {Lukin}}, \bibinfo {author}
  {\bibfnamefont {L.}~\bibnamefont {Jiang}},\ and\ \bibinfo {author}
  {\bibfnamefont {H.}~\bibnamefont {Zhou}},\ }\bibfield  {title} {\bibinfo
  {title} {Constant-overhead fault-tolerant quantum computation with
  reconfigurable atom arrays},\ }\href
  {https://doi.org/10.1038/s41567-024-02479-z} {\bibfield  {journal} {\bibinfo
  {journal} {Nature Physics}\ }\textbf {\bibinfo {volume} {20}},\ \bibinfo
  {pages} {1084} (\bibinfo {year} {2024})}\BibitemShut {NoStop}%
\bibitem [{\citenamefont {Bluvstein}\ \emph {et~al.}(2022)\citenamefont
  {Bluvstein}, \citenamefont {Levine}, \citenamefont {Semeghini}, \citenamefont
  {Wang}, \citenamefont {Ebadi}, \citenamefont {Kalinowski}, \citenamefont
  {Keesling}, \citenamefont {Maskara}, \citenamefont {Pichler}, \citenamefont
  {Greiner}, \citenamefont {Vuletić},\ and\ \citenamefont
  {Lukin}}]{Bluvstein_2022}%
  \BibitemOpen
  \bibfield  {author} {\bibinfo {author} {\bibfnamefont {D.}~\bibnamefont
  {Bluvstein}}, \bibinfo {author} {\bibfnamefont {H.}~\bibnamefont {Levine}},
  \bibinfo {author} {\bibfnamefont {G.}~\bibnamefont {Semeghini}}, \bibinfo
  {author} {\bibfnamefont {T.~T.}\ \bibnamefont {Wang}}, \bibinfo {author}
  {\bibfnamefont {S.}~\bibnamefont {Ebadi}}, \bibinfo {author} {\bibfnamefont
  {M.}~\bibnamefont {Kalinowski}}, \bibinfo {author} {\bibfnamefont
  {A.}~\bibnamefont {Keesling}}, \bibinfo {author} {\bibfnamefont
  {N.}~\bibnamefont {Maskara}}, \bibinfo {author} {\bibfnamefont
  {H.}~\bibnamefont {Pichler}}, \bibinfo {author} {\bibfnamefont
  {M.}~\bibnamefont {Greiner}}, \bibinfo {author} {\bibfnamefont
  {V.}~\bibnamefont {Vuletić}},\ and\ \bibinfo {author} {\bibfnamefont
  {M.~D.}\ \bibnamefont {Lukin}},\ }\bibfield  {title} {\bibinfo {title} {A
  quantum processor based on coherent transport of entangled atom arrays},\
  }\href {https://doi.org/10.1038/s41586-022-04592-6} {\bibfield  {journal}
  {\bibinfo  {journal} {Nature}\ }\textbf {\bibinfo {volume} {604}},\ \bibinfo
  {pages} {451} (\bibinfo {year} {2022})}\BibitemShut {NoStop}%
\bibitem [{\citenamefont {Lukin}\ \emph {et~al.}(2001)\citenamefont {Lukin},
  \citenamefont {Fleischhauer}, \citenamefont {C\^{o}t\'{e}}, \citenamefont {Duan},
  \citenamefont {Jaksch}, \citenamefont {Cirac},\ and\ \citenamefont
  {Zoller}}]{Lukin2001Blockade}%
  \BibitemOpen
  \bibfield  {author} {\bibinfo {author} {\bibfnamefont {M.~D.}\ \bibnamefont
  {Lukin}}, \bibinfo {author} {\bibfnamefont {M.}~\bibnamefont {Fleischhauer}},
  \bibinfo {author} {\bibfnamefont {R.}~\bibnamefont {C\^{o}t\'{e}}}, \bibinfo {author}
  {\bibfnamefont {L.~M.}\ \bibnamefont {Duan}}, \bibinfo {author}
  {\bibfnamefont {D.}~\bibnamefont {Jaksch}}, \bibinfo {author} {\bibfnamefont
  {J.~I.}\ \bibnamefont {Cirac}},\ and\ \bibinfo {author} {\bibfnamefont
  {P.}~\bibnamefont {Zoller}},\ }\bibfield  {title} {\bibinfo {title} {Dipole
  blockade and quantum information processing in mesoscopic atomic ensembles},\
  }\href {https://doi.org/10.1103/PhysRevLett.87.037901} {\bibfield  {journal}
  {\bibinfo  {journal} {Phys. Rev. Lett.}\ }\textbf {\bibinfo {volume} {87}},\
  \bibinfo {pages} {037901} (\bibinfo {year} {2001})}\BibitemShut {NoStop}%
\bibitem [{\citenamefont {Tong}\ \emph {et~al.}(2004)\citenamefont {Tong},
  \citenamefont {Farooqi}, \citenamefont {Stanojevic}, \citenamefont
  {Krishnan}, \citenamefont {Zhang}, \citenamefont {C\^ot\'e}, \citenamefont
  {Eyler},\ and\ \citenamefont {Gould}}]{Tong_2004}%
  \BibitemOpen
  \bibfield  {author} {\bibinfo {author} {\bibfnamefont {D.}~\bibnamefont
  {Tong}}, \bibinfo {author} {\bibfnamefont {S.~M.}\ \bibnamefont {Farooqi}},
  \bibinfo {author} {\bibfnamefont {J.}~\bibnamefont {Stanojevic}}, \bibinfo
  {author} {\bibfnamefont {S.}~\bibnamefont {Krishnan}}, \bibinfo {author}
  {\bibfnamefont {Y.~P.}\ \bibnamefont {Zhang}}, \bibinfo {author}
  {\bibfnamefont {R.}~\bibnamefont {C\^ot\'e}}, \bibinfo {author}
  {\bibfnamefont {E.~E.}\ \bibnamefont {Eyler}},\ and\ \bibinfo {author}
  {\bibfnamefont {P.~L.}\ \bibnamefont {Gould}},\ }\bibfield  {title} {\bibinfo
  {title} {Local blockade of Rydberg excitation in an ultracold gas},\ }\href
  {https://doi.org/10.1103/PhysRevLett.93.063001} {\bibfield  {journal}
  {\bibinfo  {journal} {Phys. Rev. Lett.}\ }\textbf {\bibinfo {volume} {93}},\
  \bibinfo {pages} {063001} (\bibinfo {year} {2004})}\BibitemShut {NoStop}%
\bibitem [{\citenamefont {Wilk}\ \emph {et~al.}(2010)\citenamefont {Wilk},
  \citenamefont {Ga\"etan}, \citenamefont {Evellin}, \citenamefont {Wolters},
  \citenamefont {Miroshnychenko}, \citenamefont {Grangier},\ and\ \citenamefont
  {Browaeys}}]{Wilk2010PRL}%
  \BibitemOpen
  \bibfield  {author} {\bibinfo {author} {\bibfnamefont {T.}~\bibnamefont
  {Wilk}}, \bibinfo {author} {\bibfnamefont {A.}~\bibnamefont {Ga\"etan}},
  \bibinfo {author} {\bibfnamefont {C.}~\bibnamefont {Evellin}}, \bibinfo
  {author} {\bibfnamefont {J.}~\bibnamefont {Wolters}}, \bibinfo {author}
  {\bibfnamefont {Y.}~\bibnamefont {Miroshnychenko}}, \bibinfo {author}
  {\bibfnamefont {P.}~\bibnamefont {Grangier}},\ and\ \bibinfo {author}
  {\bibfnamefont {A.}~\bibnamefont {Browaeys}},\ }\bibfield  {title} {\bibinfo
  {title} {Entanglement of two individual neutral atoms using Rydberg
  blockade},\ }\href {https://doi.org/10.1103/PhysRevLett.104.010502}
  {\bibfield  {journal} {\bibinfo  {journal} {Phys. Rev. Lett.}\ }\textbf
  {\bibinfo {volume} {104}},\ \bibinfo {pages} {010502} (\bibinfo {year}
  {2010})}\BibitemShut {NoStop}%
\bibitem [{\citenamefont {Evered}\ \emph {et~al.}(2023)\citenamefont {Evered},
  \citenamefont {Bluvstein}, \citenamefont {Kalinowski}, \citenamefont {Ebadi},
  \citenamefont {Manovitz}, \citenamefont {Zhou}, \citenamefont {Li},
  \citenamefont {Geim}, \citenamefont {Wang}, \citenamefont {Maskara},
  \citenamefont {Levine}, \citenamefont {Semeghini}, \citenamefont {Greiner},
  \citenamefont {Vuleti{\'{c}}},\ and\ \citenamefont {Lukin}}]{Evered2023}%
  \BibitemOpen
  \bibfield  {author} {\bibinfo {author} {\bibfnamefont {S.~J.}\ \bibnamefont
  {Evered}}, \bibinfo {author} {\bibfnamefont {D.}~\bibnamefont {Bluvstein}},
  \bibinfo {author} {\bibfnamefont {M.}~\bibnamefont {Kalinowski}}, \bibinfo
  {author} {\bibfnamefont {S.}~\bibnamefont {Ebadi}}, \bibinfo {author}
  {\bibfnamefont {T.}~\bibnamefont {Manovitz}}, \bibinfo {author}
  {\bibfnamefont {H.}~\bibnamefont {Zhou}}, \bibinfo {author} {\bibfnamefont
  {S.~H.}\ \bibnamefont {Li}}, \bibinfo {author} {\bibfnamefont {A.~A.}\
  \bibnamefont {Geim}}, \bibinfo {author} {\bibfnamefont {T.~T.}\ \bibnamefont
  {Wang}}, \bibinfo {author} {\bibfnamefont {N.}~\bibnamefont {Maskara}},
  \bibinfo {author} {\bibfnamefont {H.}~\bibnamefont {Levine}}, \bibinfo
  {author} {\bibfnamefont {G.}~\bibnamefont {Semeghini}}, \bibinfo {author}
  {\bibfnamefont {M.}~\bibnamefont {Greiner}}, \bibinfo {author} {\bibfnamefont
  {V.}~\bibnamefont {Vuleti{\'{c}}}},\ and\ \bibinfo {author} {\bibfnamefont
  {M.~D.}\ \bibnamefont {Lukin}},\ }\bibfield  {title} {\bibinfo {title}
  {High-fidelity parallel entangling gates on a neutral-atom quantum
  computer},\ }\href {https://doi.org/10.1038/s41586-023-06481-y} {\bibfield
  {journal} {\bibinfo  {journal} {Nature}\ }\textbf {\bibinfo {volume} {622}},\
  \bibinfo {pages} {268} (\bibinfo {year} {2023})}\BibitemShut {NoStop}%
\bibitem [{\citenamefont {Singh}\ \emph {et~al.}(2022)\citenamefont {Singh},
  \citenamefont {Anand}, \citenamefont {Pocklington}, \citenamefont {Kemp},\
  and\ \citenamefont {Bernien}}]{Singh2022PRX}%
  \BibitemOpen
  \bibfield  {author} {\bibinfo {author} {\bibfnamefont {K.}~\bibnamefont
  {Singh}}, \bibinfo {author} {\bibfnamefont {S.}~\bibnamefont {Anand}},
  \bibinfo {author} {\bibfnamefont {A.}~\bibnamefont {Pocklington}}, \bibinfo
  {author} {\bibfnamefont {J.~T.}\ \bibnamefont {Kemp}},\ and\ \bibinfo
  {author} {\bibfnamefont {H.}~\bibnamefont {Bernien}},\ }\bibfield  {title}
  {\bibinfo {title} {Dual-element, two-dimensional atom array with
  continuous-mode operation},\ }\href
  {https://doi.org/10.1103/PhysRevX.12.011040} {\bibfield  {journal} {\bibinfo
  {journal} {Phys. Rev. X}\ }\textbf {\bibinfo {volume} {12}},\ \bibinfo
  {pages} {011040} (\bibinfo {year} {2022})}\BibitemShut {NoStop}%
\bibitem [{\citenamefont {Anand}\ \emph {et~al.}(2024)\citenamefont {Anand},
  \citenamefont {Bradley}, \citenamefont {White}, \citenamefont {Ramesh},
  \citenamefont {Singh},\ and\ \citenamefont {Bernien}}]{anand2024dualspecies}%
  \BibitemOpen
  \bibfield  {author} {\bibinfo {author} {\bibfnamefont {S.}~\bibnamefont
  {Anand}}, \bibinfo {author} {\bibfnamefont {C.~E.}\ \bibnamefont {Bradley}},
  \bibinfo {author} {\bibfnamefont {R.}~\bibnamefont {White}}, \bibinfo
  {author} {\bibfnamefont {V.}~\bibnamefont {Ramesh}}, \bibinfo {author}
  {\bibfnamefont {K.}~\bibnamefont {Singh}},\ and\ \bibinfo {author}
  {\bibfnamefont {H.}~\bibnamefont {Bernien}},\ }\bibfield  {title} {\bibinfo
  {title} {A dual-species Rydberg array},\ }\href
  {https://doi.org/10.1038/s41567-024-02638-2} {\bibfield  {journal} {\bibinfo
  {journal} {Nature Physics}\ }\textbf {\bibinfo {volume} {20}},\ \bibinfo
  {pages} {1744} (\bibinfo {year} {2024})}\BibitemShut {NoStop}%
\bibitem [{\citenamefont {Beterov}\ and\ \citenamefont
  {Saffman}(2015)}]{Beterov2015PRA}%
  \BibitemOpen
  \bibfield  {author} {\bibinfo {author} {\bibfnamefont {I.~I.}\ \bibnamefont
  {Beterov}}\ and\ \bibinfo {author} {\bibfnamefont {M.}~\bibnamefont
  {Saffman}},\ }\bibfield  {title} {\bibinfo {title} {Rydberg blockade,
  F\"orster resonances, and quantum state measurements with different atomic
  species},\ }\href {https://doi.org/10.1103/PhysRevA.92.042710} {\bibfield
  {journal} {\bibinfo  {journal} {Phys. Rev. A}\ }\textbf {\bibinfo {volume}
  {92}},\ \bibinfo {pages} {042710} (\bibinfo {year} {2015})}\BibitemShut
  {NoStop}%
\bibitem [{\citenamefont {Ebadi}\ \emph {et~al.}(2022)\citenamefont {Ebadi},
  \citenamefont {Keesling}, \citenamefont {Cain}, \citenamefont {Wang},
  \citenamefont {Levine}, \citenamefont {Bluvstein}, \citenamefont {Semeghini},
  \citenamefont {Omran}, \citenamefont {Liu}, \citenamefont {Samajdar},
  \citenamefont {Luo}, \citenamefont {Nash}, \citenamefont {Gao}, \citenamefont
  {Barak}, \citenamefont {Farhi}, \citenamefont {Sachdev}, \citenamefont
  {Gemelke}, \citenamefont {Zhou}, \citenamefont {Choi}, \citenamefont
  {Pichler}, \citenamefont {Wang}, \citenamefont {Greiner}, \citenamefont
  {Vuletić},\ and\ \citenamefont {Lukin}}]{Ebadi_2022}%
  \BibitemOpen
  \bibfield  {author} {\bibinfo {author} {\bibfnamefont {S.}~\bibnamefont
  {Ebadi}}, \bibinfo {author} {\bibfnamefont {A.}~\bibnamefont {Keesling}},
  \bibinfo {author} {\bibfnamefont {M.}~\bibnamefont {Cain}}, \bibinfo {author}
  {\bibfnamefont {T.~T.}\ \bibnamefont {Wang}}, \bibinfo {author}
  {\bibfnamefont {H.}~\bibnamefont {Levine}}, \bibinfo {author} {\bibfnamefont
  {D.}~\bibnamefont {Bluvstein}}, \bibinfo {author} {\bibfnamefont
  {G.}~\bibnamefont {Semeghini}}, \bibinfo {author} {\bibfnamefont
  {A.}~\bibnamefont {Omran}}, \bibinfo {author} {\bibfnamefont {J.-G.}\
  \bibnamefont {Liu}}, \bibinfo {author} {\bibfnamefont {R.}~\bibnamefont
  {Samajdar}}, \bibinfo {author} {\bibfnamefont {X.-Z.}\ \bibnamefont {Luo}},
  \bibinfo {author} {\bibfnamefont {B.}~\bibnamefont {Nash}}, \bibinfo {author}
  {\bibfnamefont {X.}~\bibnamefont {Gao}}, \bibinfo {author} {\bibfnamefont
  {B.}~\bibnamefont {Barak}}, \bibinfo {author} {\bibfnamefont
  {E.}~\bibnamefont {Farhi}}, \bibinfo {author} {\bibfnamefont
  {S.}~\bibnamefont {Sachdev}}, \bibinfo {author} {\bibfnamefont
  {N.}~\bibnamefont {Gemelke}}, \bibinfo {author} {\bibfnamefont
  {L.}~\bibnamefont {Zhou}}, \bibinfo {author} {\bibfnamefont {S.}~\bibnamefont
  {Choi}}, \bibinfo {author} {\bibfnamefont {H.}~\bibnamefont {Pichler}},
  \bibinfo {author} {\bibfnamefont {S.-T.}\ \bibnamefont {Wang}}, \bibinfo
  {author} {\bibfnamefont {M.}~\bibnamefont {Greiner}}, \bibinfo {author}
  {\bibfnamefont {V.}~\bibnamefont {Vuletić}},\ and\ \bibinfo {author}
  {\bibfnamefont {M.~D.}\ \bibnamefont {Lukin}},\ }\bibfield  {title} {\bibinfo
  {title} {Quantum optimization of maximum independent set using Rydberg atom
  arrays},\ }\href {https://doi.org/10.1126/science.abo6587} {\bibfield
  {journal} {\bibinfo  {journal} {Science}\ }\textbf {\bibinfo {volume}
  {376}},\ \bibinfo {pages} {1209} (\bibinfo {year} {2022})} \BibitemShut
  {NoStop}%
\bibitem [{\citenamefont {Nguyen}\ \emph {et~al.}(2023)\citenamefont {Nguyen},
  \citenamefont {Liu}, \citenamefont {Wurtz}, \citenamefont {Lukin},
  \citenamefont {Wang},\ and\ \citenamefont {Pichler}}]{Nguyen_2023}%
  \BibitemOpen
  \bibfield  {author} {\bibinfo {author} {\bibfnamefont {M.-T.}\ \bibnamefont
  {Nguyen}}, \bibinfo {author} {\bibfnamefont {J.-G.}\ \bibnamefont {Liu}},
  \bibinfo {author} {\bibfnamefont {J.}~\bibnamefont {Wurtz}}, \bibinfo
  {author} {\bibfnamefont {M.~D.}\ \bibnamefont {Lukin}}, \bibinfo {author}
  {\bibfnamefont {S.-T.}\ \bibnamefont {Wang}},\ and\ \bibinfo {author}
  {\bibfnamefont {H.}~\bibnamefont {Pichler}},\ }\bibfield  {title} {\bibinfo
  {title} {Quantum optimization with arbitrary connectivity using Rydberg atom
  arrays},\ }\href {https://doi.org/10.1103/PRXQuantum.4.010316} {\bibfield
  {journal} {\bibinfo  {journal} {PRX Quantum}\ }\textbf {\bibinfo {volume}
  {4}},\ \bibinfo {pages} {010316} (\bibinfo {year} {2023})}\BibitemShut
  {NoStop}%
\bibitem [{\citenamefont {QuEra Computing Inc.}(2024)}]{Quera_2024}%
  \BibitemOpen
  \bibfield  {author} {\bibinfo {author} {\bibnamefont {QuEra Computing Inc.}},\ }\href@noop {} {\bibinfo {title} {Using neutral-atom arrays to
  build quantum computers}},\ \bibinfo {howpublished}
  {\url{https://www.quera.com/neutral-atom-platform}} (\bibinfo {year}
  {2024}),\ \bibinfo {note} {accessed: 2024-06-26}\BibitemShut {NoStop}%
\bibitem [{\citenamefont {Labuhn}\ \emph {et~al.}(2016)\citenamefont {Labuhn},
  \citenamefont {Barredo}, \citenamefont {Ravets}, \citenamefont
  {de~L{\'e}s{\'e}leuc}, \citenamefont {Macr{\`i}}, \citenamefont {Lahaye},\
  and\ \citenamefont {Browaeys}}]{Labuhn_2016}%
  \BibitemOpen
  \bibfield  {author} {\bibinfo {author} {\bibfnamefont {H.}~\bibnamefont
  {Labuhn}}, \bibinfo {author} {\bibfnamefont {D.}~\bibnamefont {Barredo}},
  \bibinfo {author} {\bibfnamefont {S.}~\bibnamefont {Ravets}}, \bibinfo
  {author} {\bibfnamefont {S.}~\bibnamefont {de~L{\'e}s{\'e}leuc}}, \bibinfo
  {author} {\bibfnamefont {T.}~\bibnamefont {Macr{\`i}}}, \bibinfo {author}
  {\bibfnamefont {T.}~\bibnamefont {Lahaye}},\ and\ \bibinfo {author}
  {\bibfnamefont {A.}~\bibnamefont {Browaeys}},\ }\bibfield  {title} {\bibinfo
  {title} {Tunable two-dimensional arrays of single Rydberg atoms for realizing
  quantum Ising models},\ }\href {https://doi.org/10.1038/nature18274}
  {\bibfield  {journal} {\bibinfo  {journal} {Nature}\ }\textbf {\bibinfo
  {volume} {534}},\ \bibinfo {pages} {667} (\bibinfo {year}
  {2016})}\BibitemShut {NoStop}%
\bibitem [{\citenamefont {Sheng}\ \emph {et~al.}(2022)\citenamefont {Sheng},
  \citenamefont {Hou}, \citenamefont {He}, \citenamefont {Wang}, \citenamefont
  {Guo}, \citenamefont {Zhuang}, \citenamefont {Mamat}, \citenamefont {Xu},
  \citenamefont {Liu}, \citenamefont {Wang},\ and\ \citenamefont
  {Zhan}}]{Sheng_2022}%
  \BibitemOpen
  \bibfield  {author} {\bibinfo {author} {\bibfnamefont {C.}~\bibnamefont
  {Sheng}}, \bibinfo {author} {\bibfnamefont {J.}~\bibnamefont {Hou}}, \bibinfo
  {author} {\bibfnamefont {X.}~\bibnamefont {He}}, \bibinfo {author}
  {\bibfnamefont {K.}~\bibnamefont {Wang}}, \bibinfo {author} {\bibfnamefont
  {R.}~\bibnamefont {Guo}}, \bibinfo {author} {\bibfnamefont {J.}~\bibnamefont
  {Zhuang}}, \bibinfo {author} {\bibfnamefont {B.}~\bibnamefont {Mamat}},
  \bibinfo {author} {\bibfnamefont {P.}~\bibnamefont {Xu}}, \bibinfo {author}
  {\bibfnamefont {M.}~\bibnamefont {Liu}}, \bibinfo {author} {\bibfnamefont
  {J.}~\bibnamefont {Wang}},\ and\ \bibinfo {author} {\bibfnamefont
  {M.}~\bibnamefont {Zhan}},\ }\bibfield  {title} {\bibinfo {title}
  {Defect-free arbitrary-geometry assembly of mixed-species atom arrays},\
  }\href {https://doi.org/10.1103/PhysRevLett.128.083202} {\bibfield  {journal}
  {\bibinfo  {journal} {Phys. Rev. Lett.}\ }\textbf {\bibinfo {volume} {128}},\
  \bibinfo {pages} {083202} (\bibinfo {year} {2022})}\BibitemShut {NoStop}%
\bibitem [{\citenamefont {Hu}\ \emph {et~al.}(2025)\citenamefont {Hu},
  \citenamefont {Gomez}, \citenamefont {Chen}, \citenamefont {Trowbridge},
  \citenamefont {Goldschmidt}, \citenamefont {Manchester}, \citenamefont
  {Chong}, \citenamefont {Jaffe},\ and\ \citenamefont {Yelin}}]{hu2025}%
  \BibitemOpen
  \bibfield  {author} {\bibinfo {author} {\bibfnamefont {H.-Y.}\ \bibnamefont
  {Hu}}, \bibinfo {author} {\bibfnamefont {A.~M.}\ \bibnamefont {Gomez}},
  \bibinfo {author} {\bibfnamefont {L.}~\bibnamefont {Chen}}, \bibinfo {author}
  {\bibfnamefont {A.}~\bibnamefont {Trowbridge}}, \bibinfo {author}
  {\bibfnamefont {A.~J.}\ \bibnamefont {Goldschmidt}}, \bibinfo {author}
  {\bibfnamefont {Z.}~\bibnamefont {Manchester}}, \bibinfo {author}
  {\bibfnamefont {F.~T.}\ \bibnamefont {Chong}}, \bibinfo {author}
  {\bibfnamefont {A.}~\bibnamefont {Jaffe}},\ and\ \bibinfo {author}
  {\bibfnamefont {S.~F.}\ \bibnamefont {Yelin}},\ }\href
  {https://arxiv.org/abs/2508.19075} {\bibinfo {title} {Universal dynamics with
  globally controlled analog quantum simulators}} (\bibinfo {year} {2025}),\
  \Eprint {https://arxiv.org/abs/2508.19075} {arXiv:2508.19075 [quant-ph]}
  \BibitemShut {NoStop}%
\bibitem [{\citenamefont {Iyer}\ and\ \citenamefont
  {Poulin}(2015)}]{Iyer_Poulin_2015}%
  \BibitemOpen
  \bibfield  {author} {\bibinfo {author} {\bibfnamefont {P.}~\bibnamefont
  {Iyer}}\ and\ \bibinfo {author} {\bibfnamefont {D.}~\bibnamefont {Poulin}},\
  }\bibfield  {title} {\bibinfo {title} {Hardness of decoding quantum
  stabilizer codes},\ }\href {https://doi.org/10.1109/TIT.2015.2422294}
  {\bibfield  {journal} {\bibinfo  {journal} {IEEE Transactions on Information
  Theory}\ }\textbf {\bibinfo {volume} {61}},\ \bibinfo {pages} {5209}
  (\bibinfo {year} {2015})}\BibitemShut {NoStop}%
\bibitem [{\citenamefont {Steane}(1996)}]{Steane_1996}%
  \BibitemOpen
  \bibfield  {author} {\bibinfo {author} {\bibfnamefont {A.~M.}\ \bibnamefont
  {Steane}},\ }\bibfield  {title} {\bibinfo {title} {Error correcting codes in
  quantum theory},\ }\href {https://doi.org/10.1103/PhysRevLett.77.793}
  {\bibfield  {journal} {\bibinfo  {journal} {Phys. Rev. Lett.}\ }\textbf
  {\bibinfo {volume} {77}},\ \bibinfo {pages} {793} (\bibinfo {year}
  {1996})}\BibitemShut {NoStop}%
\bibitem [{\citenamefont {Calderbank}\ and\ \citenamefont
  {Shor}(1996)}]{Calderbank_1996}%
  \BibitemOpen
  \bibfield  {author} {\bibinfo {author} {\bibfnamefont {A.~R.}\ \bibnamefont
  {Calderbank}}\ and\ \bibinfo {author} {\bibfnamefont {P.~W.}\ \bibnamefont
  {Shor}},\ }\bibfield  {title} {\bibinfo {title} {Good quantum
  error-correcting codes exist},\ }\href
  {https://doi.org/10.1103/PhysRevA.54.1098} {\bibfield  {journal} {\bibinfo
  {journal} {Phys. Rev. A}\ }\textbf {\bibinfo {volume} {54}},\ \bibinfo
  {pages} {1098} (\bibinfo {year} {1996})}\BibitemShut {NoStop}%
\bibitem [{\citenamefont {{Gottesman}}(1997)}]{Gottesman_1997}%
  \BibitemOpen
  \bibfield  {author} {\bibinfo {author} {\bibfnamefont {D.}~\bibnamefont
  {{Gottesman}}},\ }\emph {\bibinfo {title} {{Stabilizer codes and quantum
  error correction}}},\ \href@noop {} {Ph.D. thesis},\ \bibinfo  {school}
  {California Institute of Technology} (\bibinfo {year} {1997})\BibitemShut
  {NoStop}%
\bibitem [{\citenamefont {Aaronson}\ and\ \citenamefont
  {Gottesman}(2004)}]{Aaronson2004}%
  \BibitemOpen
  \bibfield  {author} {\bibinfo {author} {\bibfnamefont {S.}~\bibnamefont
  {Aaronson}}\ and\ \bibinfo {author} {\bibfnamefont {D.}~\bibnamefont
  {Gottesman}},\ }\bibfield  {title} {\bibinfo {title} {Improved simulation of
  stabilizer circuits},\ }\href {https://doi.org/10.1103/PhysRevA.70.052328}
  {\bibfield  {journal} {\bibinfo  {journal} {Phys. Rev. A}\ }\textbf {\bibinfo
  {volume} {70}},\ \bibinfo {pages} {052328} (\bibinfo {year}
  {2004})}\BibitemShut {NoStop}%
\bibitem [{\citenamefont {Calderbank}\ \emph {et~al.}(1998)\citenamefont
  {Calderbank}, \citenamefont {Rains}, \citenamefont {Shor},\ and\
  \citenamefont {Sloane}}]{calderbank1997}%
  \BibitemOpen
  \bibfield  {author} {\bibinfo {author} {\bibfnamefont {A.~R.}\ \bibnamefont
  {Calderbank}}, \bibinfo {author} {\bibfnamefont {E.~M.}\ \bibnamefont
  {Rains}}, \bibinfo {author} {\bibfnamefont {P.~W.}\ \bibnamefont {Shor}},\
  and\ \bibinfo {author} {\bibfnamefont {N.~J.~A.}\ \bibnamefont {Sloane}},\
  }\bibfield  {title} {\bibinfo {title} {Quantum error correction via codes over
  GF(4)},\ }\href {https://doi.org/10.1109/18.681315} {\bibfield  {journal}
  {\bibinfo  {journal} {IEEE Transactions on Information Theory}\ }\textbf
  {\bibinfo {volume} {44}},\ \bibinfo {pages} {1369} (\bibinfo {year}
  {1998})}\BibitemShut {NoStop}%
\bibitem [{\citenamefont {MacKay}(2003)}]{mackay2003information}%
  \BibitemOpen
  \bibfield  {author} {\bibinfo {author} {\bibfnamefont {D.~J.~C.}\ \bibnamefont {MacKay}},\ }\href {https://books.google.com/books?id=AKuMj4PN_EMC} {\emph
  {\bibinfo {title} {Information Theory, Inference, and Learning Algorithms}}}\
  (\bibinfo  {publisher} {Cambridge University Press},\ \bibinfo {year}
  {2003})\BibitemShut {NoStop}%
\bibitem [{\citenamefont {Breuckmann}\ and\ \citenamefont
  {Eberhardt}(2021)}]{Breuckmann_Eberhardt_PRX_2021}%
  \BibitemOpen
  \bibfield  {author} {\bibinfo {author} {\bibfnamefont {N.~P.}\ \bibnamefont
  {Breuckmann}}\ and\ \bibinfo {author} {\bibfnamefont {J.~N.}\ \bibnamefont
  {Eberhardt}},\ }\bibfield  {title} {\bibinfo {title} {Quantum low-density
  parity-check codes},\ }\href {https://doi.org/10.1103/PRXQuantum.2.040101}
  {\bibfield  {journal} {\bibinfo  {journal} {PRX Quantum}\ }\textbf {\bibinfo
  {volume} {2}},\ \bibinfo {pages} {040101} (\bibinfo {year}
  {2021})}\BibitemShut {NoStop}%
\bibitem [{\citenamefont {Dinur}\ \emph {et~al.}(2023)\citenamefont {Dinur},
  \citenamefont {Hsieh}, \citenamefont {Lin},\ and\ \citenamefont
  {Vidick}}]{DHLV_QLDPC_2022}%
  \BibitemOpen
  \bibfield  {author} {\bibinfo {author} {\bibfnamefont {I.}~\bibnamefont
  {Dinur}}, \bibinfo {author} {\bibfnamefont {M.-H.}\ \bibnamefont {Hsieh}},
  \bibinfo {author} {\bibfnamefont {T.-C.}\ \bibnamefont {Lin}},\ and\ \bibinfo
  {author} {\bibfnamefont {T.}~\bibnamefont {Vidick}},\ }\bibfield  {title} {\bibinfo {title} {Good quantum LDPC codes with linear time
  decoders},\ }\href {https://doi.org/10.1145/3564246.3585101} {in\ \emph
  {\bibinfo {booktitle} {Proceedings of the 55th Annual ACM Symposium on
  Theory of Computing (STOC 2023)}}\ (\bibinfo {year} {2023})\ pp.\ \bibinfo
  {pages} {905--918}}\BibitemShut {NoStop}%
\bibitem [{\citenamefont {Panteleev}\ and\ \citenamefont
  {Kalachev}(2022)}]{PKCode_2022}%
  \BibitemOpen
  \bibfield  {author} {\bibinfo {author} {\bibfnamefont {P.}~\bibnamefont
  {Panteleev}}\ and\ \bibinfo {author} {\bibfnamefont {G.}~\bibnamefont
  {Kalachev}},\ }\bibfield  {title} {\bibinfo {title} {Asymptotically good quantum and locally
  testable classical LDPC codes},\ }\href {https://doi.org/10.1145/3519935.3520017}
  {in\ \emph {\bibinfo {booktitle} {Proceedings of the 54th Annual ACM SIGACT
  Symposium on Theory of Computing (STOC 2022)}}\ (\bibinfo {year} {2022})\
  pp.\ \bibinfo {pages} {375--388}}\BibitemShut {NoStop}%
\bibitem [{\citenamefont {Leverrier}\ and\ \citenamefont
  {Zémor}(2022)}]{QTannerCode_2022}%
  \BibitemOpen
  \bibfield  {author} {\bibinfo {author} {\bibfnamefont {A.}~\bibnamefont
  {Leverrier}}\ and\ \bibinfo {author} {\bibfnamefont {G.}~\bibnamefont
  {Zémor}},\ }\bibfield  {title} {\bibinfo {title} {Quantum Tanner codes},\ }\href
  {https://doi.org/10.1109/FOCS54457.2022.00117} {in\ \emph {\bibinfo
  {booktitle} {2022 IEEE 63rd Annual Symposium on Foundations of Computer
  Science (FOCS)}}\ (\bibinfo {year} {2022})\ pp.\ \bibinfo {pages}
  {872--883}}\BibitemShut {NoStop}%
\bibitem [{\citenamefont {Gottesman}(2010)}]{gottesman2009introductionquantumerrorcorrection}%
  \BibitemOpen
  \bibfield  {author} {\bibinfo {author} {\bibfnamefont {D.}~\bibnamefont
  {Gottesman}},\ }\bibfield  {title} {\bibinfo {title} {An introduction to quantum error
  correction and fault-tolerant quantum computation},\ }\href
  {https://arxiv.org/abs/0904.2557} {in\ \emph {\bibinfo {booktitle} {Quantum
  Information Science and Its Contributions to Mathematics, Proceedings of
  Symposia in Applied Mathematics, Vol.~68}}\ (\bibinfo {publisher} {American
  Mathematical Society},\ \bibinfo {year} {2010})\ pp.\ \bibinfo {pages}
  {13--58}}\BibitemShut {NoStop}%
\bibitem [{\citenamefont {D\"{u}r}\ and\ \citenamefont
  {Briegel}(2007)}]{Dur_2007}%
  \BibitemOpen
  \bibfield  {author} {\bibinfo {author} {\bibfnamefont {W.}~\bibnamefont
  {D\"{u}r}}\ and\ \bibinfo {author} {\bibfnamefont {H.~J.}\ \bibnamefont
  {Briegel}},\ }\bibfield  {title} {\bibinfo {title} {Entanglement purification
  and quantum error correction},\ }\href
  {https://doi.org/10.1088/0034-4885/70/8/R03} {\bibfield  {journal} {\bibinfo
  {journal} {Reports on Progress in Physics}\ }\textbf {\bibinfo {volume}
  {70}},\ \bibinfo {pages} {1381} (\bibinfo {year} {2007})}\BibitemShut
  {NoStop}%
\bibitem [{\citenamefont {Hostens}\ \emph {et~al.}(2004)\citenamefont
  {Hostens}, \citenamefont {Dehaene},\ and\ \citenamefont {De Moor}}]{Hostens_2004}%
  \BibitemOpen
  \bibfield  {author} {\bibinfo {author} {\bibfnamefont {E.}~\bibnamefont
  {Hostens}}, \bibinfo {author} {\bibfnamefont {J.}~\bibnamefont {Dehaene}},\
  and\ \bibinfo {author} {\bibfnamefont {B.}~\bibnamefont {De Moor}},\
  }\bibfield  {title} {\bibinfo {title} {The equivalence of two approaches to
  the design of entanglement distillation protocols},\ }\href {https://arxiv.org/abs/quant-ph/0406017} {arXiv:quant-ph/0406017\ (\bibinfo {year} {2004})}\BibitemShut {NoStop}%
\bibitem [{\citenamefont {{Zang}}\ \emph {et~al.}(2025)\citenamefont {{Zang}},
  \citenamefont {{Chen}}, \citenamefont {{Chitambar}}, \citenamefont
  {{Suchara}},\ and\ \citenamefont {{Zhong}}}]{Zang_2024}%
  \BibitemOpen
  \bibfield  {author} {\bibinfo {author} {\bibfnamefont {A.}~\bibnamefont
  {{Zang}}}, \bibinfo {author} {\bibfnamefont {X.}~\bibnamefont {{Chen}}},
  \bibinfo {author} {\bibfnamefont {E.}~\bibnamefont {{Chitambar}}}, \bibinfo
  {author} {\bibfnamefont {M.}~\bibnamefont {{Suchara}}},\ and\ \bibinfo
  {author} {\bibfnamefont {T.}~\bibnamefont {{Zhong}}},\ }\bibfield  {title}
  {\bibinfo {title} {{No-Go Theorems for Universal Entanglement
  Purification}},\ }\href {https://doi.org/10.1103/PhysRevLett.134.190803} {\bibfield  {journal}
  {\bibinfo  {journal} {Phys. Rev. Lett.}\ }\textbf {\bibinfo {volume}
  {134}},\ \bibinfo {pages} {190803} (\bibinfo {year} {2025})}\BibitemShut
  {NoStop}%
\bibitem [{\citenamefont {Singh}\ \emph {et~al.}(2023)\citenamefont {Singh},
  \citenamefont {Bradley}, \citenamefont {Anand}, \citenamefont {Ramesh},
  \citenamefont {White},\ and\ \citenamefont {Bernien}}]{Singh_2023}%
  \BibitemOpen
  \bibfield  {author} {\bibinfo {author} {\bibfnamefont {K.}~\bibnamefont
  {Singh}}, \bibinfo {author} {\bibfnamefont {C.~E.}\ \bibnamefont {Bradley}},
  \bibinfo {author} {\bibfnamefont {S.}~\bibnamefont {Anand}}, \bibinfo
  {author} {\bibfnamefont {V.}~\bibnamefont {Ramesh}}, \bibinfo {author}
  {\bibfnamefont {R.}~\bibnamefont {White}},\ and\ \bibinfo {author}
  {\bibfnamefont {H.}~\bibnamefont {Bernien}},\ }\bibfield  {title} {\bibinfo
  {title} {Mid-circuit correction of correlated phase errors using an array of
  spectator qubits},\ }\href {https://doi.org/10.1126/science.ade5337}
  {\bibfield  {journal} {\bibinfo  {journal} {Science}\ }\textbf {\bibinfo
  {volume} {380}},\ \bibinfo {pages} {1265} (\bibinfo {year} {2023})} \BibitemShut
  {NoStop}%
\bibitem [{\citenamefont {Fang}\ \emph {et~al.}(2025)\citenamefont {Fang},
  \citenamefont {Miles}, \citenamefont {Goldwin}, \citenamefont {Lichtman},
  \citenamefont {Gillette}, \citenamefont {Bergdolt}, \citenamefont
  {Deshpande}, \citenamefont {Norrell}, \citenamefont {Huft}, \citenamefont
  {Kats},\ and\ \citenamefont {Saffman}}]{Fang_SciAdv_2025}%
  \BibitemOpen
  \bibfield  {author} {\bibinfo {author} {\bibfnamefont {C.}~\bibnamefont
  {Fang}}, \bibinfo {author} {\bibfnamefont {J.}~\bibnamefont {Miles}},
  \bibinfo {author} {\bibfnamefont {J.}~\bibnamefont {Goldwin}}, \bibinfo
  {author} {\bibfnamefont {M.}~\bibnamefont {Lichtman}}, \bibinfo {author}
  {\bibfnamefont {M.}~\bibnamefont {Gillette}}, \bibinfo {author}
  {\bibfnamefont {M.}~\bibnamefont {Bergdolt}}, \bibinfo {author}
  {\bibfnamefont {S.}~\bibnamefont {Deshpande}}, \bibinfo {author}
  {\bibfnamefont {S.~A.}\ \bibnamefont {Norrell}}, \bibinfo {author}
  {\bibfnamefont {P.}~\bibnamefont {Huft}}, \bibinfo {author} {\bibfnamefont
  {M.~A.}\ \bibnamefont {Kats}},\ and\ \bibinfo {author} {\bibfnamefont
  {M.}~\bibnamefont {Saffman}},\ }\bibfield  {title} {\bibinfo {title}
  {Interleaved dual-species arrays of single atoms using a passive optical
  element and one trapping laser},\ }\href
  {https://doi.org/10.1126/sciadv.adw4166} {\bibfield  {journal} {\bibinfo
  {journal} {Science Advances}\ }\textbf {\bibinfo {volume} {11}},\ \bibinfo
  {pages} {eadw4166} (\bibinfo {year} {2025})} \BibitemShut
  {NoStop}%
\bibitem [{\citenamefont {Guttridge}\ \emph {et~al.}(2025)\citenamefont
  {Guttridge}, \citenamefont {Hepworth}, \citenamefont {Ruttley}, \citenamefont
  {Durst}, \citenamefont {Eiles},\ and\ \citenamefont
  {Cornish}}]{Guttridge_PRL_2025}%
  \BibitemOpen
  \bibfield  {author} {\bibinfo {author} {\bibfnamefont {A.}~\bibnamefont
  {Guttridge}}, \bibinfo {author} {\bibfnamefont {T.~R.}\ \bibnamefont
  {Hepworth}}, \bibinfo {author} {\bibfnamefont {D.~K.}\ \bibnamefont
  {Ruttley}}, \bibinfo {author} {\bibfnamefont {A.~A.~T.}\ \bibnamefont
  {Durst}}, \bibinfo {author} {\bibfnamefont {M.~T.}\ \bibnamefont {Eiles}},\
  and\ \bibinfo {author} {\bibfnamefont {S.~L.}\ \bibnamefont {Cornish}},\
  }\bibfield  {title} {\bibinfo {title} {Individual assembly of two-species
  Rydberg molecules using optical tweezers},\ }\href
  {https://doi.org/10.1103/PhysRevLett.134.133401} {\bibfield  {journal}
  {\bibinfo  {journal} {Phys. Rev. Lett.}\ }\textbf {\bibinfo {volume} {134}},\
  \bibinfo {pages} {133401} (\bibinfo {year} {2025})}\BibitemShut {NoStop}%
\bibitem [{\citenamefont {Saffman}\ \emph {et~al.}(2010)\citenamefont
  {Saffman}, \citenamefont {Walker},\ and\ \citenamefont
  {M\o{}lmer}}]{Saffman2010}%
  \BibitemOpen
  \bibfield  {author} {\bibinfo {author} {\bibfnamefont {M.}~\bibnamefont
  {Saffman}}, \bibinfo {author} {\bibfnamefont {T.~G.}\ \bibnamefont
  {Walker}},\ and\ \bibinfo {author} {\bibfnamefont {K.}~\bibnamefont
  {M\o{}lmer}},\ }\bibfield  {title} {\bibinfo {title} {Quantum information
  with Rydberg atoms},\ }\href {https://doi.org/10.1103/RevModPhys.82.2313}
  {\bibfield  {journal} {\bibinfo  {journal} {Rev. Mod. Phys.}\ }\textbf
  {\bibinfo {volume} {82}},\ \bibinfo {pages} {2313} (\bibinfo {year}
  {2010})}\BibitemShut {NoStop}%
\bibitem [{\citenamefont {Saffman}\ \emph {et~al.}(2020)\citenamefont
  {Saffman}, \citenamefont {Beterov}, \citenamefont {Dalal}, \citenamefont
  {P\'aez},\ and\ \citenamefont {Sanders}}]{Saffman2020}%
  \BibitemOpen
  \bibfield  {author} {\bibinfo {author} {\bibfnamefont {M.}~\bibnamefont
  {Saffman}}, \bibinfo {author} {\bibfnamefont {I.~I.}\ \bibnamefont
  {Beterov}}, \bibinfo {author} {\bibfnamefont {A.}~\bibnamefont {Dalal}},
  \bibinfo {author} {\bibfnamefont {E.~J.}\ \bibnamefont {P\'aez}},\ and\
  \bibinfo {author} {\bibfnamefont {B.~C.}\ \bibnamefont {Sanders}},\
  }\bibfield  {title} {\bibinfo {title} {Symmetric Rydberg controlled-$Z$ gates
  with adiabatic pulses},\ }\href {https://doi.org/10.1103/PhysRevA.101.062309}
  {\bibfield  {journal} {\bibinfo  {journal} {Phys. Rev. A}\ }\textbf {\bibinfo
  {volume} {101}},\ \bibinfo {pages} {062309} (\bibinfo {year}
  {2020})}\BibitemShut {NoStop}%
\bibitem [{\citenamefont {Self}\ \emph {et~al.}(2024)\citenamefont {Self},
  \citenamefont {Benedetti},\ and\ \citenamefont {Amaro}}]{Self2024}%
  \BibitemOpen
  \bibfield  {author} {\bibinfo {author} {\bibfnamefont {C.~N.}\ \bibnamefont
  {Self}}, \bibinfo {author} {\bibfnamefont {M.}~\bibnamefont {Benedetti}},\
  and\ \bibinfo {author} {\bibfnamefont {D.}~\bibnamefont {Amaro}},\ }\bibfield
   {title} {\bibinfo {title} {Protecting expressive circuits with a quantum
  error detection code},\ }\href {https://doi.org/10.1038/s41567-023-02282-2}
  {\bibfield  {journal} {\bibinfo  {journal} {Nature Physics}\ }\textbf
  {\bibinfo {volume} {20}},\ \bibinfo {pages} {219} (\bibinfo {year}
  {2024})}\BibitemShut {NoStop}%
\bibitem [{\citenamefont {Leung}\ and\ \citenamefont
  {Shor}(2008)}]{Leung_Shor_2008}%
  \BibitemOpen
  \bibfield  {author} {\bibinfo {author} {\bibfnamefont {A.~W.}\ \bibnamefont
  {Leung}}\ and\ \bibinfo {author} {\bibfnamefont {P.~W.}\ \bibnamefont
  {Shor}},\ }\bibfield  {title} {\bibinfo {title} {Entanglement purification
  with two-way classical communication},\ }\href@noop {} {\bibfield  {journal}
  {\bibinfo  {journal} {Quantum Info. Comput.}\ }\textbf {\bibinfo {volume}
  {8}},\ \bibinfo {pages} {311–329} (\bibinfo {year} {2008})}\BibitemShut
  {NoStop}%
\bibitem [{foo()}]{footnote1}%
  \BibitemOpen
  \href@noop {} {\bibinfo {title} {For the qubit case, the isotropic state is
  locally equivalent to the Werner state by a Pauli operation.}}\BibitemShut
  {Stop}%
\bibitem [{\citenamefont {Baranes}\ \emph {et~al.}(2026)\citenamefont
  {Baranes}, \citenamefont {Cain}, \citenamefont {Ataides}, \citenamefont
  {Bluvstein}, \citenamefont {Sinclair}, \citenamefont {Vuletić}, \citenamefont
  {Zhou},\ and\ \citenamefont {Lukin}}]{baranes2025}%
  \BibitemOpen
  \bibfield  {author} {\bibinfo {author} {\bibfnamefont {G.}~\bibnamefont
  {Baranes}}, \bibinfo {author} {\bibfnamefont {M.}~\bibnamefont {Cain}},
  \bibinfo {author} {\bibfnamefont {J.~P.~B.}\ \bibnamefont {Ataides}},
  \bibinfo {author} {\bibfnamefont {D.}~\bibnamefont {Bluvstein}}, \bibinfo
  {author} {\bibfnamefont {J.}~\bibnamefont {Sinclair}}, \bibinfo {author}
  {\bibfnamefont {V.}~\bibnamefont {Vuletić}}, \bibinfo {author} {\bibfnamefont
  {H.}~\bibnamefont {Zhou}},\ and\ \bibinfo {author} {\bibfnamefont {M.~D.}\
  \bibnamefont {Lukin}},\ }\bibfield  {title} {\bibinfo {title} {Leveraging qubit loss detection in
  fault-tolerant quantum algorithms},\ }\href {https://doi.org/10.1103/ycwc-3myc}
  {\bibfield  {journal} {\bibinfo  {journal} {Phys. Rev. X}\ }\textbf
  {\bibinfo {volume} {16}},\ \bibinfo {pages} {011002} (\bibinfo {year}
  {2026})}\BibitemShut {NoStop}%
\bibitem [{\citenamefont {Perrin}\ \emph {et~al.}(2025)\citenamefont {Perrin},
  \citenamefont {Jandura},\ and\ \citenamefont
  {Pupillo}}]{perrin2025quantumerrorcorrectionresilient}%
  \BibitemOpen
  \bibfield  {author} {\bibinfo {author} {\bibfnamefont {H.}~\bibnamefont
  {Perrin}}, \bibinfo {author} {\bibfnamefont {S.}~\bibnamefont {Jandura}},\
  and\ \bibinfo {author} {\bibfnamefont {G.}~\bibnamefont {Pupillo}},\ }\bibfield  {title} {\bibinfo {title} {Quantum error correction resilient
  against atom loss},\ }\href {https://doi.org/10.22331/q-2025-10-13-1884}
  {\bibfield  {journal} {\bibinfo  {journal} {Quantum}\ }\textbf {\bibinfo
  {volume} {9}},\ \bibinfo {pages} {1884} (\bibinfo {year}
  {2025})}\BibitemShut {NoStop}%
\bibitem [{Wu\ \emph {et~al.}(2019)}]{WuStern2019}%
  \BibitemOpen
  \bibfield  {author} {\bibinfo {author} {\bibfnamefont {T.~Y.}\ \bibnamefont
  {Wu}}, \bibinfo {author} {\bibfnamefont {A.}\ \bibnamefont {Kumar}}, \bibinfo {author} {\bibfnamefont {F.}\ \bibnamefont {Giraldo}}, \ and\ \bibinfo {author} {\bibfnamefont {D.~S.}\ \bibnamefont {Weiss}},\ }\bibfield  {title} {\bibinfo {title} {Stern-Gerlach detection of neutral-atom qubits in a state-dependent optical lattice},\ }\href {https://doi.org/10.1038/s41567-019-0478-8}
  {\bibfield  {journal} {\bibinfo  {journal} {Nat.\ Phys.}\ }\textbf
  {\bibinfo {volume} {15}},\ \bibinfo {pages} {538} (\bibinfo {year}
  {2019})}\BibitemShut {NoStop}%
\bibitem [{\citenamefont {Rains}(2001)}]{Rains_2001}%
  \BibitemOpen
  \bibfield  {author} {\bibinfo {author} {\bibfnamefont {E.~M.}~\bibnamefont
  {Rains}},\ }\bibfield  {title} {\bibinfo {title} {A semidefinite program for
  distillable entanglement},\ }\href {https://doi.org/10.1109/18.959270}
  {\bibfield  {journal} {\bibinfo  {journal} {IEEE Transactions on Information
  Theory}\ }\textbf {\bibinfo {volume} {47}},\ \bibinfo {pages} {2921}
  (\bibinfo {year} {2001})}\BibitemShut {NoStop}%
\bibitem [{\citenamefont {Abdelhadi}\ \emph {et~al.}(2026)\citenamefont
  {Abdelhadi}, \citenamefont {Jochym-O'Connor}, \citenamefont {Siddhu},\ and\
  \citenamefont {Smolin}}]{AJOSS2025}%
  \BibitemOpen
  \bibfield  {author} {\bibinfo {author} {\bibfnamefont {D.}~\bibnamefont
  {Abdelhadi}}, \bibinfo {author} {\bibfnamefont {T.}~\bibnamefont
  {Jochym-O'Connor}}, \bibinfo {author} {\bibfnamefont {V.}~\bibnamefont
  {Siddhu}},\ and\ \bibinfo {author} {\bibfnamefont {J.}~\bibnamefont
  {Smolin}},\ }\bibfield  {title} {\bibinfo {title} {Adaptive channel reshaping for improved
  entanglement distillation},\ }\href {https://doi.org/10.1103/8kdf-37gp}
  {\bibfield  {journal} {\bibinfo  {journal} {Phys. Rev. Res.}\ }\textbf
  {\bibinfo {volume} {8}},\ \bibinfo {pages} {013018} (\bibinfo {year}
  {2026})}\BibitemShut {NoStop}%
\bibitem [{\citenamefont {Jiang}\ \emph {et~al.}(2007)\citenamefont {Jiang},
  \citenamefont {Taylor}, \citenamefont {S\o{}rensen},\ and\ \citenamefont
  {Lukin}}]{Jiang2007}%
  \BibitemOpen
  \bibfield  {author} {\bibinfo {author} {\bibfnamefont {L.}~\bibnamefont
  {Jiang}}, \bibinfo {author} {\bibfnamefont {J.~M.}\ \bibnamefont {Taylor}},
  \bibinfo {author} {\bibfnamefont {A.~S.}\ \bibnamefont {S\o{}rensen}},\ and\
  \bibinfo {author} {\bibfnamefont {M.~D.}\ \bibnamefont {Lukin}},\ }\bibfield
  {title} {\bibinfo {title} {Distributed quantum computation based on small
  quantum registers},\ }\href {https://doi.org/10.1103/PhysRevA.76.062323}
  {\bibfield  {journal} {\bibinfo  {journal} {Phys. Rev. A}\ }\textbf {\bibinfo
  {volume} {76}},\ \bibinfo {pages} {062323} (\bibinfo {year}
  {2007})}\BibitemShut {NoStop}%
\bibitem [{\citenamefont {Shi}\ \emph {et~al.}(2025)\citenamefont {Shi},
  \citenamefont {Patil},\ and\ \citenamefont {Guha}}]{Shi2025PRXQ}%
  \BibitemOpen
  \bibfield  {author} {\bibinfo {author} {\bibfnamefont {Y.}~\bibnamefont
  {Shi}}, \bibinfo {author} {\bibfnamefont {A.}~\bibnamefont {Patil}},\ and\
  \bibinfo {author} {\bibfnamefont {S.}~\bibnamefont {Guha}},\ }\bibfield
  {title} {\bibinfo {title} {Stabilizer entanglement distillation and efficient
  fault-tolerant encoders},\ }\href
  {https://doi.org/10.1103/PRXQuantum.6.010339} {\bibfield  {journal} {\bibinfo
   {journal} {PRX Quantum}\ }\textbf {\bibinfo {volume} {6}},\ \bibinfo {pages}
  {010339} (\bibinfo {year} {2025})}\BibitemShut {NoStop}%
\bibitem [{\citenamefont {Reimpell}\ and\ \citenamefont
  {Werner}(2005)}]{RW2005PRL}%
  \BibitemOpen
  \bibfield  {author} {\bibinfo {author} {\bibfnamefont {M.}~\bibnamefont
  {Reimpell}}\ and\ \bibinfo {author} {\bibfnamefont {R.~F.}\ \bibnamefont
  {Werner}},\ }\bibfield  {title} {\bibinfo {title} {Iterative optimization of
  quantum error correcting codes},\ }\href
  {https://doi.org/10.1103/PhysRevLett.94.080501} {\bibfield  {journal}
  {\bibinfo  {journal} {Phys. Rev. Lett.}\ }\textbf {\bibinfo {volume} {94}},\
  \bibinfo {pages} {080501} (\bibinfo {year} {2005})}\BibitemShut {NoStop}%
\bibitem [{Jang\ \emph {et~al.}(2025)}]{Jang2025}%
  \BibitemOpen
  \bibfield  {author} {\bibinfo {author} {\bibfnamefont {E.}\ \bibnamefont
  {Jang}}, \bibinfo {author} {\bibfnamefont {Y.}\ \bibnamefont {Kim}}, \bibinfo {author} {\bibfnamefont {H.}\ \bibnamefont {Kim}}, \bibinfo {author} {\bibfnamefont {S.}\ \bibnamefont {Choi}}, \bibinfo {author} {\bibfnamefont {Y.}\ \bibnamefont {Huang}}, \ and\ \bibinfo {author} {\bibfnamefont {W.~W.}\ \bibnamefont {Ro}},\ }\bibfield  {title} {\bibinfo {title} {Qubit Movement-Optimized Program Generation on Zoned Neutral Atom Processors},\ }\href {https://doi.org/10.1145/3696443.3708937}
  {in\ \emph{Proc.\ 23rd ACM/IEEE Int.\ Symp.\ on Code Generation and Optimization (CGO~'25)} (\bibinfo {year} {2025})\ pp.\ \bibinfo {pages} {459--475}}\BibitemShut {NoStop}%
\bibitem [{Lin\ \emph {et~al.}(2025)}]{Lin2024}%
  \BibitemOpen
  \bibfield  {author} {\bibinfo {author} {\bibfnamefont {W.-H.}\ \bibnamefont
  {Lin}}, \bibinfo {author} {\bibfnamefont {D.~B.}\ \bibnamefont {Tan}}, \ and\ \bibinfo {author} {\bibfnamefont {J.}\ \bibnamefont {Cong}},\ }\bibfield  {title} {\bibinfo {title} {Reuse-Aware Compilation for Zoned Quantum Architectures Based on Neutral Atoms},\ }\href {https://doi.org/10.1109/HPCA61900.2025.00021} {in\ \emph {\bibinfo
  {booktitle} {2025 IEEE International Symposium on High-Performance Computer
  Architecture (HPCA)}}\ (\bibinfo {year} {2025})\ pp.\ \bibinfo {pages}
  {127--142}}\BibitemShut {NoStop}%
\bibitem [{Bluvstein\ \emph {et~al.}(2024)}]{Bluvstein2024}%
  \BibitemOpen
  \bibfield  {author} {\bibinfo {author} {\bibfnamefont {D.}\ \bibnamefont
  {Bluvstein}}\ \emph{et~al.},\ }\bibfield  {title} {\bibinfo {title} {Logical quantum processor based on reconfigurable atom arrays},\ }\href {https://doi.org/10.1038/s41586-023-06927-3}
  {\bibfield  {journal} {\bibinfo  {journal} {Nature}\ }\textbf
  {\bibinfo {volume} {626}},\ \bibinfo {pages} {58} (\bibinfo {year}
  {2024})}\BibitemShut {NoStop}%
\bibitem [{\citenamefont {Cesa}\ and\ \citenamefont
  {Pichler}(2023)}]{Cesa2023PRL}%
  \BibitemOpen
  \bibfield  {author} {\bibinfo {author} {\bibfnamefont {F.}~\bibnamefont
  {Cesa}}\ and\ \bibinfo {author} {\bibfnamefont {H.}~\bibnamefont {Pichler}},\
  }\bibfield  {title} {\bibinfo {title} {Universal quantum computation in
  globally driven Rydberg atom arrays},\ }\href
  {https://doi.org/10.1103/PhysRevLett.131.170601} {\bibfield  {journal}
  {\bibinfo  {journal} {Phys. Rev. Lett.}\ }\textbf {\bibinfo {volume} {131}},\
  \bibinfo {pages} {170601} (\bibinfo {year} {2023})}\BibitemShut {NoStop}%
\bibitem [{\citenamefont {Petrosyan}\ \emph {et~al.}(2024)\citenamefont
  {Petrosyan}, \citenamefont {Norrell}, \citenamefont {Poole},\ and\
  \citenamefont {Saffman}}]{Norcia2024}%
  \BibitemOpen
  \bibfield  {author} {\bibinfo {author} {\bibfnamefont {D.}~\bibnamefont
  {Petrosyan}}, \bibinfo {author} {\bibfnamefont {S.}~\bibnamefont {Norrell}},
  \bibinfo {author} {\bibfnamefont {C.}~\bibnamefont {Poole}},\ and\ \bibinfo
  {author} {\bibfnamefont {M.}~\bibnamefont {Saffman}},\ }\bibfield  {title}
  {\bibinfo {title} {Fast measurements and multiqubit gates in dual-species
  atomic arrays},\ }\href {https://doi.org/10.1103/PhysRevA.110.042404}
  {\bibfield  {journal} {\bibinfo  {journal} {Phys. Rev. A}\ }\textbf
  {\bibinfo {volume} {110}},\ \bibinfo {pages} {042404} (\bibinfo {year}
  {2024})}\BibitemShut {NoStop}%
\bibitem [{Knottnerus\ \emph {et~al.}(2025)}]{Knottnerus2025}%
  \BibitemOpen
  \bibfield  {author} {\bibinfo {author} {\bibfnamefont {I.~H.~A.}\ \bibnamefont
  {Knottnerus}}, \bibinfo {author} {\bibfnamefont {Y.~C.}\ \bibnamefont {Tseng}}, \bibinfo {author} {\bibfnamefont {A.}\ \bibnamefont {Urech}}, \bibinfo {author} {\bibfnamefont {R.~J.~C.}\ \bibnamefont {Spreeuw}}, \ and\ \bibinfo {author} {\bibfnamefont {F.}\ \bibnamefont {Schreck}},\ }\bibfield  {title} {\bibinfo {title} {Parallel assembly of neutral atom arrays with an SLM using linear phase interpolation},\ }\href {https://doi.org/10.21468/SciPostPhys.19.4.118}
  {\bibfield  {journal} {\bibinfo  {journal} {SciPost Phys.}\ }\textbf
  {\bibinfo {volume} {19}},\ \bibinfo {pages} {118} (\bibinfo {year}
  {2025})}\BibitemShut {NoStop}%
\bibitem [{\citenamefont {Gu}\ \emph {et~al.}(2024)\citenamefont {Gu},
  \citenamefont {Tang}, \citenamefont {Caha}, \citenamefont {Choe},
  \citenamefont {He},\ and\ \citenamefont {Kubica}}]{Gu2024}%
  \BibitemOpen
  \bibfield  {author} {\bibinfo {author} {\bibfnamefont {S.}~\bibnamefont
  {Gu}}, \bibinfo {author} {\bibfnamefont {E.}~\bibnamefont {Tang}}, \bibinfo
  {author} {\bibfnamefont {L.}~\bibnamefont {Caha}}, \bibinfo {author}
  {\bibfnamefont {S.~H.}\ \bibnamefont {Choe}}, \bibinfo {author}
  {\bibfnamefont {Z.}~\bibnamefont {He}},\ and\ \bibinfo {author}
  {\bibfnamefont {A.}~\bibnamefont {Kubica}},\ }\bibfield  {title} {\bibinfo
  {title} {Single-shot decoding of good quantum LDPC codes},\ }\href
  {https://doi.org/10.1007/s00220-024-04951-6} {\bibfield  {journal} {\bibinfo
  {journal} {Communications in Mathematical Physics}\ }\textbf {\bibinfo
  {volume} {405}},\ \bibinfo {pages} {85} (\bibinfo {year} {2024})}\BibitemShut
  {NoStop}%
\bibitem [{Aschauer(2005)}]{Aschauer2005thesis}%
  \BibitemOpen
  \bibfield  {author} {\bibinfo {author} {\bibfnamefont {H.}\ \bibnamefont
  {Aschauer}},\ }\bibfield  {title} {\bibinfo {title} {Quantum communication in noisy environments},\ }\href {https://edoc.ub.uni-muenchen.de/3588/} {Ph.D. thesis},\ \bibinfo  {school} {Ludwig-Maximilians-Universit\"{a}t M\"{u}nchen} (\bibinfo {year} {2005})\BibitemShut {NoStop}%
\bibitem [{Jansen\ \emph {et~al.}(2022)}]{Jansen2022}%
  \BibitemOpen
  \bibfield  {author} {\bibinfo {author} {\bibfnamefont {S.}\ \bibnamefont
  {Jansen}}, \bibinfo {author} {\bibfnamefont {K.}\ \bibnamefont
  {Goodenough}}, \bibinfo {author} {\bibfnamefont {S.}\ \bibnamefont
  {de~Bone}}, \bibinfo {author} {\bibfnamefont {D.}\ \bibnamefont
  {Gijswijt}}, \ and\ \bibinfo {author} {\bibfnamefont {D.}\ \bibnamefont
  {Elkouss}},\ }\bibfield  {title} {\bibinfo {title} {Enumerating all bilocal Clifford distillation protocols through symmetry reduction},\ }\href {https://doi.org/10.22331/q-2022-05-19-715}
  {\bibfield  {journal} {\bibinfo  {journal} {Quantum}\ }\textbf
  {\bibinfo {volume} {6}},\ \bibinfo {pages} {715} (\bibinfo {year}
  {2022})}\BibitemShut {NoStop}%
\bibitem [{Krastanov\ \emph {et~al.}(2019)}]{Krastanov2019}%
  \BibitemOpen
  \bibfield  {author} {\bibinfo {author} {\bibfnamefont {S.}\ \bibnamefont
  {Krastanov}}, \bibinfo {author} {\bibfnamefont {V.~V.}\ \bibnamefont
  {Albert}}, \ and\ \bibinfo {author} {\bibfnamefont {L.}\ \bibnamefont
  {Jiang}},\ }\bibfield  {title} {\bibinfo {title} {Optimized entanglement purification},\ }\href {https://doi.org/10.22331/q-2019-02-18-123}
  {\bibfield  {journal} {\bibinfo  {journal} {Quantum}\ }\textbf
  {\bibinfo {volume} {3}},\ \bibinfo {pages} {123} (\bibinfo {year}
  {2019})}\BibitemShut {NoStop}%
\bibitem [{Miguel-Ramiro\ \emph {et~al.}(2025)}]{MiguelRamiro2025}%
  \BibitemOpen
  \bibfield  {author} {\bibinfo {author} {\bibfnamefont {J.}\ \bibnamefont
  {Miguel-Ramiro}}, \bibinfo {author} {\bibfnamefont {A.}\ \bibnamefont
  {Pirker}}, \ and\ \bibinfo {author} {\bibfnamefont {W.}\ \bibnamefont
  {D\"ur}},\ }\bibfield  {title} {\bibinfo {title} {Improving entanglement purification through coherent superposition of roles},\ }\href {https://doi.org/10.22331/q-2025-04-09-1702}
  {\bibfield  {journal} {\bibinfo  {journal} {Quantum}\ }\textbf
  {\bibinfo {volume} {9}},\ \bibinfo {pages} {1702} (\bibinfo {year}
  {2025})}\BibitemShut {NoStop}%
\bibitem [{\citenamefont {Stiebitz}\ \emph {et~al.}(2012)\citenamefont
  {Stiebitz}, \citenamefont {Scheide}, \citenamefont {Toft},\ and\
  \citenamefont {Favrholdt}}]{stiebitz2012graph}%
  \BibitemOpen
  \bibfield  {author} {\bibinfo {author} {\bibfnamefont {M.}~\bibnamefont
  {Stiebitz}}, \bibinfo {author} {\bibfnamefont {D.}~\bibnamefont {Scheide}},
  \bibinfo {author} {\bibfnamefont {B.}~\bibnamefont {Toft}},\ and\ \bibinfo
  {author} {\bibfnamefont {L.}~\bibnamefont {Favrholdt}},\ }\href
  {https://books.google.com.hk/books?id=GAduDt7c0A0C} {\emph {\bibinfo {title}
  {Graph Edge Coloring: Vizing's Theorem and Goldberg's Conjecture}}}\ (\bibinfo  {publisher} {Wiley},\ \bibinfo {year}
  {2012})\BibitemShut {NoStop}%
\bibitem [{\citenamefont {Bravyi}\ and\ \citenamefont
  {Maslov}(2021)}]{Bravyi_2021TIT}%
  \BibitemOpen
  \bibfield  {author} {\bibinfo {author} {\bibfnamefont {S.}~\bibnamefont
  {Bravyi}}\ and\ \bibinfo {author} {\bibfnamefont {D.}~\bibnamefont
  {Maslov}},\ }\bibfield  {title} {\bibinfo {title} {Hadamard-free circuits
  expose the structure of the Clifford group},\ }\href
  {https://doi.org/10.1109/TIT.2021.3081415} {\bibfield  {journal} {\bibinfo
  {journal} {IEEE Transactions on Information Theory}\ }\textbf {\bibinfo
  {volume} {67}},\ \bibinfo {pages} {4546} (\bibinfo {year}
  {2021})}\BibitemShut {NoStop}%
\bibitem [{\citenamefont {Hill}\ and\ \citenamefont
  {Wootters}(1997)}]{Concurrence1997PRL}%
  \BibitemOpen
  \bibfield  {author} {\bibinfo {author} {\bibfnamefont {S.}~\bibnamefont
  {Hill}}\ and\ \bibinfo {author} {\bibfnamefont {W.~K.}\ \bibnamefont
  {Wootters}},\ }\bibfield  {title} {\bibinfo {title} {Entanglement of a pair
  of quantum bits},\ }\href {https://doi.org/10.1103/PhysRevLett.78.5022}
  {\bibfield  {journal} {\bibinfo  {journal} {Phys. Rev. Lett.}\ }\textbf
  {\bibinfo {volume} {78}},\ \bibinfo {pages} {5022} (\bibinfo {year}
  {1997})}\BibitemShut {NoStop}%
\end{thebibliography}
\end{document}